\documentclass[mnsc]{informs3}
\usepackage{booktabs} % For formal tables

\usepackage{mathtools}
\usepackage{subfig}
\usepackage{float}

\OneAndAHalfSpacedXI
\usepackage{url}            % simple URL typesetting

\usepackage{tcolorbox}
\usepackage{bbm}
\usepackage{dsfont}
\usepackage{amsmath,amsfonts,amssymb,commath,hyperref}
\usepackage{natbib}
\usepackage{caption}
\usepackage[normalem]{ulem}  % in the preamble
\usepackage[colorinlistoftodos,prependcaption,textsize=small]{todonotes}
\usepackage[shortlabels]{enumitem}
\usepackage{xspace}
\usepackage[multiple]{footmisc}

\usepackage{algorithm}
\usepackage[noend]{algpseudocode}
\newcommand{\PP}[1]{\mathbb{P}\left[#1\right]}
\newcommand{\parenthesis}[1]{\left(#1\right)}
\newcommand{\bracket}[1]{\left\{#1\right\}}

\newcommand{\taucharge}{\tau}
\newcommand{\tauservice}{t}
\newcommand{\taurep}{t}
\newcommand{\tauidle}{\overline{t}}

\newcommand{\depotset}{\bracket{D}}
\newcommand{\rtd}{s}

\newcommand{\batteryvar}{\psi}
\newcommand{\instance}{I}
\newcommand{\instances}{\mathcal{I}}
\newcommand{\size}{S}

\newcommand{\soblivious}{\widetilde{s}}
\newcommand{\woblivious}{\widetilde{w}}
\newcommand{\OPT}{V^\star}
\newcommand{\voblivious}{V^{\mathrm{o}}}
\newcommand{\energyfunction}{\mathcal{E}}

\newcommand{\reduced}{RC}

\usepackage[capitalize]{cleveref}
\usepackage[suppress]{color-edits}
\addauthor{fa}{blue}
\addauthor{sb}{purple}
\addauthor{kz}{teal}
\addauthor{temp}{red}

\usepackage{natbib}
 \bibpunct[, ]{(}{)}{,}{a}{}{,}%
 \def\bibfont{\small}%

\TheoremsNumberedThrough     % Preferred (Theorem 1, Lemma 1, Theorem 2)
\ECRepeatTheorems

\EquationsNumberedThrough    % Default: (1), (2), ...
\MANUSCRIPTNO{}

\begin{document}
\RUNAUTHOR{}
\RUNTITLE{Stay, Reposition, or Recharge: Battery-Aware Operation of Robotaxi Fleets}  
\TITLE{Stay, Reposition, or Recharge: Battery-Aware Operation of Robotaxi Fleets}

\ARTICLEAUTHORS{%
\AUTHOR{Philipp Afèche}
\AFF{University of Toronto, Toronto, Canada
\\ \EMAIL{philipp.afeche@rotman.utoronto.ca}}%, \URL{}}
\AUTHOR{Saif Benjaafar}
\AFF{University of Michigan, Ann Arbor, MI\\ \EMAIL{saifb@umich.edu}}
\AUTHOR{Jiayu (Kamessi) Zhao}
\AFF{Stanford University, Stanford, CA\\ \EMAIL{kamessi@stanford.edu}}
% Enter all authors
} 

\ABSTRACT{Electric robotaxis are rapidly expanding from pilots to scaled commercial deployment. Unlike human-driven ride-hailing vehicles, they require centralized fleet control and careful management of vehicle state of charge (SoC). We study how a fleet manager should jointly make service, repositioning, idling, and charging decisions in such systems. We develop a fluid model in which vehicles evolve across locations and battery states, and show that optimal charge-aware operations admit simple threshold structures. At each location, vehicles with SoC below a location-specific threshold should return to the depot to recharge, while only those above it may reposition. Moreover, \emph{battery sensitivity}, a notion we introduce to capture the value of an additional battery level at a location, governs repositioning decisions: low-SoC vehicles should not be sent to more battery-sensitive destinations, nor high-SoC vehicles to less battery-sensitive ones. These structural features remain robust under destination-dependent fulfillment and a nonstationary environment. We then quantify the value of charge awareness: relative to a fully charge-aware policy, a charge-oblivious policy that ignores SoC in RTD and repositioning decisions can perform arbitrarily worse. A numerical case study calibrated using 2025 NYC Taxi and Limousine Commission trip records illustrates these thresholds and shows that charge-aware control can generate substantial gains in a realistic urban network. Importantly, a little charge-awareness goes a long way: simple threshold-based heuristics recover most of the value of the fully charge-aware solution.
}

\maketitle

% % Paper body

\section{Introduction}\label{sec:introduction}

Robotaxi fleets are moving rapidly from pilots to scaled commercial deployments. Waymo, the market leader in U.S. robotaxi operations, now dispatches fully autonomous rides across ten major metropolitan markets, including the San Francisco Bay Area, Los Angeles, and Phoenix \citep{apnews2026waymo10markets}, with plans to expand to more than 20 additional cities, including international markets such as Tokyo and London \citep{techcrunch2026waymo20cities}. Meanwhile, competitors such as Amazon’s Zoox are catching up to offer robotaxi service to the public. Amazon's Zoox has opened its robotaxi service to public riders in Las Vegas and has announced plans to expand service and testing in cities such as San Francisco, Austin, and Miami \citep{reuters2026zooxexpansion}. Tesla has also begun rolling out robotaxi services, including an initial launch in Austin and subsequent expansion to Dallas and Houston \citep{reuters2026teslacapex, reuters2026tesladallashouston}. As robotaxi fleets expand, fleet managers face a new class of operational problems. Unlike traditional ride-hailing platforms such as Uber and Lyft, which primarily rely on human drivers who make decentralized decisions about whether to wait, reposition, or leave the platform, robotaxi fleet managers centrally control the movement of their vehicles. The scale of empty miles driven by robotaxis highlights the importance of such decisions: California Public Utilities Commission data reports that, as of late 2025, Waymo vehicles spent 44\% of their driving time without passengers \citep{planetizen2025waymoempty}, which includes repositioning as well as driving to charging hubs. This makes effective fleet operations crucial to the scalability, service quality, and economic viability of robotaxi systems.

\begin{figure}[H]
    \centering
    \subfloat[\centering ]{{\includegraphics[width=0.43\textwidth]{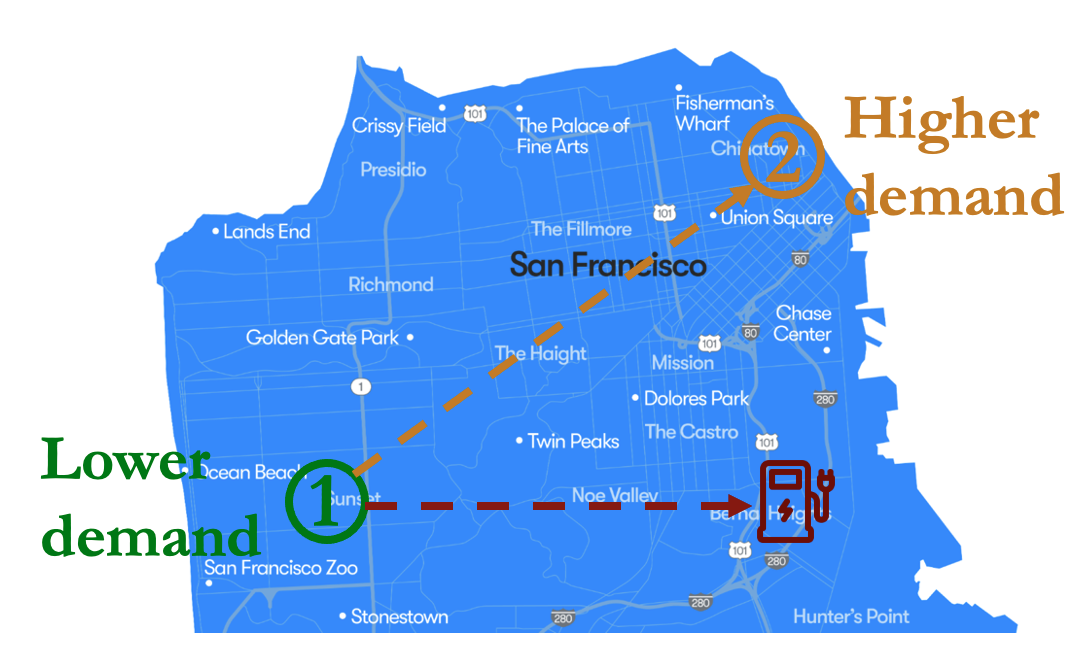}} }%
    \hspace{0.5cm}
    \subfloat[\centering ]    {{\includegraphics[width=0.4\textwidth]{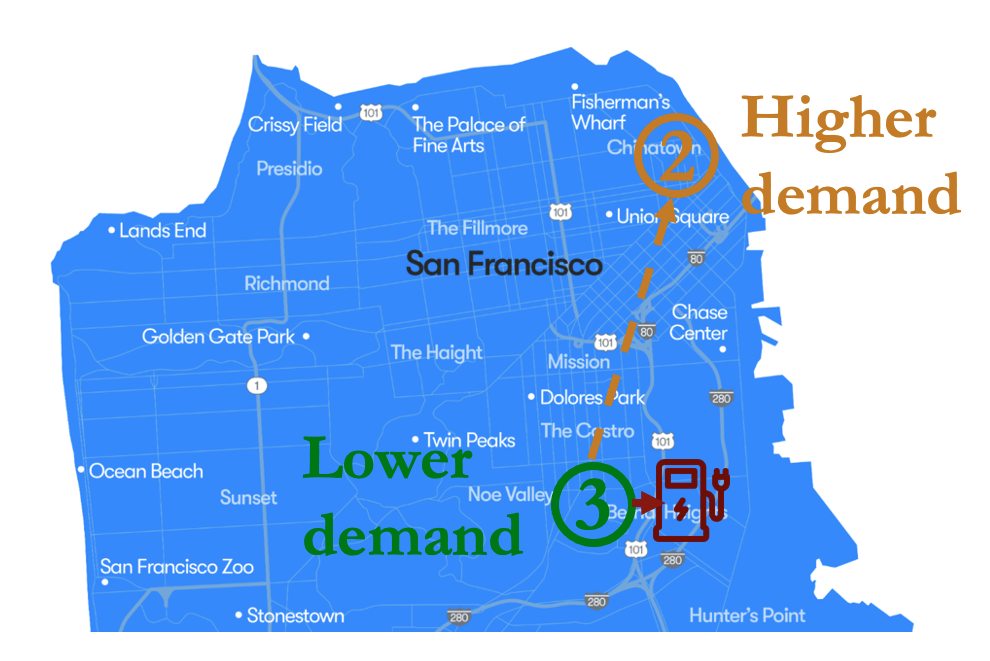} }}
    \caption{Illustration of charge-aware robotaxi operations in San Francisco. The charging icon marks Waymo's depot on Toland Street. Location~2, around the Financial District, is where demand is typically high. Locations~1 and~3, the Sunset District and Bernal Heights, both have lower demand but differ in their proximity to the depot \citep{sfchronicle2024waymo}. The dashed arrows show the two options for an empty vehicle: repositioning to location~2 or returning to the depot to recharge.}
    \label{fig:topology}
\end{figure}

Importantly, robotaxi fleets are electric. For instance, Waymo currently relies on the all-electric Jaguar I-PACE, among other electric models, and Zoox's purpose-built robotaxi is also fully electric \citep{waymo2023electricfleet,techcrunch2020zooxev}. When deployed in a robotaxi setting, these electric fleets require dedicated charging infrastructure, and fleet managers often rely on one or a few centralized depots. For example, Waymo's San Francisco operation has relied predominantly on a single depot in the southeastern part of the service region (see \cref{fig:topology}(a)), where vehicles are charged, cleaned, and repaired \citep{lee2024waymodepot}. The need to charge at centralized depots is a first-order operational consideration, as we illustrate through a simple back-of-the-envelope calculation: the Jaguar I-PACE used by Waymo has an EPA range of roughly 250 miles. When these vehicles operate continuously at urban speeds of 30--40 miles per hour (even before accounting for highway trips in markets such as the Bay Area or Phoenix), the battery can be depleted in about six hours. This implies that vehicles may need to visit depots multiple times a day, including during peak hours. Each depot visit involves not only charging but also regular cleaning, safety checks, and time to set up charging. Thus, even with DC fast chargers, bringing the battery back to 80\% can take on the order of one hour \citep{caranddriver2024ipace}. The significant frequency and duration of depot visits imply that vehicle charging needs are non-negligible, creating a fundamental tradeoff among service fulfillment, repositioning, idling, and charging. This tradeoff motivates the focus of this paper: charge-aware operations for robotaxi fleets. In particular, we show that the state of charge (SoC) of vehicles is a central operational state variable in robotaxi networks. 

The problem of how to optimally account for the SoC is not only important, but also nontrivial. In traditional ride-hailing platforms, if the platform wants to increase supply at say location~$2$, it may suffice to encourage some of the drivers currently at location~$1$ to move to~$2$. In an electric robotaxi fleet, however, the same repositioning decision depends critically on the vehicle's SoC. As illustrated in \cref{fig:topology}(a), %suppose that demand is higher at location~$2$ than at location~$1$. 
if a vehicle at location~$1$ has sufficiently high SoC, repositioning it from location~$1$ to location~$2$ may be worthwhile. A low-SoC vehicle, however, has only limited time left before it must return to the depot, possibly from a location with worse depot access, so the fleet manager may prefer to return it to the depot first and relaunch it to location~$2$. Depot proximity also plays a role: if a location is sufficiently close to the depot (see \cref{fig:topology}(b)), the fleet manager may want to take advantage of that proximity to recharge, even if the charge level is relatively high. Thus, the interactions among vehicle SoC, distance to the depot, demand patterns, and other factors make charge-aware robotaxi operations substantially more intricate than standard fleet repositioning. To capture these interactions, our model explicitly incorporates both spatial positioning and battery dynamics.

Despite its practical significance, charge-aware operations of depot-based robotaxi fleets remain unexplored in the literature. The classical repositioning literature studies how platforms reposition vehicles across locations, but does not involve charging considerations \citep{braverman2019empty, vr_benjaafar2022}. Conversely, the growing literature on electric vehicle sharing and electrified mobility focuses primarily on charging infrastructure and capacity planning; operational decisions, specifically, repositioning and recharging, ignore the vehicle SoC \citep{varma2023electric, benjaafar2025electrifying}. This paper bridges these two literature streams by studying how an AV fleet manager should make operational decisions in a charge-aware manner. By explicitly modeling the operational life cycle of an electric robotaxi fleet, including service, repositioning, idling, and charging at depots, we address three related questions. First, what are the structural features of an optimal repositioning policy? Second, under what conditions does charge awareness generate substantial operational value? Finally, how can these structural insights be translated into simple yet effective policies for managing robotaxi fleets?

\subsection{Main Results and Contributions}\label{sec:contributions}

We develop a fluid model of an electric robotaxi fleet that operates over a spatial network with general topology. Vehicles jointly evolve across locations and battery states. The model captures the core operational decisions facing a depot-based robotaxi fleet: whether vehicles should serve demand, reposition empty, idle, return to the depot for charging, or be relaunched after charging. This formulation allows us to study how SoC considerations change the nature of these vehicle movement decisions, and to derive interpretable policy structures that can be used to guide practical charge-aware control.

\textbf{Structure of charge-aware operations.}
We identify two features of charge-aware operational control that are distinct from classic repositioning models that treat vehicles as homogeneous units of capacity: early return-to-depot (RTD) and charge-aware repositioning. In particular,
the optimal solution admits simple threshold structures. In \cref{sec:early_rtd}, we study early RTD decisions and show that each location has a battery threshold such that vehicles return to the depot to recharge when their SoC is below it and never do so when their SoC is above it (\cref{thm:threshold_structure}). We further characterize how these thresholds depend on primitives such as depot access, demand, revenue, service time, and repositioning costs (\cref{prop:comparison}). In \cref{sec:reposition}, we study charge-aware repositioning. We first show that vehicles never reposition when their SoC is below the threshold identified in \cref{thm:threshold_structure} (\cref{thm:no_reposition}). Above this threshold, repositioning may be worthwhile, and the choice of destination depends on how much an additional battery level is worth at each destination. We capture this through the notion of \emph{battery sensitivity}, and characterize the primitives that determine the sensitivity, e.g., better depot access (\cref{prop:propagation_strong}). We then show that battery sensitivity governs repositioning decisions: low-SoC vehicles should not be sent to more battery-sensitive
destinations, nor high-SoC vehicles to less battery-sensitive ones (\cref{thm:threshold2}). We conduct a numerical case study using empirical data from \citet{nyctlc_trip_records} to illustrate these structural features in a realistic network (\cref{sec:thresholds_visualization,sec:battery_sensitivity}). Moreover, these structural results extend to fulfillment decisions and a nonstationary environment (\cref{sec:ext_fulfillment,sec:ext_nonstationary}).

\textbf{Value of charge awareness.}
We quantify the performance loss from ignoring vehicle SoC. In \cref{sec:importance}, we compare a fully charge-aware solution with a charge-oblivious policy that forbids early RTD and charge-aware repositioning. Using a minimal two-location instance needed to isolate the mechanism, we show that the performance gap can be unbounded (\cref{prop:stylized_monotonicity}). The gap is driven by a fundamental trade-off: a charge-oblivious policy either under-utilizes high-charge vehicles by letting them idle, or over-sends low-charge vehicles to locations that result in costly detours to the depot. We find that this inefficiency is amplified by asymmetric depot access, longer idling time, and limited fleet supply. Furthermore, this inefficiency remains unbounded even if 
only one of the two forms of charge awareness is precluded. In our numerical case study in \cref{sec:sensitivity}, we find that the fully charge-aware solution improves profit by at least $10.5\%$ relative to a charge-oblivious proxy. We further conduct sensitivity analyses to identify the market features that affect the value of charge awareness, including battery range, charging capacity, depot location, and the number of depots. Despite the large performance gap, one does not need full charge awareness to capture most of the gains. In \cref{sec:heuristics}, we leverage the structural results from \cref{sec:results} to design simple heuristics that close more than $80\%$ of the gap between the charge-oblivious proxy and the fully charge-aware solution.

\subsection{Related Work}\label{sec:rel_work}

Our paper connects three literature streams that study shared mobility operations along different dimensions: vehicle repositioning decisions, electrical vehicle infrastructure and operations, and various economic and operational implications of autonomous vehicles.

\textbf{Vehicle repositioning.}
The first stream studies vehicle repositioning in shared mobility and on-demand rental systems. \citet{vr_nair2011} provide an early stochastic optimization model for fleet management in vehicle-sharing systems. \citet{braverman2019empty} study empty-car routing in ride-hailing systems using a closed queueing network and establish steady-state convergence to a fluid limit. \citet{banerjee2022pricing} develop a unified approximation framework for pricing, matching, and rebalancing in shared vehicle systems, with explicit non-asymptotic performance guarantees for finite fleets. \citet{vr_benjaafar2022} formulate dynamic inventory repositioning in on-demand rental networks as a Markov decision process and establish structural properties of the optimal policy, including no-repositioning regions. A parallel line of work studies decentralized systems in which self-interested drivers decide whether and where to reposition, so that the platform steers supply only indirectly, e.g., through spatial pricing \citep{vr_bimpikis2019spatial} or demand admission control \citep{vr_afeche2023ridehailing}. Other papers enrich this centralized problem by accounting for spatial demand uncertainty \citep{vr_he2020robust}, jointly optimizing pricing and relocation \citep{vr_chen2024realtime}, and examining the environmental implications of strategic repositioning \citep{keskin2024green}. We refer readers to \citet{vr_zhang2023review} for a recent review. These papers share with ours a focus on spatial imbalance and repositioning. However, they do not incorporate charging: the decision of whether to reposition a vehicle depends on where it is, but not on its SoC.

A related branch of this literature studies centralized control and self-repositioning in autonomous or partially autonomous mobility systems. \citet{vr_pavone2012} propose a robotic solution to mobility-on-demand rebalancing and show that an optimal rebalancing policy can be obtained from a fluid linear program. \citet{vr_zhang2016} develop a queueing-theoretic model of autonomous mobility-on-demand systems and derive rebalancing controls for self-driving fleets. More recent work considers related centralized-control problems in mixed or autonomous mobility systems: \citet{benjaafar2025impact} and \citet{av_xie2023} study real-time mobility-on-demand management with mixed autonomy, coordinating centrally controlled AVs with decentralized human drivers, while \citet{tang2025spatial} examine how AV spatial flexibility affects fleet composition and spatial deployment in shared-mobility systems. These papers are close to ours in their emphasis on centralized control, spatial imbalance, and self-repositioning vehicles. The key difference is that they ignore the charging dimension: a repositioning decision changes not only the spatial distribution of vehicles but also their future charging opportunities and needs. This leads to different repositioning solution structures, such as early return-to-depot decisions and charge-dependent repositioning rules.

\textbf{Electric vehicle infrastructure and operations.}
The second stream studies electric-vehicle sharing and charging infrastructure. One branch focuses on planning questions, including where charging facilities should be located and how fleet size and charging capacity should be jointly dimensioned. For example, \citet{benjaafar2025electrifying} develop dimensioning rules for fleet size and the number of chargers at each hub using a closed queueing network, while \citet{paganini2022optimization} study the spatial deployment and sizing of charging facilities. Related work examines battery-swapping systems, including infrastructure deployment, battery inventory requirements, and implications for EV adoption and environmental performance \citep{mak2013infrastructure,avci2015electric,qi2023scaling}. None of these papers study charge-aware vehicle repositioning or recharging decisions.

The closest papers in this stream integrate charging with fleet operational decisions. \citet{varma2023electric} derive lower bounds on the fleet and charging capacity required to meet a target service level and develop a power-of-$d$ dispatch policy that yields nearly tight asymptotic bounds. Their analysis assumes that chargers are uniformly distributed across the service region, whereas we study how heterogeneous access to centralized charging depots shapes charge-aware control decisions. \citet{ev_he2017service,ev_he2021} respectively study service-region design and charging infrastructure planning while accounting for vehicle relocation and charging. However, there are important differences between these papers and ours. In particular, \citet{ev_he2017service} do not model vehicle-level SoC. \citet{ev_he2021} do consider SoC, but do not analytically characterize the structure of optimal control policies. A related literature develops optimization and learning algorithms for coordinating vehicle assignment, relocation, and charging in electric autonomous fleets \citep{ev_rossi2020,ev_boewing2020,ev_chenliu2023,ev_singhal2024}. These papers focus on developing computationally tractable methods for integrated system design and fleet control, whereas we analytically characterize solution structures, including location-specific return-to-depot thresholds and charge-aware repositioning rules, and quantify the value of these controls.

% The closest papers in this stream are those that integrate charging with fleet operational decisions. \citet{ev_he2017service, ev_he2021} respectively study service-region design and charging infrastructure planning while accounting for vehicle relocation. Relatedly, a computational literature develops optimization and learning algorithms for the joint assignment, relocation, and charging of electric autonomous fleets \citep{ev_rossi2020, ev_boewing2020, ev_chenliu2023, ev_singhal2024}. These works differ from ours in two respects. First, they assume that charging opportunities are decentralized: chargers are distributed across the service region or available at demand locations. When chargers are close to demand, charging is primarily a time-capacity tradeoff. In contrast, robotaxi fleets often rely on one or a few centralized depots, so charging also has a prominent spatial dimension: one must jointly decide where and when a vehicle leaves the service network for charging. Second, they focus on the computational tractability of the associated optimization problem rather than on structural analytical results. 

\textbf{Economic and operational implications of autonomous vehicles.}
A growing literature in operations management focuses on the economic and operational implications of AV technology. These include papers that study how competing platforms' access to AVs affects platform profits and the welfare of drivers and riders \citep{av_siddiq2022}, implications of AV market configuration on consumer welfare \citep{lian2024capturing}, ownership and contracting structures that shape AV deployment incentives \citep{av_freund2025}, and the interaction between human-driver recruitment and AV fleet decisions \citep{av_castro2024}. None of these papers models the spatial repositioning and charging of an AV fleet, which is the focus of our paper. Other related works in the space consider AV-enabled traffic-flow improvements \citep{baron2021introducing}, highway operations and lane-design policies for AV traffic \citep{mirzaeian2020queueing}, and remote-driver pooling as a human-in-the-loop form of automation \citep{benjaafar2024human}.
\section{Model and Problem Formulation}\label{sec:model}

We consider a mobility service that operates a fleet of electric robotaxis over a set of demand locations. Trip requests arrive continuously over time at each location and can be served by vehicles in the same location. Each trip takes a positive amount of time to complete. At the end of a trip, the fleet manager decides whether to keep the vehicle idle, reposition it to another location, or send it to the depot for charging. Because vehicles are electric, they are characterized not only by their locations, but also by their SoC levels. Vehicles consume energy as they serve trips or move across locations, and they must eventually return to the depot to recharge once their SoC becomes sufficiently low. After charging, vehicles can be relaunched from the depot to any location or remain parked there. Consequently, the fleet manager must continuously determine both, where vehicles should operate across the network and when they should be be sent to charge. The goal of the fleet manager is to make these decisions so as to maximize expected profit (revenue net of repositioning costs). We adopt a continuous-time deterministic fluid model to capture the dynamics of such a system, assuming that demand is stationary. Such models are widely used in the repositioning literature and provide a tractable approximation for stochastic systems with large fleets.\footnote{\citet{braverman2019empty} show that, in a closed queueing network model of empty-car routing, the policy induced by the fluid solution is asymptotically optimal in a large-market regime. One can similarly show that our fluid model provides a first-order description of the corresponding stochastic system; we omit the proof for brevity.} Under the fluid formulation, instead of tracking individual vehicles, the fleet manager chooses steady-state flow rates describing how vehicles transition across locations and charge states. We next introduce the model primitives, followed by the mathematical formulation of the flow optimization problem.

\subsection{Model Primitives}

\smallskip
The robotaxi fleet operates $\size$ vehicles over a set of demand locations $\mathcal N=\{1,\dots,n\}$ and a depot $D$ that fulfills vehicle charging needs (we consider a setting with multiple depots in \cref{sec:numerics}). 

\noindent\textbf{Demand.} Each location $i \in \mathcal N$ is characterized by an exogenous request arrival rate $\lambda_i \ge 0$ and destination probabilities $p_{ij}\in[0,1]$, with $p_{ii}>0$ and $\sum_{j} p_{ij}=1 \;\forall i\in\mathcal N$. %Following \citet{braverman2019empty}, we impose a mild regularity condition that $p_{ii}>0$ for all $i\in\mathcal N$  to avoid periodicity in the induced system. 
We assume that the demand routing matrix $(p_{ij})_{i,j\in\mathcal N}$ is irreducible. A served trip $(i,j) \in\mathcal N\times\mathcal N$ generates a revenue $r_{ij}\ge 0$, net of any operating costs. Repositioning an empty vehicle from $i$ to $j$ incurs a cost $c_{ij} > 0$ for $i, j \in \mathcal{N} \cup \depotset$. The constant $\tauservice_{ij}>0$ denotes both the trip time from location $i$ to location $j$ for trips involving passengers and for empty trips due to repositioning. We impose a standard triangle inequality assumption on repositioning costs and times: for any pairwise distinct $i,j,\ell \in \mathcal N \cup \depotset$, $c_{i\ell} < c_{ij} + c_{j\ell}$ and $\taurep_{i\ell} < \taurep_{ij} + \taurep_{j\ell}.$ Note that we treat prices and demand as exogenous to isolate the impact of the operational levers of vehicle repositioning on system performance. Such a model specification is consistent with specifications elsewhere in the literature; see for example \citep{braverman2019empty, vr_benjaafar2022}. We consider stationary system primitives to approximate operating periods during which market conditions are relatively stable, e.g., peak hours from 3--9PM, and we defer a discussion of the nonstationary environment to \cref{sec:ext_nonstationary}.
\smallskip

% \kzcomment{Clarify relationship between ``modes'' and ``states'' in the final pass}

\noindent \textbf{Vehicle operational modes.} At any point in time, a vehicle is in one of five modes: (i) serving a trip, (ii) repositioning, (iii) idle at a demand location, (iv) being charged at the depot or (v) parked at the depot. Modes (i)--(iii) reduce the SoC, whereas mode (iv) increases it and mode (v) leaves the SoC unchanged. 
\smallskip

\noindent \textbf{Vehicle SoC dynamics.} We model the evolution of vehicle SoC over time as follows. To operate on a finite state space, we discretize SoC into levels $k\in\mathcal K=\{0,1,\dots,K\}$, where a larger $k$ corresponds to a higher charge level. The state of a vehicle is given by $(i,k)$, which includes not only its location~$i$ but also its charge level $k$. Notice that $\mathcal K$ represents the operational SoC range rather than the physical battery limits. In practice, platforms may choose this range based on considerations such as battery health and service reliability. For example, a fleet manager may launch robotaxis at $80\%$ SoC rather than full charge (in which case $k=K$ corresponds to $80\%$ SoC), and may recall vehicles for charging at a safety threshold of $20\%$ SoC ($k=0$ corresponds to $20\%$ SoC) to avoid the risk of stranding vehicles with empty batteries. In this work, we take this operational SoC range as exogenously given after such considerations have been made.

For ease of exposition, we adopt the simplifying assumption that each task under modes (i)-(iii) reduces SoC by one level from $k$ to $k-1$.\footnote{We can extend our model to allow heterogeneous energy consumption across tasks; for example, a task of duration $t$ may reduce the SoC from $k$ to $k - \energyfunction(t)$, where $\energyfunction(\cdot)$ is non-decreasing and concave. We omit this parametrization for simplicity and discuss how our results extend to such a setting in Appendix~\ref{app:duration_energy}.} In practice, battery depletion may be slower while idling than while driving; our model naturally extends to such a setting by allowing a distinct time $\tauidle_{ii} > \taurep_{ii}$ for the battery to drop by one level while idling at $i$, under which all structural results in \cref{sec:results,sec:importance} continue to hold. At the depot, charging a vehicle (mode (iv)) incurs a fixed set-up time $\batteryvar_1>0$ and an additional time $\batteryvar_2\geq 0$ for each SoC level restored.\footnote{We assume for simplicity that the charging rate is uniform across the SoC range (see \citet{varma2023electric} for a similar treatment). Although charging speeds may vary near physical limits, they are roughly constant within the operational range of AVs (e.g., 20--80\% SoC). Appendix~\ref{app:nonlinear} examines nonlinear charging functions $\tau_k$ and confirms that our qualitative insights are robust.} For instance, a vehicle arriving at the depot with SoC $k<K$ requires time $\taucharge_k := \batteryvar_1+\batteryvar_2(K-k)$ to reach full charge $K$. We abstract from charging-capacity constraints in the base model and examine their implications in Appendix~\ref{app:depot-cap}. A vehicle can be dispatched (i.e., repositioned) from the depot to any demand location or remain parked at the depot (mode (v)), in which case no battery is consumed. Throughout, bold symbols denote the vector or matrix collections of the corresponding primitives. The tuple $\parenthesis{n, \size, K, \boldsymbol{\lambda},\boldsymbol{p},\boldsymbol{r},\boldsymbol{c},\boldsymbol{\tauservice},\boldsymbol{\batteryvar}}$ defines an instance $\instance$ of the model. \smallskip

\noindent \textbf{Fleet control decisions.} The fleet manager controls vehicle flow to maximize the profit per unit time. For each vehicle that arrives at location $i$ with SoC $k$, the fleet manager decides whether to serve a trip request, return to the depot (RTD) for charging, or reposition empty to another demand location. If none of these actions is taken, the vehicle remains idle at location $i$. For each vehicle at the depot, the fleet manager decides at what SoC level to stop charging and whether to dispatch it to a demand location or leave it parked at the depot.

\subsection{Problem Formulation}\label{sec:charge_aware}

%Under a fully charge-aware solution, control decisions depend explicitly on both location $i$ and charge level $k$. 
In a fluid model, deciding vehicle actions is equivalent to choosing flow rates for how vehicles transition across location–charge states. Let $f_{ik}\ge0$ denote the rate at which vehicles in state $(i,k)$ serve demand, for $i\in\mathcal N$ and $k\in\mathcal K$. Since customer destinations are distributed according to $p_{ij}$, the rate at which demand from location $i$ to $j$ is served with vehicles with SoC $k$ equals  $f_{ik}p_{ij}$.\footnote{In the main body of the paper, we allow fulfillment decisions to depend on the origin location and battery level, but not on the trip destination. This follows a convention in the repositioning literature and reflects the practical consideration that platforms typically want to avoid discriminating across trip destinations. In \cref{sec:ext_fulfillment}, we show that our main results are robust when fulfillment decisions are allowed to depend on destinations.} Let $\rtd_{ik}\ge 0$, for $i\in\mathcal N$ and $k\in\mathcal K$, denote the rate at which vehicles in state $(i,k)$ return to the depot for charging and let $w_{ijk}\ge 0$ denote the rate at which vehicles in state $(i,k)$ reposition to a different location $j\in\mathcal N$. The rest of the vehicles remain at location $i$. After time $\taurep_{ii}$, these vehicles transition to state $(i,k-1)$; the auxiliary variable $w_{iik}$ captures the rate at which this happens. Then, at the depot, $w_{Djk}\ge 0$ denotes the rate at which vehicles with SoC $k$ are launched to location $j \in \mathcal{N}$. The flow rate vectors $(\boldsymbol f,\boldsymbol w,\boldsymbol \rtd)$ therefore summarize the fleet manager's decision variables.

% ---------- Complete LP ----------
The fleet manager aims to maximize profit, i.e., revenue minus costs. Revenue is generated when vehicles fulfill demand. Costs are incurred when vehicles reposition among demand locations and the depot.\footnote{We do not explicitly model monetary charging cost; however, charging is not free in our model due to the opportunity cost of vehicle time (e.g., charging time $\tau_k$). As such, the structural results in~\cref{sec:results} naturally extend to a setting where there is an electricity cost to restore each battery level.} The fleet manager's problem can be formally stated as follows:
\begin{subequations}\label{lp:fluid}
\begin{align}
\max_{\boldsymbol f,\boldsymbol w,\boldsymbol{\rtd}}\quad
&
\sum_{i,j\in\mathcal N}\sum_{k \in \mathcal{K}} r_{ij}\,\big(f_{ik}\,p_{ij}\big)
\;-\;
\sum_{i\in\mathcal{N},\, j \in \mathcal{N}}\sum_{k \in \mathcal{K}} c_{ij}\,w_{ijk}
\;-\;
\sum_{j \in \mathcal{N}}\sum_{k \in \mathcal{K}} c_{Dj}\,w_{Djk}
\;-\;
\sum_{i\in\mathcal N}\sum_{k \in \mathcal{K}} c_{iD}\,\rtd_{ik}
\label{obj}
\\
\text{s.t.}\quad
&
\sum_{k\in \mathcal{K}} f_{ik}\;\le\;\lambda_{i},
\qquad\qquad\qquad\qquad\qquad\qquad\qquad\qquad\qquad\quad\;\; \forall\, i\in\mathcal N
\label{c:demand}
\\
&
\sum_{j\in\mathcal N}f_{j,k+1}\,p_{ji}+\sum_{j\in\mathcal N \cup \depotset}w_{ji,k+1}
=\;
f_{ik}
+\sum_{j\in\mathcal N}\!w_{ijk}
+\rtd_{ik},
\;\qquad\;\; \forall i\in\mathcal N,\; k\in\{0,\dots,K-1\}
\label{c:node_balance}
\\
&
w_{ij0}=w_{Dj0}=w_{jiK}=0,\quad f_{i0}=f_{iK}=0,\quad \rtd_{iK}=0,
\qquad\quad\;\;\;  \forall\, i,j \in\mathcal N
\label{eq:node_balance_boundary}
\\
&
\sum_{i\in\mathcal N} \sum_{k\in \mathcal{K}} \rtd_{ik}\;=\;\sum_{j\in\mathcal N} \sum_{k\in \mathcal{K}}w_{Djk}, &
\label{c:depot_map}
\\
&\sum_{i,j\in\mathcal N}\sum_{k\in \mathcal{K}} \big(f_{ik}\,p_{ij}\big)\,\tauservice_{ij}
+
\sum_{i \in \mathcal{N}, j \in \mathcal{N}}\sum_{k\in \mathcal{K}} w_{ijk}\,\taurep_{ij}
+
\sum_{j \in \mathcal{N}}\sum_{k\in \mathcal{K}} w_{Djk}\,\taurep_{Dj}
+
\sum_{i\in\mathcal N} \sum_{k\in \mathcal{K}} \rtd_{ik} \taurep_{iD} \notag \\
&  \qquad \qquad \qquad \qquad \;\; +
\batteryvar_1
\sum_{i\in\mathcal N}\sum_{k\in \mathcal{K}}
\rtd_{ik}
\;+\;
\batteryvar_2
\left(
\sum_{j\in\mathcal N}\sum_{k\in \mathcal{K}}
k\,w_{Djk}
-
\sum_{i\in\mathcal N}\sum_{k\in \mathcal{K}}
k\,\rtd_{ik}
\right)
\;\;\leq \size,
\label{c:vehicle_time}
\\
&
\boldsymbol f,\boldsymbol w,\boldsymbol{\rtd}\ge 0.
%\label{c:nonnegative}
\end{align}
\end{subequations}

We denote an optimal solution to \eqref{lp:fluid} by $(\boldsymbol f^\star,\boldsymbol w^\star,\boldsymbol{\rtd}^\star)$. When making reference to a particular problem instance $\instance$, we denote the optimal objective value of \eqref{lp:fluid} by $\OPT(\instance)$.

We now describe the constraints that characterize feasible solutions. Constraint \eqref{c:demand} requires that the fulfillment rate at each origin does not exceed the arrival rate. %Since $f_{ik}$ represents the fulfillment rate at state $(i,k)$, summing across charge levels yields the total fulfillment rate at~$i$.
Constraint \eqref{c:node_balance} imposes flow conservation across all non-depot location--charge states. For every $i\in\mathcal N$ and $k\in\{0,\dots,K-1\}$, total inflow into state $(i,k)$ must equal total outflow from that state. Inflows arise from trip completions and repositioning arrivals from level $k+1$, while outflows correspond to fulfillment, repositioning, and return-to-depot flows at level $k$. The boundary conditions in \eqref{eq:node_balance_boundary} imply that vehicles with empty charge do not serve demand, reposition among non-depot locations, or get launched from the depot, and that vehicles at the top charge level are only relaunched from the depot. Constraint \eqref{c:depot_map} imposes flow conservation at the depot, requiring that the total rate at which vehicles depart the depot equals the total rate at which vehicles arrive there for charging. We note that there always exists an optimal solution in which vehicles are launched only at full charge (see \cref{prop:launch_full_charge} in Appendix~\ref{app:proof_model}). We thus assume without loss of optimality that $w_{Djk}=0 \;\forall j\in\mathcal N, k<K$ throughout the remainder of the paper. Finally, constraint \eqref{c:vehicle_time} imposes the fleet-size constraint. By Little's law, the total vehicle time spent across all activities, computed as each activity's flow rate multiplied by its sojourn time, must not exceed the fleet size $\size$. This includes time spent serving trips, repositioning between or idling at non-depot locations, launching from or returning to the depot, and charging at the depot. Any slack in \eqref{c:vehicle_time} represents vehicles parked at the depot, which need not be modeled explicitly because parking is costless and unconstrained in the base model.

% % ---------- Complete LP ----------
% \smallskip
% \noindent\textbf{The optimization problem.} A feasible solution to the fluid model is a set of flow rates $(\boldsymbol f,\boldsymbol w, \boldsymbol{\rtd})$ that satisfies constraints \eqref{c:demand}--\eqref{c:vehicle_time}. Among all feasible solutions, the fleet manager seeks one that solves the following Linear Program (LP):
% \[
% \begin{aligned}
% \max_{\boldsymbol{f}, \boldsymbol{w}, \boldsymbol{\rtd}}\quad & \eqref{obj} \\
% \text{s.t.}\quad 
% & \eqref{c:demand},\ \eqref{c:node_balance},\ \eqref{eq:node_balance_boundary},\ \eqref{c:depot_map},\ \eqref{c:vehicle_time},\ \boldsymbol{f}, \boldsymbol{w}, \boldsymbol{\rtd} \geq 0.
% \end{aligned}
% \]
% We denote an optimal solution to the program above by $(\boldsymbol f^\star,\boldsymbol w^\star, \boldsymbol{\rtd^\star})$. Moreover, for instance $\instance$, we denote the optimal objective value of the program by $\OPT(\instance)$.

\section{Structure of an Optimal Solution}\label{sec:results}

In this section, we characterize the structure of an optimal solution to problem \eqref{lp:fluid}. We show that optimal operating decisions are \textit{charge-aware}, i.e., they depend on a vehicle's SoC. In the next two subsections, we study optimal RTD and repositioning decisions, respectively, and show that both exhibit simple threshold structures in vehicle SoC.

\subsection{The Structure of Optimal RTD}\label{sec:early_rtd}

We first examine when vehicles should return to the depot for charging. The following theorem characterizes this decision through a location-specific threshold in the vehicle's SoC.

\begin{theorem}
\label{thm:threshold_structure}
Consider any optimal solution $(\boldsymbol f^\star,\boldsymbol w^\star,\boldsymbol{\rtd}^\star)$. For each location $i\in\mathcal N$, there exists a threshold $\kappa_i^\ast\in\mathcal K$ such that vehicles do not return to the depot if their SoC $k>\kappa_i^\ast$, i.e., $\rtd^\ast_{ik}=0$ for all $k>\kappa_i^\ast$. If the optimal solution satisfies strict complementary slackness, then vehicles RTD with a positive rate at every SoC $k\leq\kappa_i^\ast$, i.e., $\rtd^\ast_{ik}>0$ for all $k\leq\kappa_i^\ast$.
\end{theorem}

\Cref{thm:threshold_structure} shows that RTD is confined to the lower end of the SoC range. The optimal location-specific thresholds balance the following considerations. On the one hand, returning a vehicle at high SoC forces it to recharge more frequently. Every depot visit is both costly and requires a charge-independent travel time and charge setup time, during which the vehicle is out of service. RTD at high SoC therefore reduces the effective fleet capacity. RTD at lower SoC protects against this loss. On the other hand, RTD at the minimum SoC ($k=0$), when vehicles \textit{must} recharge to restore availability, may be suboptimal, because the RTD travel cost and time are location-dependent. A vehicle with lower SoC has fewer trips left before it must RTD, possibly from a location where doing so is more costly or time-consuming. Therefore, RTD at higher SoC from a more ``favorable" location protects against RTD at lower SoC, from a potentially less favorable location.

\Cref{thm:threshold_structure} suggests (and numerical results confirm it) that it can be optimal to return a vehicle to the depot before its SoC reaches the terminal level $k=0$, a policy we call \textit{early RTD}. Formally, the proof compares, at each location~$i$ and SoC $k$, the continuation value of keeping a vehicle in operation with the value of RTD and resuming operation with full charge. We show that the continuation value increases sufficiently in the SoC so that RTD is less attractive at SoC levels above some threshold. The resulting location-specific threshold $\kappa_i^\ast$ provides a simple characterization of optimal RTD decisions. This threshold structure continues to hold when tasks involve heterogeneous battery consumption (see Appendix~\ref{app:duration_energy}).

We next examine how the optimal RTD thresholds vary across locations. The following proposition identifies the primitives that govern this threshold. Throughout, we say that two locations $i,i'\in\mathcal N$ are \emph{identical in all primitives} if every primitive remains unchanged when \(i\) is replaced by \(i'\) in any one of its location indices.

\begin{proposition}
\label{prop:comparison}
Consider two locations $i,i'\in\mathcal N$. The RTD thresholds satisfy
$\kappa_i^\ast\ge\kappa_{i'}^\ast$ if $i$ and $i'$ are identical in all
primitives except that one or more of the following conditions hold:
\begin{enumerate}[label=(\roman*)]
    \item (Cheaper RTD) $c_{iD}\le c_{i'D}$;
    \item (Faster RTD) $\tau_{iD}\le\tau_{i'D}$;
    \item (Lower local demand)
          $\lambda_i\le\lambda_{i'}$;
    \item (Lower trip revenue)
          $r_{ij}\le r_{i'j}$ for all $j\in\mathcal N$;
    \item (Longer travel times)
          $\tau_{ij}\ge\tau_{i'j}$ for all $j\in\mathcal N$; or
    \item (Higher repositioning costs)
          $c_{ij}\ge c_{i'j}$ for all $j\in\mathcal N$.
\end{enumerate}
\end{proposition}

A higher threshold $\kappa_i^\ast$ means that vehicles are sent back for charging earlier. \cref{prop:comparison} shows that locations for which returning to the depot is more convenient (either because doing so is cheaper or faster) or on-road opportunities are less valuable will have (weakly) higher RTD thresholds. Conversely, if a location involves more costly or time-consuming repositioning to the depot, or offers stronger demand or more profitable service opportunities, the fleet manager is more conservative about RTD and sets a lower threshold. These comparative statics follow from how the underlying primitives shift the tradeoff between the continuation and RTD values at a location. In \cref{sec:thresholds_visualization}, we show that early RTD is prevalent in our empirical setting and that the resulting thresholds exhibit the patterns predicted by \cref{prop:comparison}.

\subsection{The Structure of Optimal Repositioning}\label{sec:reposition}

We next turn to optimal repositioning decisions. We first show that one should not reposition vehicles from location $i$ with SoC below the threshold $\kappa_i^\ast$ introduced in \cref{thm:threshold_structure}.

\begin{theorem}\label{thm:no_reposition}
Consider any optimal solution $(\boldsymbol f^\star,\boldsymbol w^\star,\boldsymbol{\rtd}^\star).$ Vehicles at location $i$ do not reposition to other demand locations if their SoC $k<\kappa_i^\ast$, i.e., $w^\ast_{ijk}=0$ for all $i,j\in\mathcal N$ and $k<\kappa_i^\ast$.
\end{theorem}

The above result can be explained as follows. Below $\kappa_i^\ast$, an empty repositioning move consumes scarce remaining charge and may leave the vehicle at another demand location with little SoC remaining. Because such a low-SoC vehicle can generate only limited downstream value before it must return to the depot, the repositioning move does not justify its charge and time costs. The optimal solution therefore forbids these inefficient empty moves at low SoC.

Above the threshold $\kappa_i^\ast$, repositioning may be worthwhile, but \cref{thm:no_reposition} does not specify where vehicles should or should not be sent. The optimal destination choice may itself depend on SoC, because an additional unit of charge may create different amounts of downstream value at different locations. We capture this cross-location comparison using the notion of \emph{battery sensitivity}, which measures the marginal value of one additional battery level at a location.

\begin{definition}[Battery sensitivity]\label{def:battery_sensitivity}
Let $\pi^\ast_{ik}$ denote an optimal value of the dual variable for the flow-balance constraint \eqref{c:node_balance} at state $(i,k)$, representing the marginal value of a vehicle at location $i$ with SoC $k$.\footnote{When the optimal dual solution is not unique, we use the selection described in Appendix~\ref{app:dual}.} The \textit{battery sensitivity} at location $i\in\mathcal N$ and SoC $k\in\{1,\ldots,K-1\}$ is
\[
\Delta_{ik}\ :=\ \pi^\ast_{ik}-\pi^\ast_{i,k-1}.
\]
A larger $\Delta_{ik}$ means that an additional battery level creates more operational value at location $i$.
\end{definition}

We next identify what makes a location more or less battery-sensitive, and then show how the resulting battery-sensitivity order determines optimal repositioning decisions. For the rest of this section, we focus on locations $i$ and $i'$ at which the local vehicle supply falls short of demand, i.e., $\sum_{k\in \mathcal{K}} f^\star_{ik}\;<\;\lambda_{i}$ and $\sum_{k\in \mathcal{K}} f^\star_{i'k}\;<\;\lambda_{i'}$, since repositioning is relevant only to such locations.\footnote{Using the same triangle-inequality argument as in \cref{thm:no_reposition}, one can show that it is strictly suboptimal to reposition vehicles from some location $j$ to $i$ only to subsequently reposition them away from $i$.}

\begin{proposition}\label{prop:propagation_strong}
Consider two locations $i,i'\in\mathcal N$. Then, we have $\Delta_{ik}\le \Delta_{i'k}$ for all $k<K$, if $i$ and $i'$ are identical in all primitives except that RTD from $i$ is cheaper, i.e., $c_{iD}\le c_{i'D}$, and/or faster, i.e., $\taurep_{iD}\le \taurep_{i'D}$.
\end{proposition}

\Cref{prop:propagation_strong} shows that better depot access%
% (see, e.g., \cref{fig:charge_aware_repositioning})
, in the form of a cheaper or faster return-to-depot option, reduces a location's battery sensitivity, and that this ordering holds across the entire SoC range. Intuitively, when the depot is easy to access, vehicles can replenish their batteries easily, so one additional unit of charge creates only limited incremental value. In contrast, when returning to the depot is costly or time-consuming, an additional unit of charge allows the vehicle to postpone an expensive depot trip and stay productive in the service network for longer, implying a higher battery sensitivity.

Under additional vehicle action restrictions, \cref{prop:propagation} in Appendix~\ref{app:section_3_corollary} shows that a location's battery sensitivity also increases with stronger local demand, higher trip revenue, shorter service times, lower repositioning costs, or shorter repositioning times. We complement \Cref{prop:propagation_strong} and \cref{prop:propagation} with a data-driven case study in \cref{sec:battery_sensitivity}.

% \begin{figure}[h]
% \centering{{\includegraphics[width=0.4\textwidth]{charge_aware_repositioning.png} }}
%     \caption{Illustration of the charge-aware repositioning decision from location~$j$ to potential destinations~$i$ and~$i'$ (dashed arrows). The two destinations are the same except that $i$ is closer to the depot than $i'$. }
%     \label{fig:charge_aware_repositioning}
% \end{figure}

Having identified what makes a destination more or less battery-sensitive, we next show which destinations to avoid at different vehicle charge levels.

\begin{theorem}
\label{thm:threshold2}
In an optimal solution $(\boldsymbol{f}^\ast,\boldsymbol{w}^\ast, \boldsymbol{s}^\ast)$, consider locations $i,i'\in\mathcal N$ with $\Delta_{ik}<\Delta_{i'k}$ for all $k\in\{1,\ldots,K-1\}$.
Then, for every $j\in\mathcal N\setminus\bracket{i,i'}$, there exists a threshold
$\bar{\kappa}_j\in\mathcal{K}$ such that:
\begin{enumerate}[label=(\roman*)]
    \item vehicles do not reposition to $i'$ below the threshold, i.e., $w^\ast_{ji'k}=0\;\forall k<\bar\kappa_j$; and
    \item vehicles do not reposition to $i$ above the threshold, i.e., $w^\ast_{jik}=0\;\forall k>\bar\kappa_j$.
\end{enumerate}
\end{theorem}

\Cref{thm:threshold2} compares two locations $i$ and $i'$, where $i$ is less battery-sensitive than~$i'$ for all $k\in\{1,\ldots,K-1\}$. Intuitively, additional charge creates greater downstream value at the more battery-sensitive location $i'$. It is therefore inefficient to send a low-SoC vehicle to $i'$, where its limited charge cannot fully realize that value, or to send a high-SoC vehicle to the less battery-sensitive location $i$, where its additional charge is less valuable. In other words, protecting the operational value of charge requires matching vehicle SoC to the battery sensitivity of the destination. \Cref{thm:threshold2} formalizes this intuition through the threshold $\bar{\kappa}_j$, which determines which of the two destinations to avoid at different battery levels.

%\Cref{thm:threshold2} is established by comparing the marginal values of repositioning a vehicle from $j$ to~$i$ versus~$i'$ across SoC levels. 
The battery-sensitivity condition can be relaxed in two useful ways that we formalize in Appendix~\ref{app:section_3_corollary}. First, if the ordering is weak rather than strict, \cref{cor:weak_order_implication} yields a weaker monotonicity result that allows ties between destinations. Second, if the ordering holds only over a contiguous subset of SoC levels, \cref{cor:segment_single_crossing} establishes the corresponding destination threshold locally within that range.

\begin{remark}
In \cref{sec:ext_fulfillment}, we show that the structural results established in this section for repositioning trips, i.e., $\boldsymbol{w}$, naturally extend to served trips, i.e., $\boldsymbol{f}$. This follows because served trips can be interpreted as revenue-generating repositioning moves.

\end{remark}

Taken together, the results in this section show that optimal RTD and repositioning decisions are both charge-aware, yet can be characterized by simple threshold structures. As we demonstrate in \cref{sec:heuristics}, these structural results can be translated into simple and effective charge-aware heuristics. These results also suggest that policies that ignore the SoC may forgo substantial value, which we examine in the next section.

\section{The Value of Charge-Awareness}\label{sec:importance}

Since an optimal policy requires the fleet manager to condition operational decisions on a vehicle’s battery level — thereby introducing a more complex decision rule and additional operational effort—a natural question is how much value such sophistication actually brings. In this section, we distinguish policies by the extent to which decisions depend on battery level, %{\color{magenta} or state of charge,} 
 (SoC), a concept we refer to as \emph{charge-awareness}, and compare policies with different levels of charge-awareness.

The policy considered in Sections \ref{sec:model}-\ref{sec:results} is fully \emph{charge-aware}, as it allows the fleet manager to condition decisions freely on $(i,k)$, vehicle location and vehicle charge level. %enabling the fleet manager to tailor demand fulfillment and vehicle idling, repositioning, and charging decisions to the vehicle’s remaining charge. 
In \cref{sec:charge_oblivious}, we introduce a \textit{charge-oblivious} benchmark where decisions are agnostic of the charging levels. In \cref{sec:comparison}, we identify operational features that amplify the performance gain of the charge-aware over the charge-oblivious solution and show that the worst-case gain can be unbounded. Finally, in \cref{sec:partial} we isolate the value of exercising only early RTD or charge-aware repositioning, but not both. We highlight under what conditions each control matters most.

% \tr{Questions about organization and scope: \\
% (1) Separate theoretical from numerical analysis? They alternate in the current version.\\
% (2) Alternatively, a separate subsection for the ``partially'' oblivious policies (announced in intro) and presented in Corollaries 1 and 2.\\
% (3) Re. numerical analysis, let's agree what key points we want to highlight (I think the text should be accordingly more explicit). Some thoughts: (i) I think better highlighting the magnitude of thez value in relation to the parameters may be helpful. (ii) Confirming theoretical result of monotonicity should NOT be a goal. (iii) Not sure about the value of illustrating non-monotone effects. My sense is this discussion could be improved (e.g., to explain what is interesting/important, also in contrast to the monotone effects). (iv) Is there value to illustrating the value of RTD-only or repo-only charge-awareness, in comparison to "full" charge awareness?}  

\subsection{Benchmark: The Charge-Oblivious Policy}\label{sec:charge_oblivious}

We benchmark the fully charge-aware solution against a policy that suppresses the two charge-aware features identified in \cref{sec:results}.

\begin{definition}
\label{def:charge_oblivious}
A solution $(\boldsymbol f,\boldsymbol w,\boldsymbol{\rtd})$ is said to be \textit{charge-oblivious} if:
\begin{enumerate}[label=(\roman*)]
\item %\sout{(No early RTD)} 
Vehicles return to the depot only at the terminal state $k=0$, i.e.,
\begin{equation}\label{eq:obl_no_RTD}
\rtd_{ik} = 0,
\qquad \forall i\in\mathcal N,\; k=1,\dots,K-1,
\end{equation}
\item At each location $i$, the distribution of repositioning flows to destinations $j\in\mathcal N$ is independent of the charge level $k\geq 1$, i.e., there exist fractions $\{\woblivious_{ij},\; j\in\mathcal N\}$ such that \footnote{The ratio condition applies whenever $\sum_{l\in\mathcal N}w_{ilk}>0$; at a state with no such flow, it imposes no restriction.}
\begin{equation}\label{eq:obl_simplex}
\frac{w_{ijk}}{\sum_{l\in\mathcal N} w_{ilk}}=\woblivious_{ij},
\qquad
\woblivious_{ij}\ge 0,
\qquad
\sum_{j\in\mathcal N} \woblivious_{ij} = 1,
\qquad
\forall i,j\in\mathcal N,\; k\in \bracket{1,\ldots,K{-}1}.
\end{equation}
\end{enumerate}
\end{definition}
% Although the main body of the paper focuses primarily on empty-vehicle movements, it is natural to impose the same charge-oblivious restriction on service decisions. Accordingly, we require that, at each location $i$, the fraction of available vehicles assigned to service be independent of charge level. That is, there exists $\foblivious_i\in[0,1]$ such that
% \begin{equation}\label{eq:obl_induced}
% \frac{f_{ik}}{f_{ik}+\sum_{l\in\mathcal N} w_{ilk}}
% =\foblivious_i,
% \qquad
% \forall i\in\mathcal N,\; k\in \bracket{1,\ldots,K{-}1}.
% \end{equation}
The constraints %${\color{magenta} imposed by}} 
\mbox{\eqref{eq:obl_no_RTD}}--\mbox{\eqref{eq:obl_simplex}} restrict the feasible flow rates $(\boldsymbol f,\boldsymbol w,\boldsymbol \rtd)$. Adding these constraints to problem (1a-1f) yields the benchmark charge-oblivious flow optimization problem: 

% \kzedit{To isolate the value of charge-aware repositioning and RTD, we fix fulfillment at $\boldsymbol f=\boldsymbol f^\star$ from the fully charge-aware solution and optimize the repositioning and RTD flows $(\boldsymbol w,\boldsymbol\rtd)$ subject to \eqref{eq:obl_no_RTD}--\eqref{eq:obl_simplex}.} 

%\sout{\noindent\textbf{The charge-oblivious program.} %A \emph{charge-oblivious solution} is any set of flow rates $(\boldsymbol f,\boldsymbol w, \boldsymbol{\rtd})$ together with battery-independent flow rates $(\boldsymbol{\foblivious},\boldsymbol{\woblivious})$ that satisfy the feasibility constraints \eqref{c:demand}--\eqref{c:vehicle_time} and the charge-oblivious structures \eqref{eq:obl_no_RTD}--\eqref{eq:obl_simplex} 
%One can find an optimal charge-oblivious solution by solving:}
% Original program retained below.
\[
\begin{aligned}
\max_{\boldsymbol{f}, \boldsymbol{w},\boldsymbol{\rtd},\boldsymbol{\woblivious}}\quad & \eqref{obj} \\
\text{s.t.}\quad 
& \eqref{c:demand}-\eqref{c:vehicle_time},\
\eqref{eq:obl_no_RTD}-
\eqref{eq:obl_simplex},\ \boldsymbol{f}, \boldsymbol{w}, \boldsymbol{\rtd} \geq 0 .
\end{aligned}
\]
% \kzedit{\[
% \begin{aligned}
% \max_{\boldsymbol{w},\boldsymbol{\rtd},\boldsymbol{\woblivious}}\quad & \eqref{obj} \\
% \text{s.t.}\quad
% & \eqref{c:demand}-\eqref{c:vehicle_time},\
% \eqref{eq:obl_no_RTD}-\eqref{eq:obl_simplex},\\
% & \boldsymbol f=\boldsymbol f^\star,\quad
% \boldsymbol w,\boldsymbol\rtd\geq0.
% \end{aligned}
% \]}
We denote the optimal objective value of this benchmark by $\voblivious(\instance)$ for an instance $I$. This program is generally nonconvex because \eqref{eq:obl_simplex} is bilinear.

\subsection{Comparison between Charge-Aware and Charge-Oblivious Policies}\label{sec:comparison}

Since the charge-oblivious problem optimizes over a subset of the policies considered in the fully charge-aware problem, we have $\voblivious(\instance) \leq \OPT(\instance)$ for any instance $\instance$. The central questions are: (1) How large can the performance gain, $\OPT(\instance)/\voblivious(\instance),$ be? (2) What operational features contribute to this gain? To keep the analysis tractable, we focus on a family of instances with two demand locations and one depot. These are the simplest instances needed to highlight the drivers behind the value of charge-awareness. In \cref{sec:sensitivity}, we corroborate the analytical findings in this section with numerical results for a full network calibrated with empirical data.

\begin{figure}[h]
\centering
    \subfloat[\centering ]{{\includegraphics[width=0.4\textwidth]{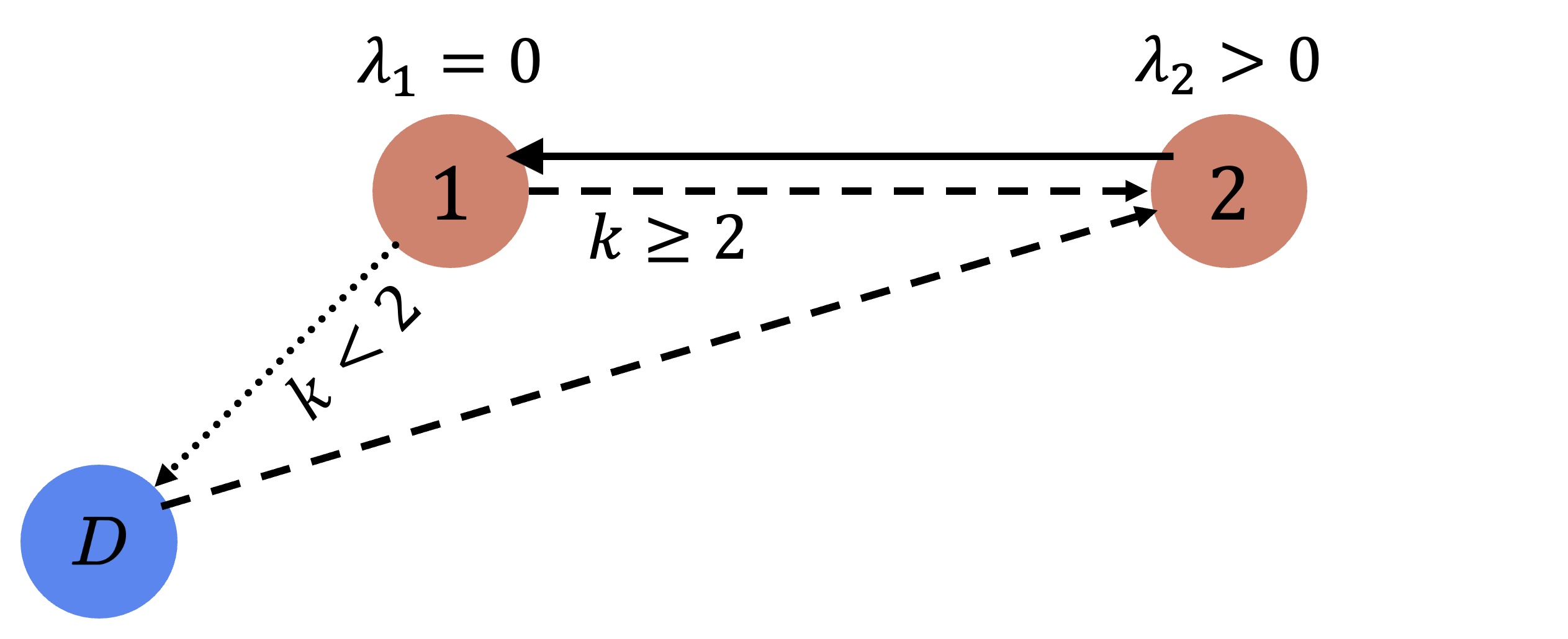}} }%
    \hspace{0.5cm}
    \subfloat[\centering ]    {{\includegraphics[width=0.48\textwidth]{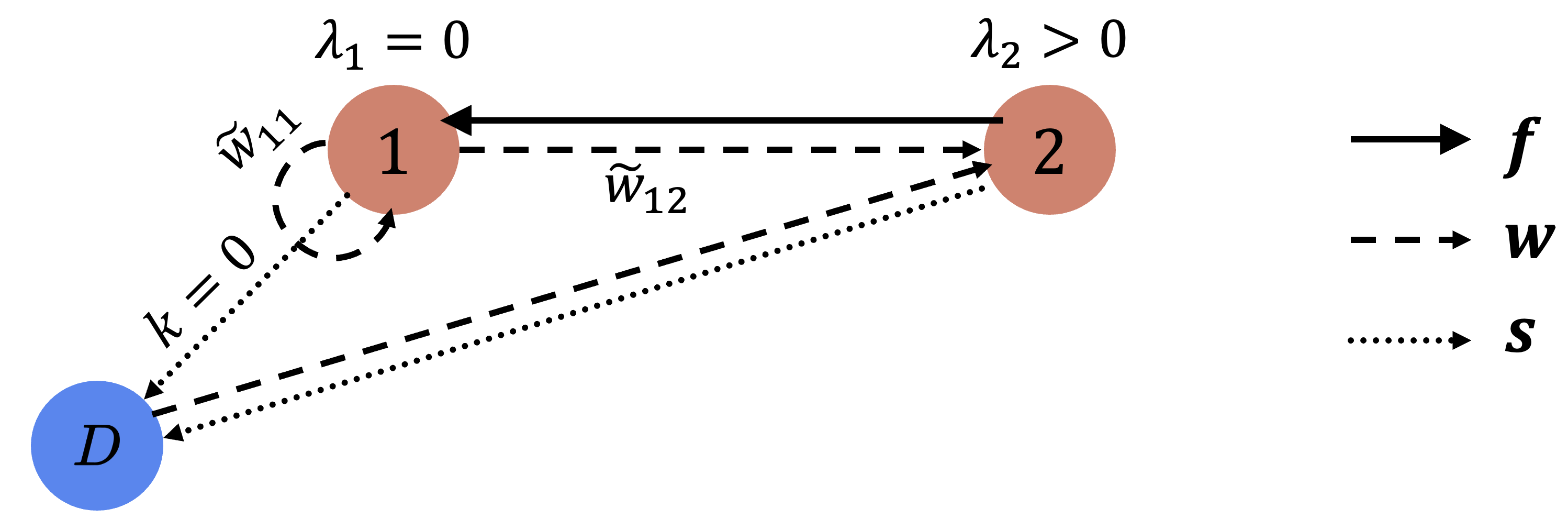} }}
    \caption{Family of instances with two locations. Flows of vehicles that serve demand $(\mathbf{f})$, reposition $(\mathbf{w})$ and RTD $(\mathbf{s})$ under charge-aware (left) and charge-oblivious (right) policy.}
    \label{fig:3location}
\end{figure}

\paragraph{A family of instances with two locations.} Consider a family of instances $\mathcal{I}$ with depot $D$ and $\mathcal{N} = \bracket{1,2}.$ All demand is for trips that originate at location $2$ and terminate at location $1$:%\\\tr{PA: Minor point, I think it is 'more logical' to have demand go from 1 to 2, not vice versa.}
\[
\lambda_1 = 0,\qquad\lambda_2>0,\qquad p_{21}=1.
\]
We use the shorthand notation $r := r_{21}>0$ for the trip revenue, $\tauidle := \taurep_{11} > 0$ for the charge-depletion time at location 1, and allow heterogeneous repositioning costs and times on the relevant arcs $(1,2), (2,D)$ and $(1,D)$; we omit for simplicity the arcs $D\to 1$ (launching to location~$1$) and $2\to 2$ (idling at location~$2$). We assume $K \geq 3$, to ensure vehicles can serve demand before RTD. We further assume that the charging time to full, denoted by $\tau_0$, is fixed and independent of the starting SoC, i.e., $\tau_k=\tau_0\;\forall k < K$. This implies the number of vehicles being charged grows linearly in RTD frequency. Therefore, this assumption magnifies the downside of early RTD and yields a conservative estimate of the value of charge awareness. %\kzedit{The latter is for ease of exposition and is conservative: if charging time decreases with the starting SoC, early RTD also saves charging time, so the value of charge awareness only increases.}%\tr{[PA:why fixed charging time?/briefly note why findings are robust?]} \kzcomment{added}

\cref{fig:3location} illustrates the operation of the charge-aware and charge-oblivious policies. Under optimal charge-aware operations, no vehicles RTD from location $2$; vehicles at location $1$ reposition to $2$ if their SoC is sufficiently high ($k \geq 2$) and RTD otherwise; and no vehicles idle at either demand location. A charge-oblivious policy hurts performance as it cannot follow these operational rules: 
first, \eqref{eq:obl_no_RTD} rules out RTD at positive SoC, so vehicles at location~$1$ with SoC $k\ge 1$ have only two options, idle at $1$ or reposition to $2$; second, \eqref{eq:obl_simplex} forces this decision to be charge-independent, with probabilities $\tilde{w}_{11}$ and $\tilde{w}_{12}=1-\tilde{w}_{11}$, respectively. Idling at 1 is suboptimal at every SoC, whereas  repositioning to $2$ at the minimal positive SoC ($k=1$) yields wasteful ``RTD with a detour'' because vehicles \textit{must} RTD upon arrival at $2$. The optimal charge-oblivious policy trades off these wasteful capacity uses in setting $\tilde{w}_{12}$, the probability of repositioning to $2$: raising $\tilde{w}_{12}$ increases idling at $1$ but decreases the risk of RTD with a detour via $2$.

\Cref{prop:stylized_monotonicity} highlights how key operational features magnify these wasteful capacity uses, and therefore, the value of charge-awareness.

\begin{proposition}\label{prop:stylized_monotonicity} For the family of instances $\mathcal{I}$, the performance gain ${\OPT(\instance)}/{\voblivious(\instance)}$ is (weakly)
%\begin{enumerate}[label=(\roman*)]
(i) increasing in the RTD cost and travel time from location~$2$, $c_{2D}$ and $\taurep_{2D}$, respectively; (ii) increasing in the charge-depletion time at location~$1$, $\tauidle$; and (iii) decreasing in fleet size, $S$.\\
Furthermore, the worst-case performance gain is unbounded, i.e.,  $\sup_{\instance \in \instances}{\OPT(\instance)}/{\voblivious(\instance)} = +\infty.$
\end{proposition}

A higher cost $c_{2D}$ and/or time $\taurep_{2D}$ (\cref{prop:stylized_monotonicity}(i)) magnify the loss from RTD with  detour via location $2$ for %\sout{from over-sending} 
low-SoC vehicles. %\sout{to location~$2$.} 
A higher charge depletion time, $\tauidle$,  (\cref{prop:stylized_monotonicity}(ii)), %\sout{by contrast,} 
magnifies the loss from idling %leaving 
high-SoC vehicles at location~$1$. Both capacity losses are especially harmful for service levels and revenues %\sout{consequential}
if the fleet has less slack capacity (\cref{prop:stylized_monotonicity}(iii)). %under when the fleet has less slack to absorb inefficient vehicle cycles  
In worst-case instances, these inefficiencies can be so severe that $\voblivious(\instance)=0$, whereas $\OPT(\instance)>0$, and therefore yield an unbounded performance gain of charge-awareness. \cref{prop:stylized_monotonicity}(i) also suggests more broadly that spatial asymmetry in depot access may harm the performance of a charge-oblivious policy. Indeed, as we show in \cref{cor:robustness} (Appendix~\ref{app:section_4_corollary}) for a 2-location network, this asymmetry alone can significantly harm performance, even under \textit{perfectly balanced demand}, i.e., with no need for repositioning.

\begin{figure}%[h]
    \centering
    \subfloat[\centering RTD cost asymmetry (for $\tauidle=3$)]{{\includegraphics[width=0.45\textwidth]{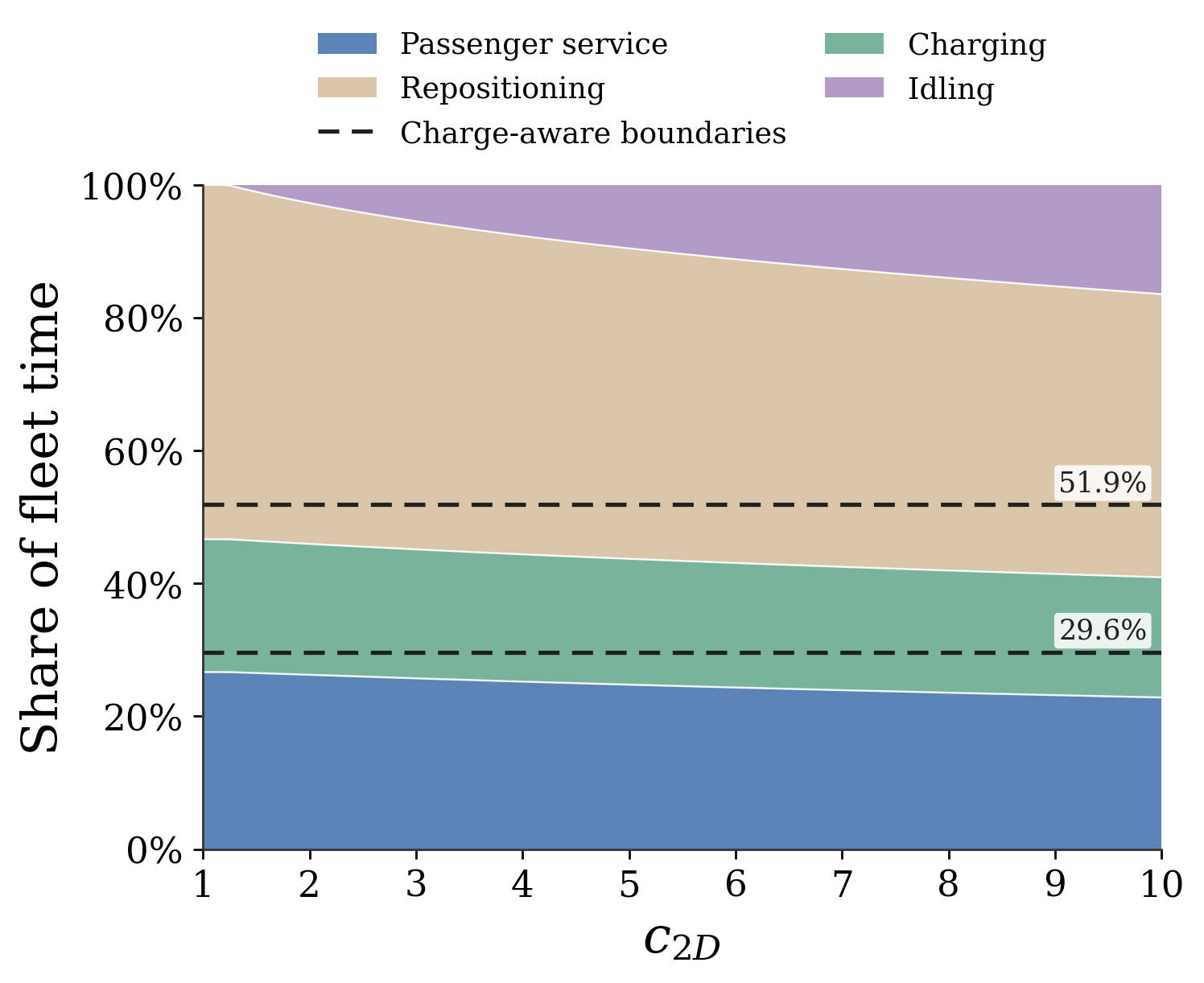}} }%
    \subfloat[\centering Charge-depletion time (for $c_{2D}=10$)]{{\includegraphics[width=0.45\textwidth]{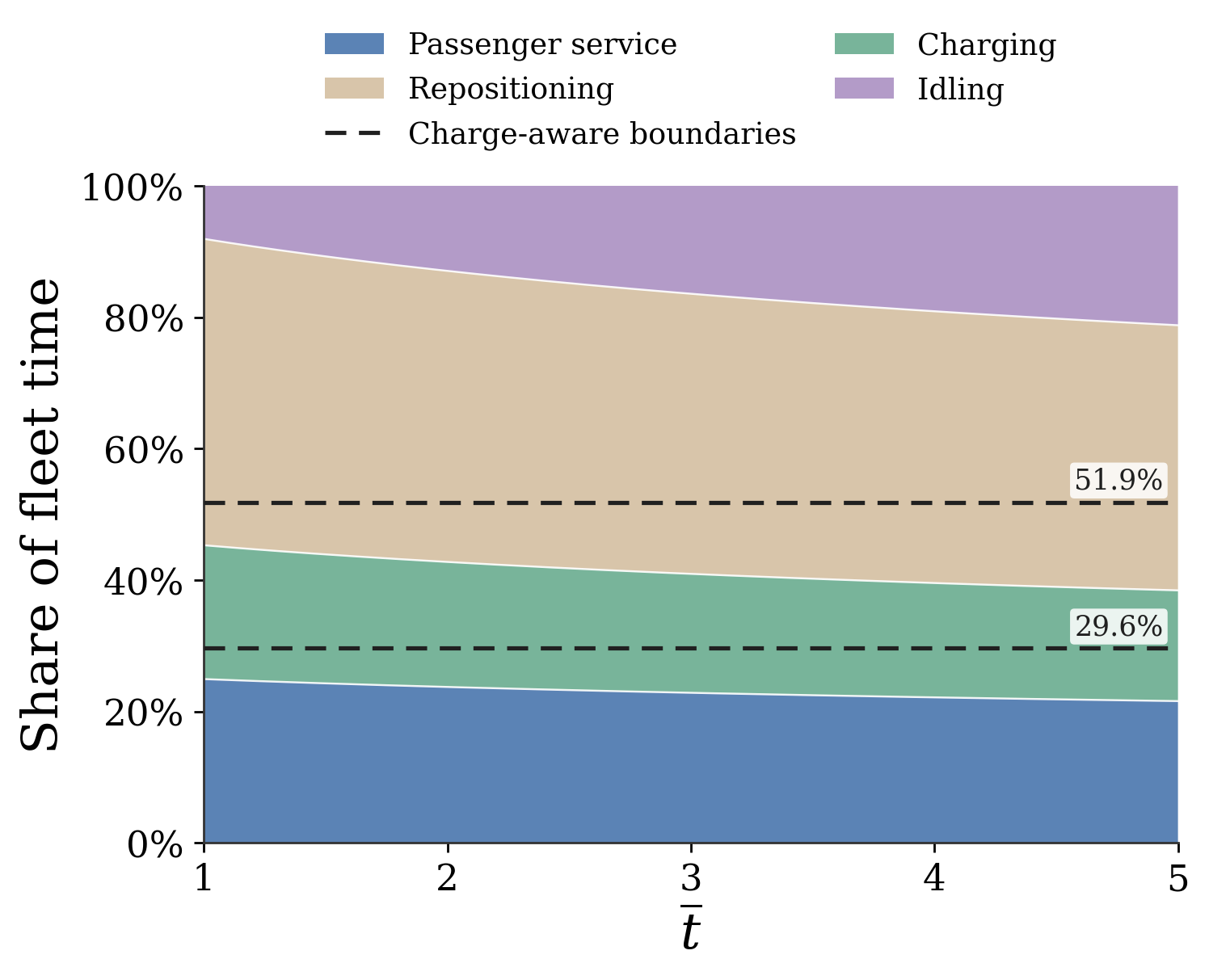}} }
    % \\
    % \subfloat[\centering Fleet size]{{\includegraphics[width=0.45\textwidth]{02_cost_by_fleet.pdf}} }
    % \subfloat[\centering Trip revenue]{{\includegraphics[width=0.45\textwidth]{03_cost_by_revenue.pdf}} }%
    \caption{Fleet utilization profile: for optimal charge-oblivious policy as a function of $c_{2D}$ and $\tauidle$; for optimal charge-aware policy (dashed lines), $29.6\%$ for passenger service and $51.9\%$ for service plus charging (independent of $c_{2D}$ or $\tauidle$).\\ $S=20$, $\lambda_2=10$, $r=5$, $K=9$, $\tau_0=3$, $c_{12}=c_{1D}=\taurep_{12}=\taurep_{21}=\taurep_{1D}=1$, $\taurep_{2D}=1.5$, $c_{D2}=\taurep_{D2}=2.5$}
    \label{fig:fleet_decomposition}
\end{figure}

We illustrate \cref{prop:stylized_monotonicity}(i) and (ii) with a numerical study. We vary $c_{2D}$ and $\tauidle$ as detailed below and fix $S=20$, $\lambda_2=10$, $r=5$, $K=9$, $\tau_0=3$, $c_{12}=c_{1D}=\taurep_{12}=\taurep_{21}=\taurep_{1D}=1$, $\taurep_{2D}=1.5$, and $\taurep_{D2}=2.5$. Under the optimal charge-aware policy, each vehicle completes cycles with a duration of $13.5$ time units between consecutive arrivals at the depot: $3$ units charging, $4$ units serving demand, $3$ units repositioning from $1$ to $2$, and $3.5$ units traveling from the depot and back. The fleet thus spends $29.6\%$ of its time serving passengers, and $S=20$ vehicles fulfill around $59.3\%$ of the demand.
In comparison, for $c_{2D}=2.5$ and $\bar{t}=3$ the optimal charge-oblivious policy ($\woblivious_{12}\approx 0.95$) completes on average fewer passenger trips ($3.98$) per cycle, over a longer cycle with mean length of $15.34$ time units, $13.6\%$ longer than under the charge-aware policy: vehicles idle at location~$1$ for $4.2\%$ of the time, and $81.8\%$ of charging cycles end with an RTD detour via location~$2$. Consequently, the fleet spends only $26.0\%$ of the time serving passengers and fulfills only $52.0\%$ of the demand (see \cref{fig:fleet_decomposition}(a)).

\begin{figure}%[h]
    \centering
    \subfloat[\centering RTD cost asymmetry]{{\includegraphics[width=0.45\textwidth]{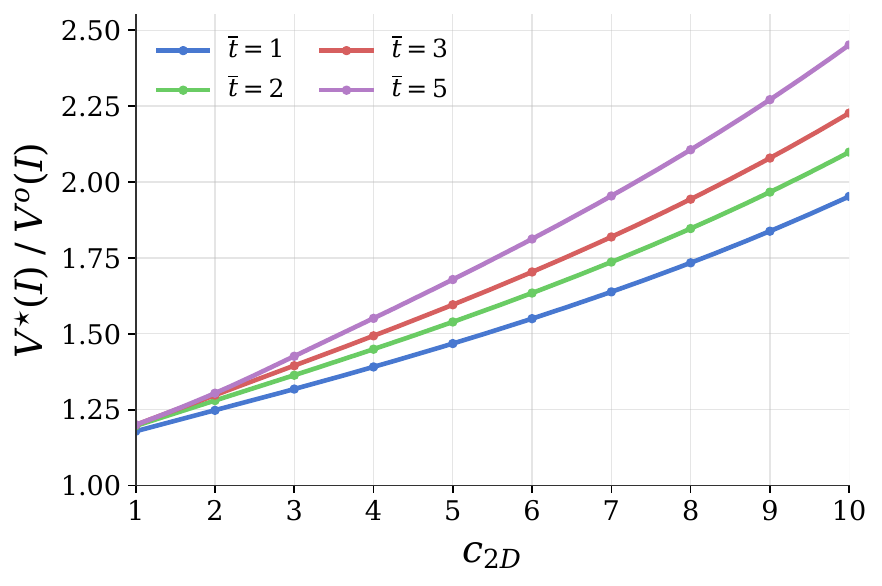}} }%
    \subfloat[\centering Charge-depletion time]{{\includegraphics[width=0.45\textwidth]{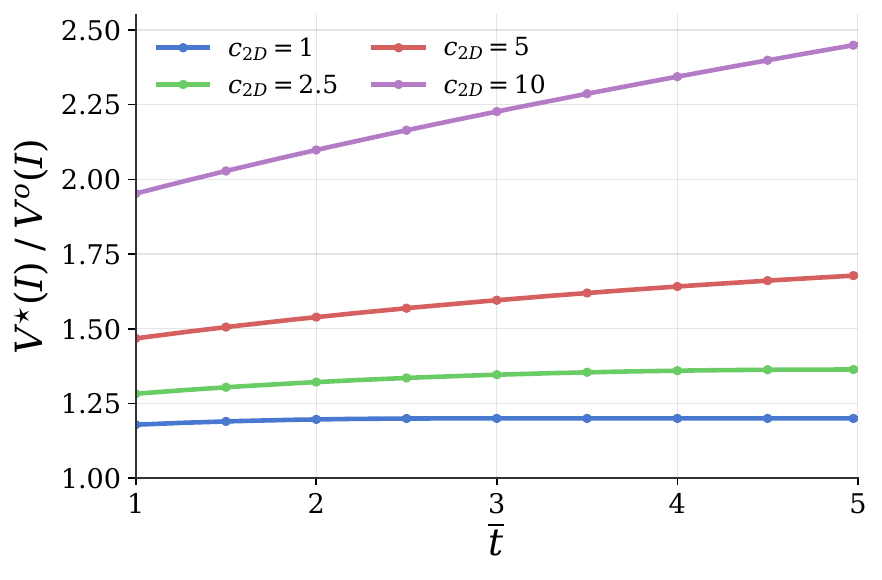}} }
    % \\
    % \subfloat[\centering Fleet size]{{\includegraphics[width=0.45\textwidth]{02_cost_by_fleet.pdf}} }
    % \subfloat[\centering Trip revenue]{{\includegraphics[width=0.45\textwidth]{03_cost_by_revenue.pdf}} }%
    \caption{\centering Performance gain ${\OPT(\instance)}/{\voblivious(\instance)}$ as a function of RTD cost asymmetry and charge-depletion time. \\ $S=20$, $\lambda_2=10$, $r=5$, $K=9$, $\tau_0=3$, $c_{12}=c_{1D}=\taurep_{12}=\taurep_{21}=\taurep_{1D}=1$, $\taurep_{2D}=1.5$, $c_{D2}=\taurep_{D2}=2.5$}
    \label{fig:stylized_monotonic}
\end{figure}

\cref{fig:fleet_decomposition} shows that the optimal charge-oblivious policy utilizes the fleet less efficiently, as $c_{2D}$ and/or $\tauidle$ increase: The time spent serving demand drops to $22.8\%$ and $21.6\%$, as shown in \cref{fig:fleet_decomposition}(a) and (b), respectively.
%compared with $29.6\%$ under the charge-aware policy.
\cref{fig:stylized_monotonic}(a)--(b) show how these inefficiencies due to costly depot access and longer charge-depletion time jointly amplify the performance gains of charge-awareness. %\tr{[PA: Personally I think we can skip the "plateau discussion" and end Section 4.2 HERE. The fact that charge-awareness keeps increasing and can DOUBLE the value, as Figure 4 shows, is making the key point. Since "unbounded" gain is (IMHO) more of a "pie in the sky" benefit (at reasonable parameter values, so there is no need to discuss it. If referees want to see that, fine.] \kzcomment{The following part was previously used to highlight the plateau. Under this more realistic set of instances, we only see a plateau in panel~(b) and not in (a), so I've abridged the discussion accordingly. Happy to truncate further if we feel that it adds little content.} \kzedit{In panel~(b), once the charge-depletion time $\tauidle$ becomes sufficiently long, the charge-oblivious solution repositions all vehicles to $2$ rather than idling them at $1$, so the performance gap plateaus; a larger depot asymmetry $c_{2D}$ raises the plateau because it makes repositioning costly as well. Panel~(a) shows a similar effect as one varies $c_{2D}$ under different $\tauidle$, although the plateau would arise much later.} While omitted for brevity, additional experiments show that smaller fleet size and lower trip revenue further raise the plateau.} 

\subsection{Partial Charge-Awareness}\label{sec:partial}

We next assess the value of exercising only one type of charge-aware control. To this end, we consider two partially charge-aware benchmarks. The first benchmark allows charge-aware repositioning only: we impose only the no-early-RTD constraint \eqref{eq:obl_no_RTD} and let $\voblivious_1(\instance)$ denote the optimal value of the resulting problem. The second benchmark allows early RTD only, i.e., vehicles may RTD at SoC $k>0$: we impose only the charge-independent repositioning constraint \eqref{eq:obl_simplex} and let $\voblivious_2(\instance)$ denote the optimal value.\footnote{Without \eqref{eq:obl_no_RTD}, we need to augment \eqref{eq:obl_simplex} so that the fraction of vehicles that RTD from each location is also independent of the SoC level, like the fractions that idle or reposition to each destination. %\kzcomment{we likely do not need the rest - keeping it here for our own understanding}%
Specifically, at each location $i\in\mathcal N$, there exist $\soblivious_i\ge 0$ and $\woblivious_{ij}\ge 0$, $j\in\mathcal N$, with $\soblivious_i+\sum_{j\in\mathcal N}\woblivious_{ij}=1$, such that $\rtd_{ik}=\soblivious_i\,(\rtd_{ik}+\sum_{l\in\mathcal N}w_{ilk})$ and $w_{ijk}=\woblivious_{ij}\,(\rtd_{ik}+\sum_{l\in\mathcal N}w_{ilk})$ for all $j\in\mathcal N$ and $k\in\bracket{1,\ldots,K{-}1}$.} We show that neither type of control alone bounds the worst-case performance loss relative to the fully charge-aware solution, but our numerical experiments indicate that either one can substantially reduce it.

% Previous statement (retained for reference):
% In the family of instances $\mathcal{I}$, 
% \begin{enumerate}
%     \item under only charge-aware repositioning, the performance gain ${\OPT(\instance)}/{\voblivious_1(\instance)}$ satisfies \cref{prop:stylized_monotonicity}(i)-(iii);
%     \item under only early RTD, the performance gain ${\OPT(\instance)}/{\voblivious_2(\instance)}$ satisfies \cref{prop:stylized_monotonicity}(i).
% \end{enumerate}
% Moreover, both worst-case performance gains are unbounded, i.e., $\sup_{\instance \in \instances}{\OPT(\instance)}/{\voblivious_1(\instance)} = \sup_{\instance \in \instances}{\OPT(\instance)}/{\voblivious_2(\instance)} = +\infty.$
\begin{corollary}%[Charge-Aware Repositioning]
\label{cor:robustness_reposition}\label{cor:robustness_rtd}
In the family of instances $\mathcal{I}$:
\begin{enumerate}[label=(\arabic*)]
    \item (Charge-aware repositioning only) ${\OPT(\instance)}/{\voblivious_1(\instance)}$ satisfies \cref{prop:stylized_monotonicity}(i)--(iii);
    \item (Early RTD only) ${\OPT(\instance)}/{\voblivious_2(\instance)}$ satisfies \cref{prop:stylized_monotonicity}(i).
\end{enumerate}
Furthermore, both worst-case gains are unbounded, i.e., $\sup_{\instance \in \instances}{\OPT(\instance)}/{\voblivious_\ell(\instance)} = +\infty$ for $\ell\in\bracket{1,2}$.
\end{corollary}

% This result shows that, even with charge-aware repositioning, the performance gap follows the same comparative-static patterns and remains unbounded in the worst case. Intuitively, even with charge-aware repositioning, a low-SoC vehicle at location $1$ still has only two options: reposition to $2$ or remain idle at $1$. Repositioning to $2$ may force the vehicle to make a costly return from location~$2$ to the depot, whereas idling wastes vehicle time.

Intuitively, each partially charge-aware benchmark lifts only one of the two restrictions \eqref{eq:obl_no_RTD}--\eqref{eq:obl_simplex}, and therefore still faces a tradeoff between the two wasteful capacity uses discussed in \cref{sec:comparison}. Under charge-aware repositioning only (Part~(1) of Corollary \ref{cor:robustness_rtd}), the fleet manager can target repositioning by SoC but cannot RTD for $k>0$: a low-SoC vehicle at location~$1$ must still either idle at $1$, wasting vehicle time, or reposition to $2$ and RTD with a detour. The gain ${\OPT(\instance)}/{\voblivious_1(\instance)}$ therefore exhibits the same comparative statics as in \cref{prop:stylized_monotonicity}. Under early RTD only (Part~(2) of Corollary \ref{cor:robustness_rtd}), vehicles at location~$1$ can RTD rather than idle, which weakens the impact of $\tauidle$. However, which vehicles reposition to $2$ versus RTD cannot depend on the SoC; therefore, repositioning a large fraction sends low-SoC vehicles to location~$2$, where they RTD with a detour, whereas returning a large fraction recalls high-SoC vehicles unnecessarily. The gain ${\OPT(\instance)}/{\voblivious_2(\instance)}$ therefore remains is increasing in $c_{2D}$ and $\taurep_{2D}$, as in \cref{prop:stylized_monotonicity}(i), but it may be non-monotone in the fleet size $S$ (unlike in \cref{prop:stylized_monotonicity}(iiii)) because early RTD also changes the length of vehicle cycles. Under either type of only partially charge-aware control, the wasteful use of capacity can be severe enough that the worst-case gain is unbounded.

% \kzdelete{We next relax the no-early-RTD restriction \eqref{eq:obl_no_RTD} while keeping \eqref{eq:obl_simplex} in the charge-oblivious benchmark. Let $\voblivious_2(\instance)$ denote the optimal value of the resulting relaxed program.}

% \kzdelete{\begin{corollary}\label{cor:robustness_rtd}
% Under early RTD but not charge-aware repositioning, the performance gap ${\OPT(\instance)}/{\voblivious_2(\instance)}$ satisfies the properties in \cref{prop:stylized_monotonicity}(i)-(ii). Moreover, the worst-case performance gap $\sup_{\instance \in \instances}{\OPT(\instance)}/{\voblivious_2(\instance)} = +\infty.$
% \end{corollary}}

% \kzdelete{\Cref{cor:robustness_rtd} characterizes the performance gap under early RTD alone. Because early RTD allows low-SoC vehicles to return to the depot rather than idle at location~$1$ (the latter is always suboptimal; see \cref{lem:pi-monotone}(iii)), $\tauidle$ no longer affects $\voblivious_2(\instance)$. Allowing early RTD also changes vehicle-cycle lengths, so the performance gap need not vary monotonically with fleet size $S$; by contrast, it remains weakly increasing in $c_{2D}$ and $\taurep_{2D}$. Importantly, early RTD alone does not resolve the dilemma at location~$1$: repositioning a large fraction of vehicles to location~$2$ risks sending low-SoC vehicles there too; conversely, returning a large fraction of vehicles to the depot may unnecessarily recall high-SoC ones.} %This dilemma again generates an unbounded worst-case performance gap.

\begin{figure}%[h]
    \centering
    \subfloat[\centering RTD cost asymmetry (for $\tauidle=3$)]{{\includegraphics[width=0.45\textwidth]{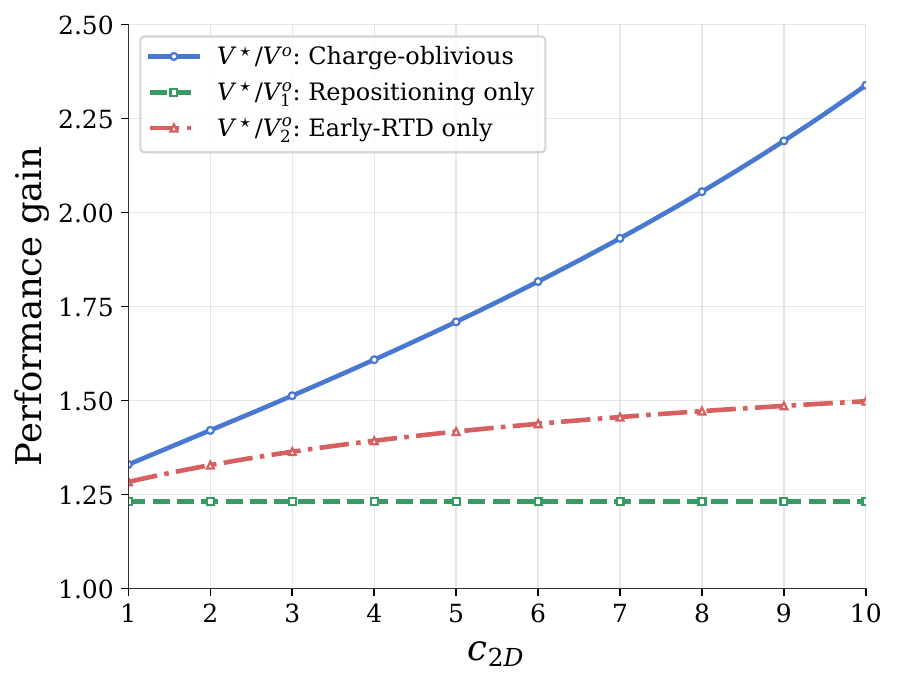}} }%
    \subfloat[\centering Charge-depletion time (for $c_{2D}=2$)]{{\includegraphics[width=0.45\textwidth]{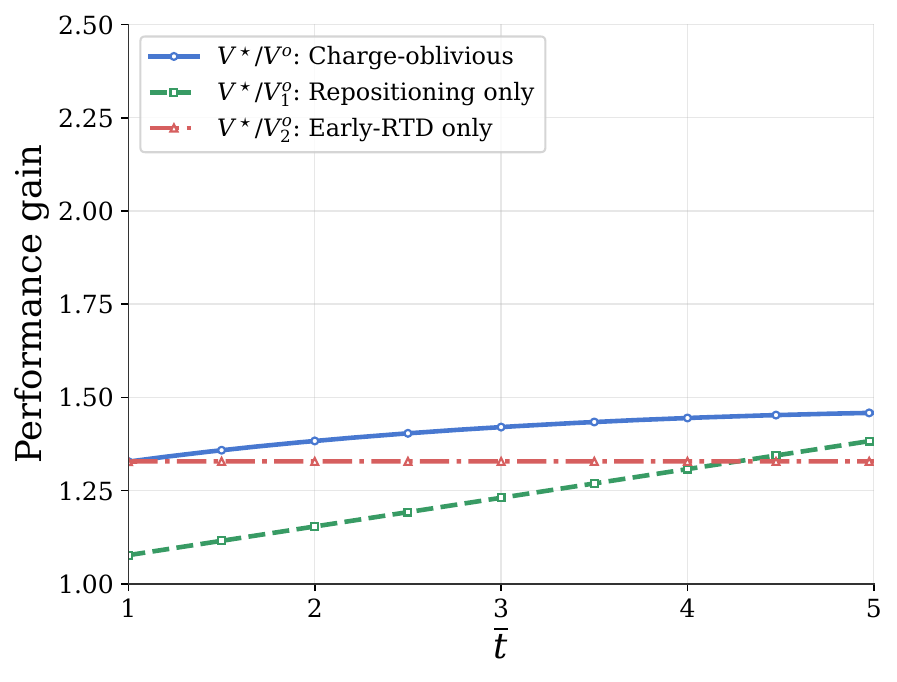}} }
    \caption{\centering 
    Performance gain ${\OPT(\instance)}/{\voblivious(\instance)}$, ${\OPT(\instance)}/{\voblivious_1(\instance)}$ and ${\OPT(\instance)}/{\voblivious_2(\instance)}$ as functions of $c_{2D}$ and $\tauidle$. \\
    $S=20$, $\lambda_2=10$, $r=5$, $K=9$, $\tau_0=3$, $c_{12}=c_{1D}=\taurep_{12}=\taurep_{21}=\taurep_{1D}=1$, $\taurep_{2D}=1.5$, $c_{D2}=\taurep_{D2}=2.5$}
    %Panel~(a) fixes $\tauidle=3$, panel~(b) fixes $c_{2D}=2$, and other parameters are as in \cref{fig:stylized_monotonic}.
    
    \label{fig:stylized_partial}
\end{figure}

We illustrate these results with a numerical study. \Cref{fig:stylized_partial} reports the performance gain of the fully charge-aware solution relative to the charge-oblivious benchmark and the two partially charge-aware benchmarks, as a function of the RTD cost asymmetry and the charge-depletion time. Either type of charge-aware control alone recovers a substantial share of the loss of the charge-oblivious policy. Relative to early RTD only, the additional gain of charge-aware repositioning increases in $c_{2D}$ (panel~(a)), because low-SoC vehicles continue to RTD with a detour via location~$2$, but is insensitive to $\tauidle$ (panel~(b)). Conversely, relative to charge-aware repositioning only, the additional gain of early RTD only increases in $\tauidle$ (panel~(b)), because low-SoC vehicles idle at location~$1$ until their batteries are depleted, but is nearly insensitive to $c_{2D}$ (panel~(a)). This highlights the complementarity of the two types of charge-aware control: each is most valuable precisely where the other falls short, and capturing the full value of charge-awareness requires~both.

Together, the results in this section suggest that charge-awareness is not merely a modeling refinement, but a first-order operational lever in electric robotaxi systems. We next generalize these insights beyond the stylized setting by examining a network calibrated with empirical data.

% Later in \cref{sec:sensitivity}, we show that its value remains significant across a wide range of realistic instances and identify the market conditions under which the performance gap is most pronounced. Moreover, we examine how much of this value can be retained through simpler charge-aware heuristics in \cref{sec:heuristics}. These heuristics vary both the degree to which decisions use battery information, as do the partially charge-aware policies studied here, and the granularity of the resulting decision rules. Whereas this section establishes the worst-case losses, the numerical results show that even limited charge-awareness yields substantial gains in realistic settings.

% \input{sec5_old}

\section{A Numerical Case Study}\label{sec:numerics}

We calibrate our model using a large-scale data set of trip records from the New York City Taxi \& Limousine Commission (TLC) to examine features of the optimal solution and the value of charge awareness.

\subsection{Data and Calibration}\label{sec:data_calibration}

\textbf{Data.} We use the 2025 High Volume For-Hire Vehicle (HVFHV) trip records published by the TLC \citep{nyctlc_trip_records}, which cover trips dispatched by high-volume ride-hailing companies, including Uber and Lyft, across the five boroughs of New York City. The records report pickup and drop-off taxi zones, timestamps, and itemized fares at the trip level. We focus on Manhattan: based on the TLC classification, we aggregate the service region into $64$ Manhattan zones and add one external charging depot $D$ in Queens, reflecting the practical difficulty of locating a large charging depot near Manhattan's demand center. Because the TLC records contain only realized trips, we use them as a proxy for latent demand (realized plus unfulfilled requests), grouping trip records by day of the week, hour of the day, and ordered origin--destination pairs $(i,j)$. For each operating window, we estimate the hourly demand rate $\lambda_i$, destination probability $p_{ij}$, and trip revenue $r_{ij}$ from the corresponding empirical trip counts, destination shares, and average fares. For each pair of zones, we compute the centroid-to-centroid distance and set the repositioning cost $c_{ij}$ to $\$0.7$ per mile, in line with the IRS business mileage rate.

\textbf{Calibration.} We operate vehicles within the 20--80\% SoC range and discretize it into $K=24$ levels. Assuming that a vehicle operates for at most six hours after being launched, each level corresponds to $15$ minutes of driving. Each depot visit takes one hour, which covers charging as well as fixed overhead such as plugging in, cleaning, and safety checks. We set the fleet size to $\size=7{,}000$ and consider two representative operating windows: peak hours (Mondays, 3{:}00--9{:}00~PM) and off-peak hours (Mondays, 12{:}00--6{:}00~AM). %; results are similar for other times of the week. 
To instantiate the LP in~\eqref{lp:fluid}, we estimate the demand primitives $(\boldsymbol{\lambda},\boldsymbol{p},\boldsymbol{r})$ separately for each window; the remaining primitives, such as the fleet size, are common to both windows. Unless otherwise specified, we report results for peak hours. Further calibration details are provided in Appendix~\ref{app:calibration}.

\subsection{Structural Features of Charge-Aware Control}\label{sec:thresholds_visualization}\label{sec:battery_sensitivity}

\begin{figure}[!t]
    \centering
    \subfloat[\centering Peak hours]{{\includegraphics[width=0.28\textwidth]{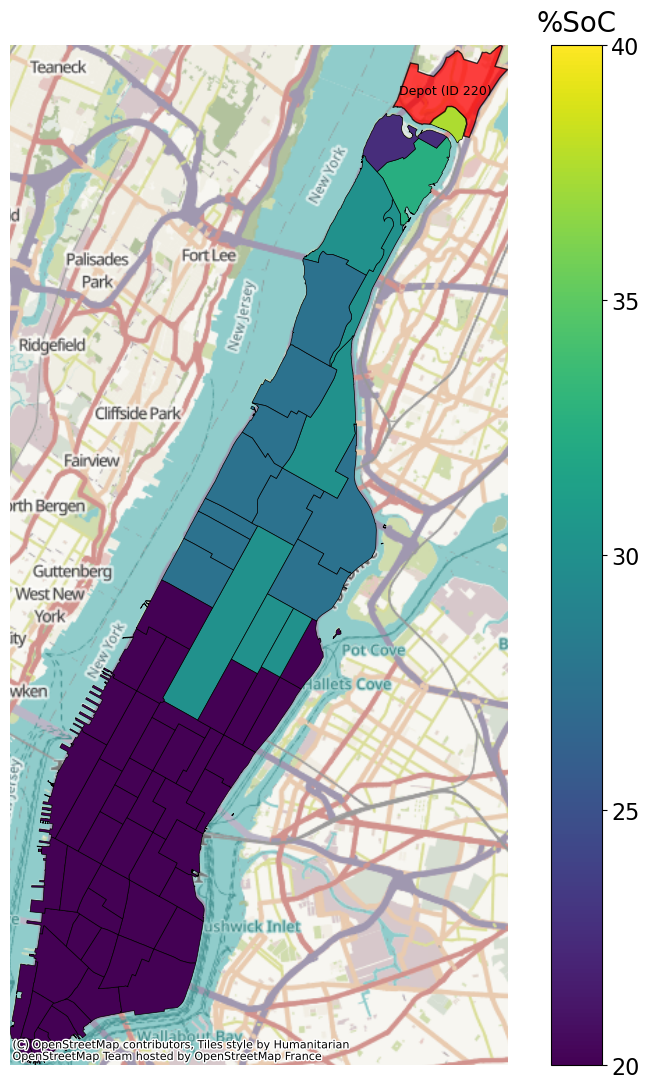} }}%
    \hspace{0cm}
    \subfloat[\centering Off-peak hours]{{\includegraphics[width=0.28\textwidth]{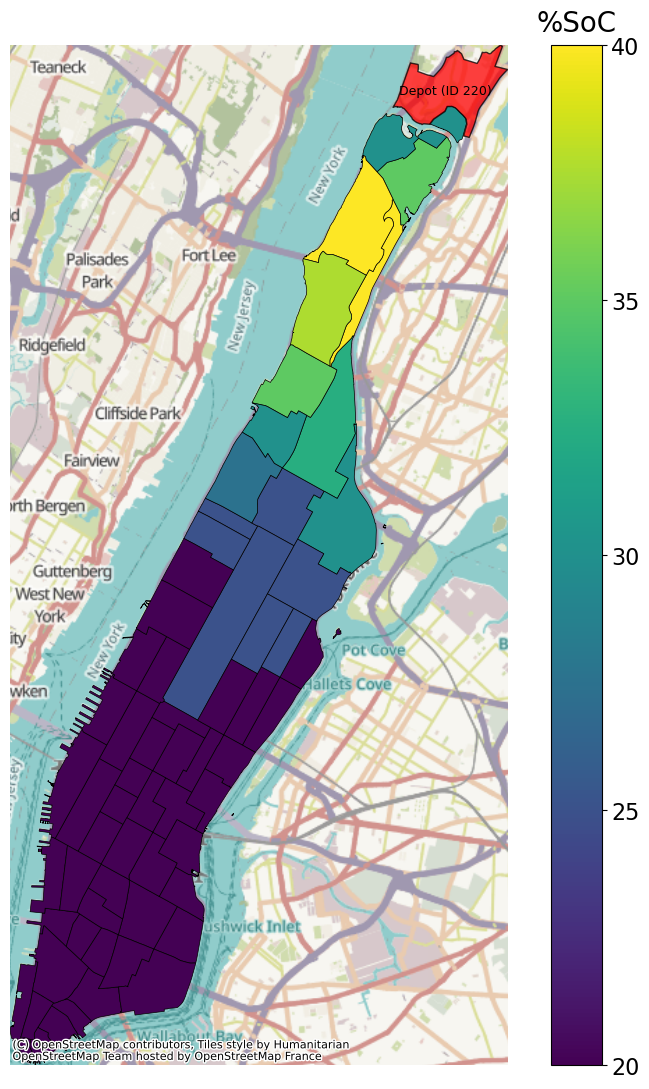} }}%
    \hspace{0cm}
    \subfloat[\centering Difference]{{\includegraphics[width=0.293\textwidth]{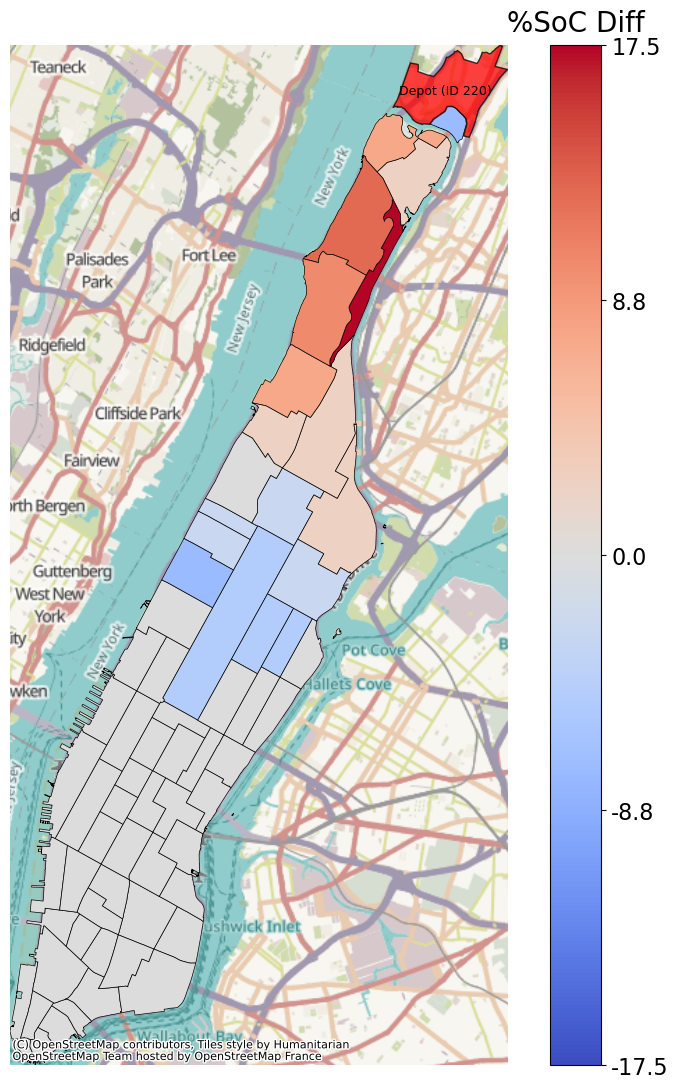} }}
    \caption{RTD thresholds during peak hours, off-peak hours, and their difference (labeled \textit{diff}). Thresholds are shown as SoC percentages, which range from 20\% to 80\%.}
    \label{fig:visualization_RTD}
\end{figure}

The optimal solutions exhibit the threshold structures of \cref{sec:results}, with the following spatial and temporal features.

\textbf{RTD thresholds are highest near the depot.} Consistent with \cref{thm:threshold_structure}, each zone has an RTD threshold $\kappa_i^\star\in\{0,1,\dots,K\}$, where a larger threshold means that vehicles at zone $i$ return to the depot at a higher SoC. As \cref{fig:visualization_RTD}(a)--(b) show, downtown zones, which are farther from the depot and have stronger revenue opportunities, have the lowest thresholds, and vehicles there return to the depot only at the 20\% safety level. The highest thresholds arise in uptown zones that are closer to the depot and weaker in revenue. By \cref{prop:comparison}, these are precisely the features that favor early RTD.

\textbf{RTD thresholds rise when demand is thin.} During off-peak hours, the system is less supply-constrained: $4.7\%$ of the fleet remains parked at the depot, whereas during peak hours all vehicles are deployed and only $69.6\%$ of demand is served. With this slack capacity, the RTD thresholds rise by up to $17.5\%$ in nine upper-Manhattan zones near the depot and remain unchanged or decrease by at most $7.5\%$ elsewhere (\cref{fig:visualization_RTD}(c)). Intuitively, a vehicle is less valuable when demand is thin (see \cref{prop:demand_shadow_price} in Appendix~\ref{app:proof_numerics}), so recharging it early forgoes little revenue. This effect is strongest in zones close to the depot, where RTD is cheap. Note that here we treat peak and off-peak hours as two separate stationary regimes and do not model the transition between them. When demand is nonstationary, the exact threshold levels may need to be adjusted over time, but the charge-aware features remain relevant, as we show in \cref{sec:ext_nonstationary}.

% These findings also inform charge-aware policy design in nonstationary environments. Our stationary model can be interpreted as applying to a fixed planning window, such as a peak or off-peak period; we do not model the transition between peak and off-peak, as dynamic control under nonstationary demand is a distinct problem with known solutions (see, e.g., \citet{vr_chen2024realtime}). Nevertheless, we expect the structural features identified in our model to remain relevant in a non-stationary setting, and the corresponding thresholds may vary over time as demand conditions evolve. For instance, during a low-demand period preceding a peak, the manager may return vehicles earlier so that they enter the peak {\color{magenta} period} fully charged. During the peak hours, in contrast, the opportunity cost of taking vehicles out of service is greater, so RTD may be delayed.

\textbf{Scarce depot capacity eliminates early RTD.} Our base model does not limit depot capacity. When we cap the number of vehicles that can be charged at the depot at any given time at $C$, early RTD disappears for $C/\size<13\%$. In such settings, the chargers are fully occupied, yet part of the fleet remains unused (e.g., $4.5\%$ at $C/\size=13\%$), so every charging slot is used to keep vehicles on the road and there is no room for proactive charging. Early RTD reappears once the entire fleet is deployed, and the solution coincides with the unconstrained one once $C$ exceeds the charging load (i.e., the number of vehicles being charged at the depot at any given time) of the unconstrained solution, which is about $1{,}000$ ($C/\size\approx14.3\%$). In Appendix~\ref{app:depot-cap}, we show that the threshold structure in \cref{thm:threshold_structure} continues to hold under this constraint, but the shadow cost of depot occupancy lowers the optimal thresholds $\boldsymbol{\kappa}^\star$.

\begin{figure}[!t]
    \centering
    \captionsetup[subfloat]{farskip=0pt}
    \begin{minipage}[c]{0.55\textwidth}
        \centering
        \subfloat[\centering 25\% SoC]{{\includegraphics[width=0.49\linewidth]{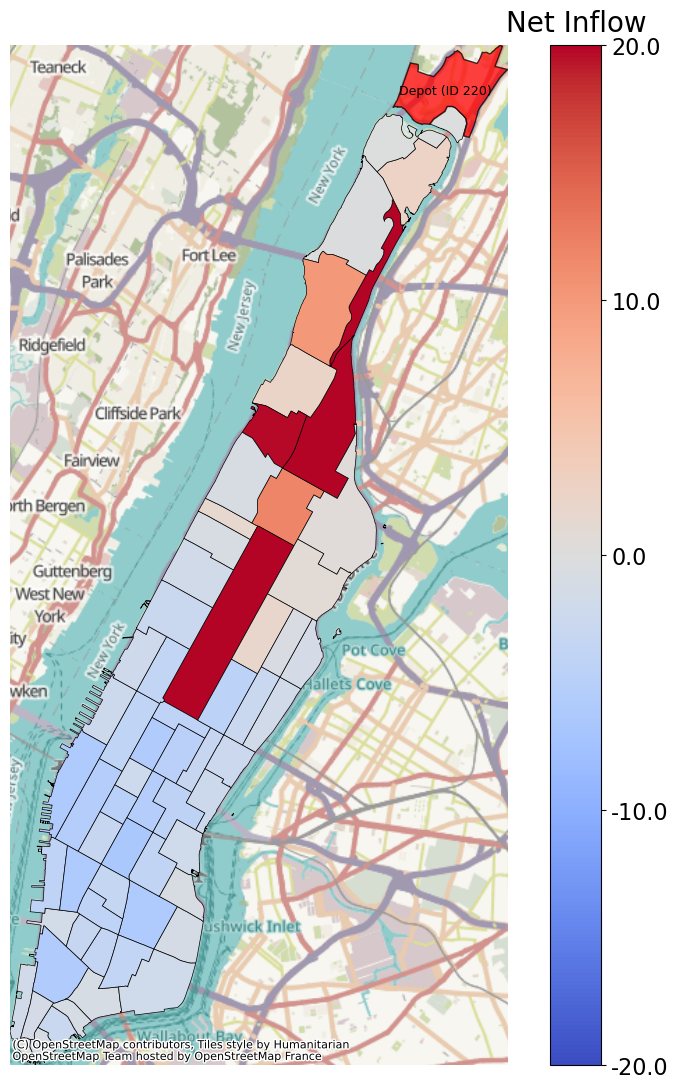}}}%
        \hfill
        \subfloat[\centering 75\% SoC]{{\includegraphics[width=0.49\linewidth]{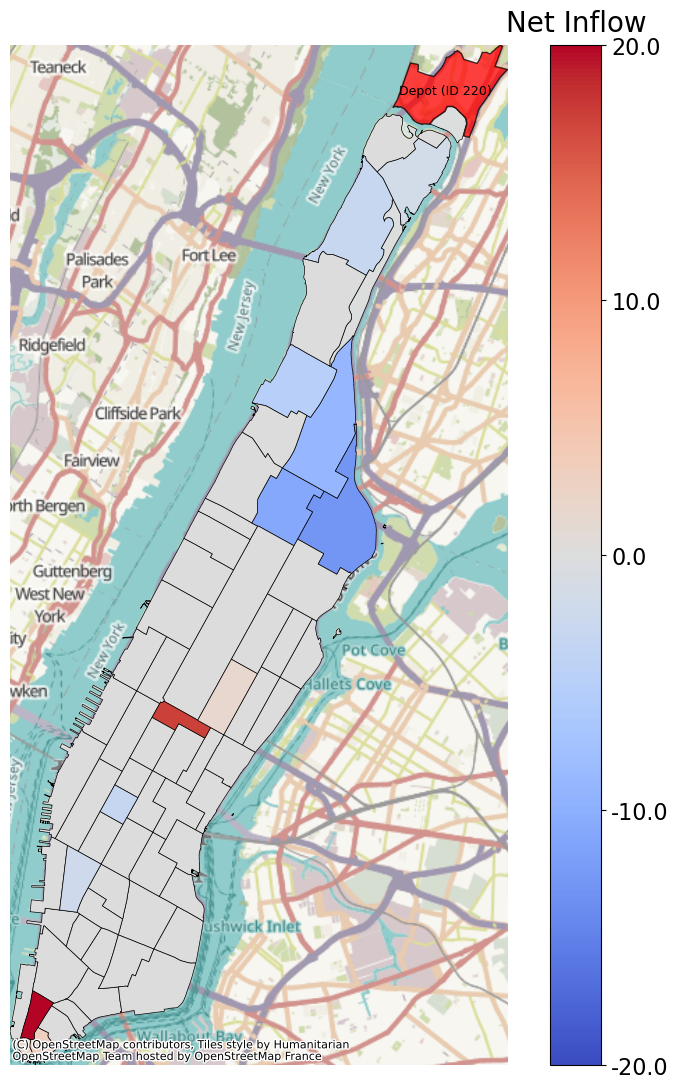}}}%
    \end{minipage}\hfill
    \begin{minipage}[c]{0.40\textwidth}
        \centering
        \includegraphics[width=\linewidth]{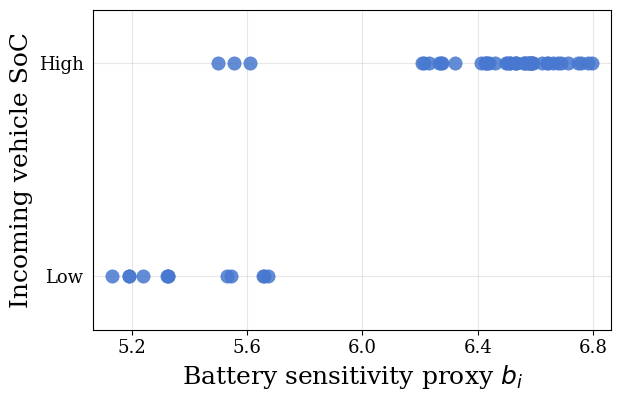}
    \end{minipage}
    \par\smallskip
    \begin{minipage}[t]{0.55\textwidth}
        \vspace{0pt}\captionsetup{type=figure}
        \caption{Net repositioning inflow at low vs. high SoC. Red indicates net inflow; blue means net outflow.}
        \label{fig:visualization_flow}
    \end{minipage}\hfill
    \begin{minipage}[t]{0.40\textwidth}
        \vspace{0pt}\captionsetup{type=figure}
        \caption{Battery sensitivity proxy and the SoC of incoming repositioned vehicles.}
        \label{fig:battery_sensitivity}
    \end{minipage}
\end{figure}

\textbf{Battery sensitivity guides charge-aware repositioning.} \cref{fig:visualization_flow} visualizes which locations receive or send out vehicles at a given SoC level: red zones have positive net inflow and blue zones have net outflow. The two plots for vehicles at low (25\%) and high (75\%) SoC reveal distinct patterns. At low SoC, uptown zones near the depot receive net inflow (red), whereas downtown zones far from the depot have net outflow (blue); for vehicles with high SoC, the pattern reverses. These patterns arise from the differences in battery sensitivity $\Delta_{ik}$ across zones. To demonstrate this, we denote the average battery sensitivity of zone $i$ by $b_i$ (formally, $b_i$ is the slope of $\pi^\star_{ik}$ with respect to $k$), and label a zone \textit{Low} if its repositioning inflow at $25\%$ SoC exceeds that at $75\%$ SoC, and \textit{High} otherwise. As \cref{fig:battery_sensitivity} shows, every Low zone has $b_i<5.7$, and every zone with $b_i>5.7$ is High. That is, consistent with \cref{thm:threshold2}, low-SoC vehicles are directed only toward the least battery-sensitive zones, which are mostly uptown near the depot. The clean separation also suggests that a single cutoff on $b_i$ suffices to identify the destinations that low-SoC vehicles should avoid, which we exploit in \cref{sec:heuristics}.

\subsection{Value of Charge Awareness}\label{sec:sensitivity}

\textbf{Charge awareness substantially increases profit.} Under the baseline calibration, charge awareness increases profit by at least $10.5\%$ (approximately \$13{,}000 per hour) during peak hours and at least $30.1\%$ (approximately \$6{,}100 per hour) during off-peak hours; see Appendix~\ref{app:approx} for construction of upper bound on charge-oblivious profit. If peak conditions occur for 30 hours per week and the remaining hours are treated as off-peak, this translates into roughly \$64 million in additional annual profit. As shown in \cref{fig:diff} in Appendix~\ref{app:approx}, the gains concentrate in downtown Manhattan: through proactive RTD and SoC-dependent repositioning moves, the charge-aware solution preserves charge and vehicle availability where they are most valuable; this is at the cost of only negligible revenue losses in other parts of the city.

\begin{figure}[!t]%
    \centering
    \subfloat[\centering Battery range]{{\includegraphics[width=0.48\textwidth]{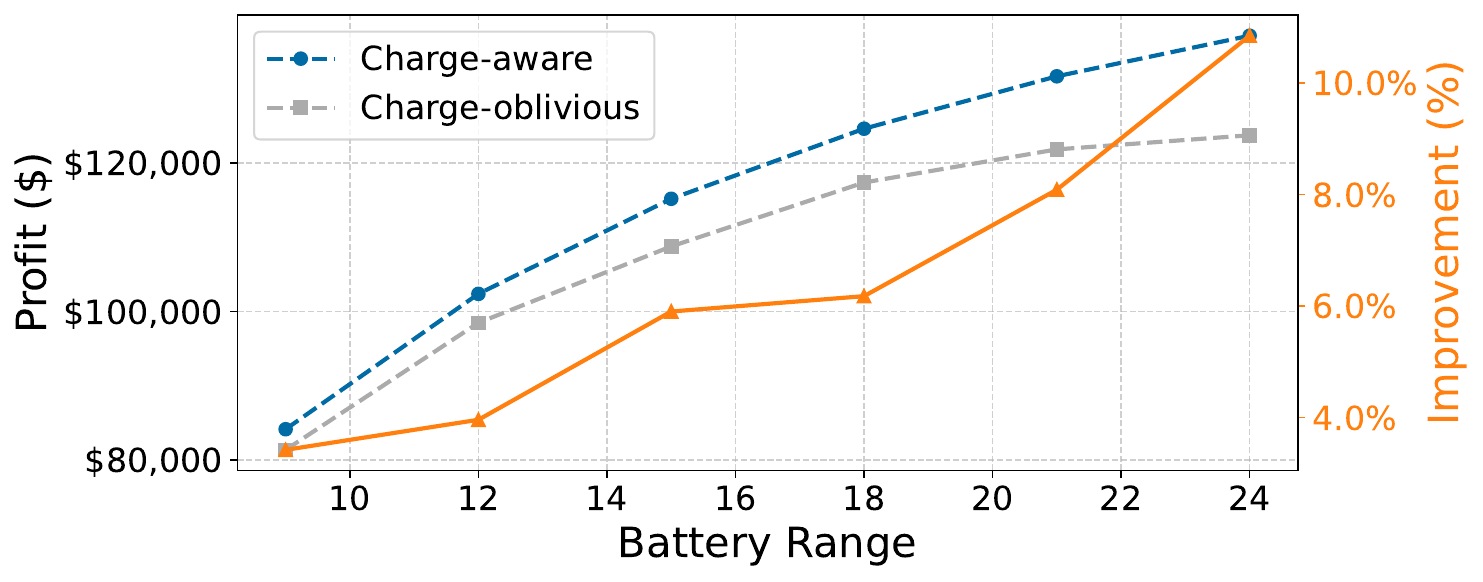}}}
    \hspace{0.5cm}
    \subfloat[\centering Fleet size]{{\includegraphics[width=0.48\textwidth]{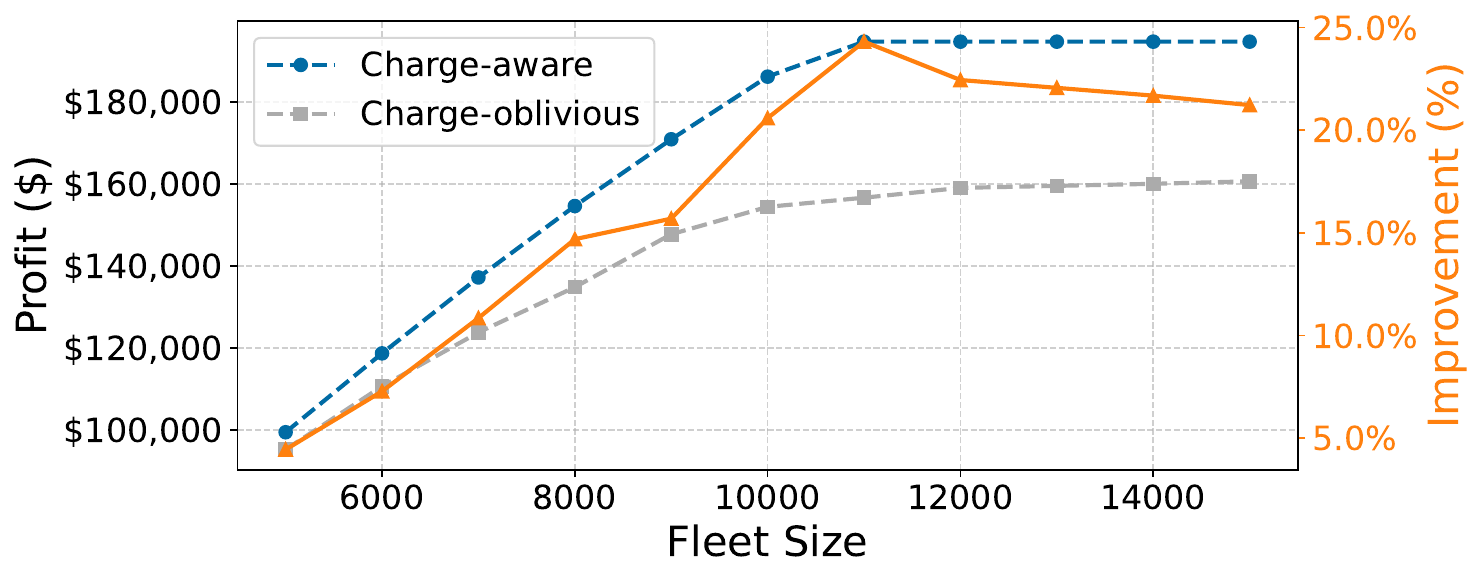}}}
    \vspace{0em}
    \subfloat[\centering Depot capacity]{{\includegraphics[width=0.48\textwidth]{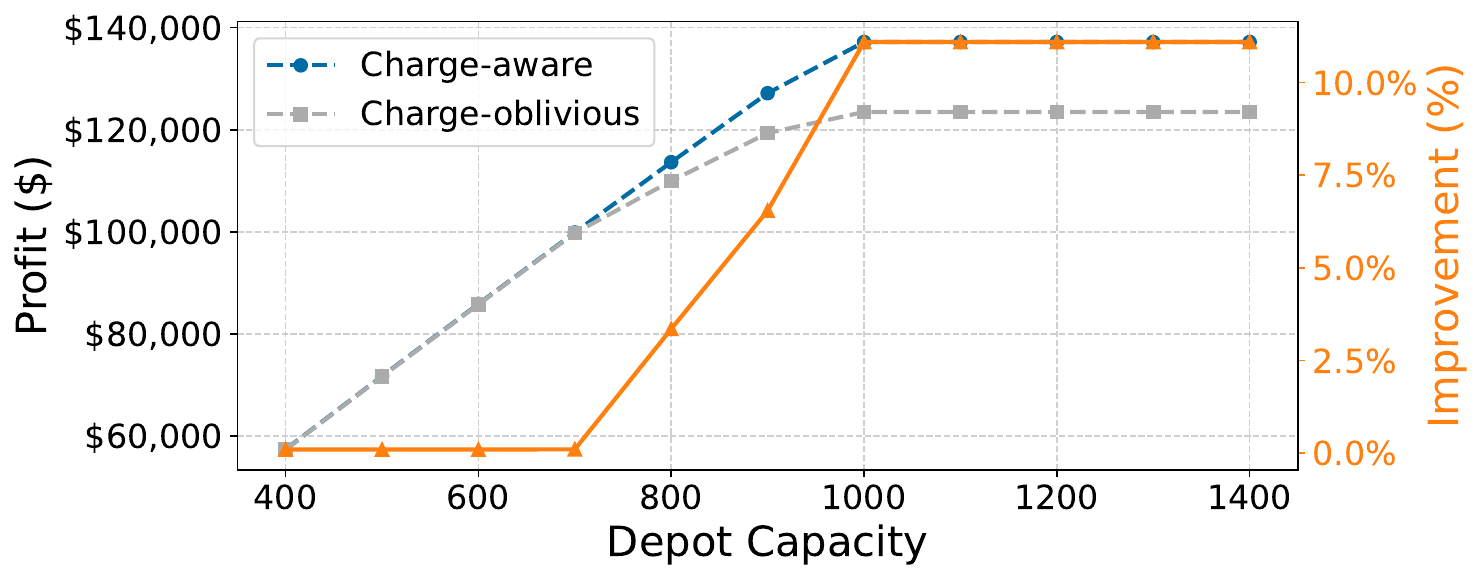}}}
    \hspace{0.5cm}
    \subfloat[\centering Number of depots]{{\includegraphics[width=0.48\textwidth]{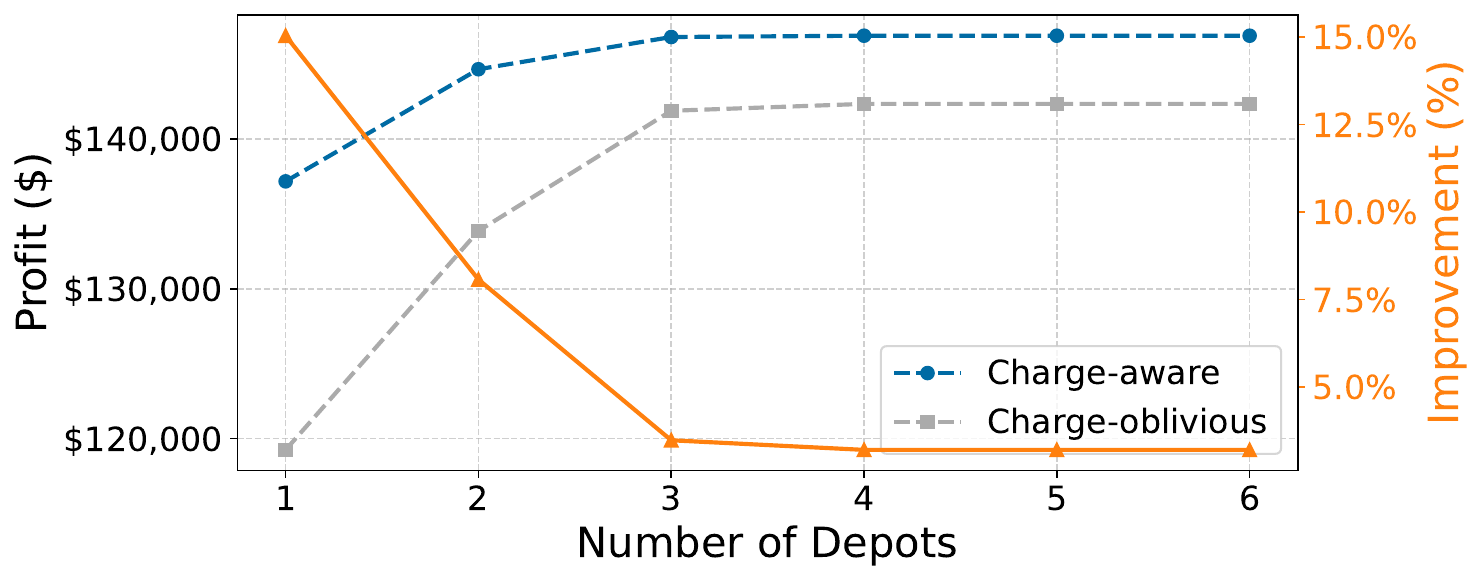}}}
    \caption{Charge-aware solution and charge-oblivious proxy under varying exogenous parameters.}
    \label{fig:value_comparison}
\end{figure}

\textbf{Features that govern the value of charge awareness.} We next conduct a sensitivity analysis to identify the features that govern the value of charge awareness, as illustrated in \Cref{fig:value_comparison}. For computational scalability, we use a tractable proxy $\widetilde V^o(\instance)$ for the nonconvex charge-oblivious problem. This proxy yields charge-awareness gains numerically close to those in the baseline comparisons above; Appendix~\ref{app:approx} describes its construction and documents this comparison. We summarize the main findings below.
\begin{itemize}
    \item A longer battery range increases the value of charge awareness because it allows the charge-aware solution more room to differentiate RTD and repositioning decisions by SoC.
    \item The effect of fleet size is non-monotone. With a very small fleet, both solutions are severely supply-constrained, leaving little room for charge awareness to create value. As vehicles are added, the charge-aware solution uses the extra capacity more efficiently, so the gap widens. Once the system becomes demand-constrained, slack vehicles absorb the inefficient moves of the charge-oblivious solution, and the gap narrows, consistent with \cref{prop:stylized_monotonicity}(iv).
    \item A larger depot capacity widens the gap, because it makes proactive RTD feasible; the gap plateaus once capacity no longer binds.
    \item Additional depots increase the profit of both solutions, with strongly diminishing returns, and narrow the gap, which nevertheless remains above $2.5\%$ with six depots. Charge awareness is also more valuable when the depot is farther from demand centers (see Appendix~\ref{app:multiple_depots}). Because building many depots in dense urban areas is likely prohibitively expensive, robotaxi fleets will probably rely on a few centralized depots, often far from demand centers, which is precisely when charge awareness is most valuable.
\end{itemize}

\subsection{Performance of Charge-Aware Heuristics}\label{sec:heuristics}

We next construct simple policies based on the threshold structures in \cref{sec:results}. Complementing \cref{sec:partial}, which varies which levers depend on SoC, we also restrict how finely decisions can differ across locations and SoC levels. The \textit{RTD heuristic}, motivated by \cref{thm:threshold_structure}, allows early RTD only at zones where the charge-aware solution uses it ($\kappa_i^\star>0$) and only at or below a common threshold $\kappa_1$. The \textit{repositioning heuristic}, motivated by \cref{thm:threshold2}, rules out repositioning vehicles with SoC below $\kappa_2$ to high-sensitivity destinations ($b_i>\delta$).\footnote{A charge-aware fulfillment heuristic constructed in the same way is analyzed in \cref{sec:ext_fulfillment}.} Both heuristics use SoC only through these binary admissibility rules; conditional on them, decisions are homogenized across SoC levels, and $(\kappa_1,\kappa_2,\delta)$ are chosen by grid search to maximize steady-state profit (see Appendix~\ref{app:constructing_heuristics} for the formal definitions).

\begin{figure}[h]
    \centering
    \includegraphics[width=0.8\textwidth]{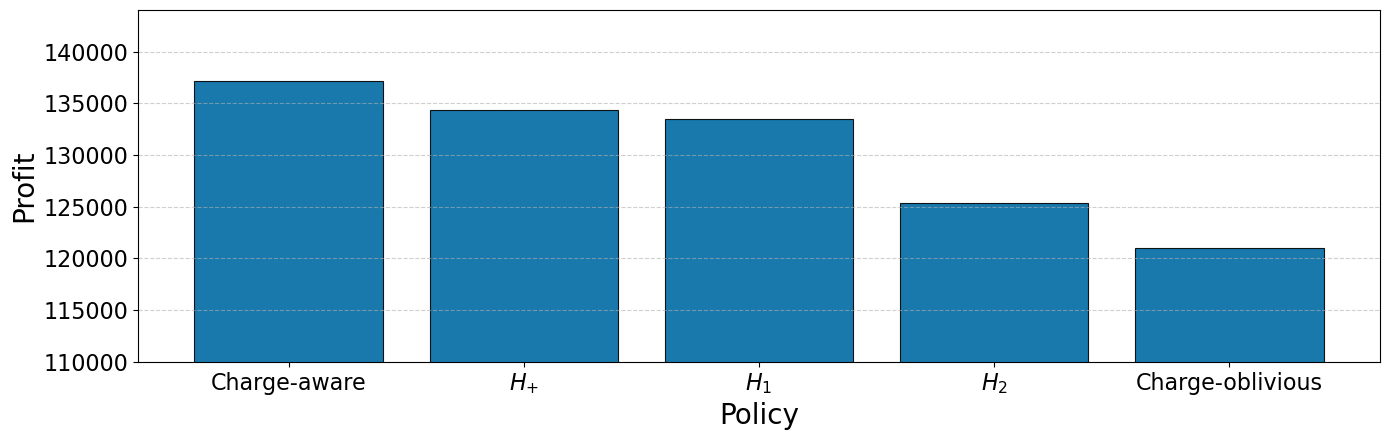}
    \caption{Profit comparison across policies with different charge-aware levers and solution granularity.}
    \label{fig:policy_profit}
\end{figure}

We compare policy $H_{+}$, which uses both heuristics, $H_1$, which uses only the repositioning heuristic and suppresses early RTD as in \eqref{eq:obl_no_RTD}, and $H_2$, which uses only the RTD heuristic with charge-independent repositioning as in \eqref{eq:obl_simplex}. For a policy $\pi$ with objective value $V^{\pi}(\instance)$, we measure the fraction of the charge-awareness gap it closes by $\parenthesis{V^{\pi}(\instance)-\widetilde V^o(\instance)}/\parenthesis{\OPT(\instance)-\widetilde V^o(\instance)}$. As \cref{fig:policy_profit} shows, $H_{+}$ closes $82.9\%$ of the gap between the charge-oblivious proxy and the charge-aware solution. In fact, $H_1$ alone already closes $79.1\%$ of the gap, while $H_2$ closes about $28.4\%$ of the gap. These results suggest that a small amount of charge awareness goes a long way: the structural results developed in previous sections can be translated into simple heuristics that capture most of the operational benefit of charge awareness.

\section{Extensions}\label{sec:extensions}

In this section, we discuss how our results generalize under two extensions: destination-dependent fulfillment (\cref{sec:ext_fulfillment}) and a nonstationary environment (\cref{sec:ext_nonstationary}).

\subsection{Charge-Aware Fulfillment}\label{sec:ext_fulfillment}

We have thus far focused on charge-aware control of empty vehicles, i.e., repositioning and RTD decisions $(\boldsymbol w,\boldsymbol{\rtd})$. We now examine fulfillment decisions and consider the extension, as discussed in \cref{sec:charge_aware}, in which $\boldsymbol f$ can also be destination-dependent. We show that the structural results for RTD and empty repositioning continue to hold under such an extension. Moreover, the optimal fulfillment decision itself admits a threshold structure analogous to charge-aware repositioning.

\textbf{Destination-dependent fulfillment.} In \eqref{lp:fluid}, the fulfillment decision depends on the origin location and battery level, but not on the destination of the request. We now let the fleet manager condition fulfillment decisions on the destination as well. Specifically, we replace $f_{ik}$ by $f_{ijk}$, where
$f_{ijk}$ denotes the rate at which vehicles in state $(i,k)$ fulfill demand from origin $i$ to destination $j$. Let $\lambda_{ij}:=\lambda_i p_{ij}$ denote the exogenous arrival rate of requests from $i$ to $j$. Then, the demand feasibility constraint becomes
\begin{equation}
\sum_{k\in \mathcal{K}} f_{ijk}\;\le\;\lambda_{ij},
\qquad \forall\, i,j\in\mathcal N.
\label{c:demand_match}
\end{equation}
All other constraints are modified accordingly by replacing the aggregate fulfillment flow $f_{ik}p_{ij}$ with the destination-specific fulfillment flow $f_{ijk}$ in the objective, flow-balance constraints, and fleet-size constraint.

We show in \cref{thm:threshold_structure_match,thm:threshold2_match} in Appendix~\ref{app:proof_fulfillment} that the threshold structures in \cref{thm:threshold_structure,thm:no_reposition} continue to hold under this extension. Intuitively, destination-dependent fulfillment alters the number of empty vehicles in each state, but it does not affect the structure of the problem.

\textbf{Structure of charge-aware fulfillment.} We now turn to the structure of the destination-dependent fulfillment decision itself. In particular, we show that the fulfillment decision $\boldsymbol{f}$ follows the same battery-sensitivity threshold structure as the repositioning decision $\boldsymbol{w}$ in \cref{thm:threshold2}.

\begin{theorem}
\label{thm:threshold_f}
In an optimal solution $(\boldsymbol{f}^\ast,\boldsymbol{w}^\ast,\boldsymbol{\rtd}^\ast)$, consider locations $i,i'\in\mathcal N$ with $\Delta_{ik}<\Delta_{i'k}$ for all $k\in\{1,\ldots,K-1\}$. Then, for every $j\in\mathcal N\setminus\bracket{i,i'}$, there exists a threshold $\widetilde{\kappa}_j\in\mathcal K$ such that:
\begin{enumerate}[label=(\roman*)]
    \item vehicles do not serve trips to $i'$ below the threshold, i.e., $f^\ast_{ji'k}=0\;\forall k<\widetilde\kappa_j$;
    \item vehicles do not serve trips to $i$ above the threshold, i.e., $f^\ast_{jik}=0\;\forall k>\widetilde\kappa_j$.
\end{enumerate}
\end{theorem}

\cref{thm:threshold_f} highlights that charge-aware fulfillment plays a role similar to charge-aware repositioning: it routes higher-SoC vehicles toward destinations where an additional battery level is more valuable, while reserving lower-SoC vehicles for less battery-sensitive destinations. Intuitively, the destination-specific fulfillment variable $f_{ijk}$ can be viewed as a revenue-generating, demand-capacitated repositioning arc from $i$ to $j$. Thus, the structural properties of repositioning decisions naturally extend to fulfillment decisions.

\textbf{Fulfillment versus repositioning.} When both fulfillment and repositioning are destination-dependent, the next result shows that fulfillment always takes precedence if demand is available.

\begin{proposition}
\label{prop:service_dominates_reposition} In an optimal solution $(\boldsymbol{f}^\ast,\boldsymbol{w}^\ast,\boldsymbol{s}^\ast)$, consider an origin-destination pair $(i,j)$ such that $r_{ij}+c_{ij}>0$. If demand from $i$ to $j$ is not fully served, i.e., $\sum_{k\in\mathcal K} f^\ast_{ijk}<\lambda_{ij}$, then vehicles do not reposition from $i$ to $j$ at any SoC, i.e., $w^\ast_{ijk}=0$ for all $k\in\mathcal K$.
\end{proposition}

Intuitively, fulfillment and empty repositioning from $i$ to $j$ lead to the same downstream state $(j,k-1)$ and consume the same vehicle time, but fulfillment earns revenue $r_{ij}$ and avoids the repositioning cost $c_{ij}$. When $r_{ij}+c_{ij}>0$, serving an available request from $i$ to $j$ thus strictly dominates moving an empty vehicle along the same arc; if the optimal solution does not exhaust the demand for $(i,j)$, it cannot be optimal to reposition empty vehicles from $i$ to $j$.

We also numerically examine the impact of charge-aware fulfillment on the case study in \cref{sec:numerics}. Once the fleet manager can condition fulfillment decisions on destinations, empty repositioning accounts for only $4.3\%$ of total vehicle time (as compared with $25.8\%$ previously), leaving little room for charge-aware repositioning. Indeed, a charge-aware fulfillment heuristic alone closes more than $96.6\%$ of the gap between the charge-oblivious proxy and the fully charge-aware solution. This is consistent with \cref{prop:service_dominates_reposition}: fulfillment takes precedence over repositioning. In practice, however, service-quality considerations may limit the fleet manager's control over fulfillment decisions, in which case charge-aware repositioning remains the main operational lever.

\subsection{Nonstationary Environment}\label{sec:ext_nonstationary}

Our base model assumes a fixed planning window, such as a peak or off-peak period, during which demand is approximately constant. In this subsection, we show that the threshold structures of \cref{sec:results} generalize to a finite-horizon fluid model with time-varying primitives.

\textbf{Setting.} We retain the primitives of \cref{sec:model}, except that (i) time is discrete, with periods $t=1,\ldots,T$ and integral duration for each task, and (ii) demand rates, revenues, and costs follow known sequences $\lambda_{ij}(t)$, $r_{ij}(t)$, and $c_{ij}(t)$, and (iii) fulfillment decisions are destination-dependent, as in \cref{sec:ext_fulfillment}. In period $t$, let $\boldsymbol X(t)$ collect the available vehicle masses $x_{ik}(t)$, $i\in\mathcal N$, $k\in\{0,\ldots,K-1\}$, the depot masses $x_{Dk}(t)$ that can be launched at level $k\in\{1,\ldots,K\}$, and the committed masses $y_{ik}(t,d)$ (resp.\ $y_{Dk}(t,d)$) that are still executing an action started before $t$ and become available in state $(i,k)$ (resp.\ for launch at level $k$) at the beginning of period $t+d$, $d\in\{1,\ldots,\bar d\}$, where $\bar d$ is the longest action duration. The initial state $\boldsymbol X(1)=\boldsymbol X^0$ is given and contains $\size$ units of vehicle mass. $\boldsymbol f(t),\boldsymbol w(t)$ and $\boldsymbol{\rtd}(t)$ denote the decisions in period $t$; they are nonnegative and satisfy
\begin{equation}\label{eq:ns_feasibility}
\begin{aligned}
\sum_{j\in\mathcal N}f_{ijk}(t)+\sum_{j\in\mathcal N}w_{ijk}(t)+\rtd_{ik}(t)
 &=x_{ik}(t), && i\in\mathcal N,\quad 0\le k<K,\\
\sum_{k=0}^{K-1}f_{ijk}(t)&\le\lambda_{ij}(t), &&i,j\in\mathcal N,\\
\sum_{j\in\mathcal N}w_{Djk}(t)&\le x_{Dk}(t), &&k\in\{1,\ldots,K\},
\end{aligned}
\end{equation}
with $f_{ij0}(t)=w_{ij0}(t)=0$; non-depot controls at $k=K$ are zero. In this model we need to additionally specify a a charge target after RTD: $\rtd_{ik\ell}(t)\ge0$ is the mass at $(i,k)$ that returns to the depot in period $t$ to be charged to level $\ell\in\{\max\{k,1\},\ldots,K\}$, and $\rtd_{ik}(t):=\sum_{\ell}\rtd_{ik\ell}(t)$. The transition map $\Phi_t$ follows the convention that an action started at $t$ with duration $d$ finishes at the beginning of $t+d$. For instance, service mass $f_{ijk}(t)$ and repositioning mass $w_{ijk}(t)$ reach $(j,k-1)$ after $\tauservice_{ij}$ periods; launch mass $w_{Djk}(t)$ reaches $(j,k-1)$ after $\taurep_{Dj}$ periods; RTD mass $\rtd_{ik\ell}(t)$ becomes available for launch at level $\ell$ at the beginning of period $t+\taurep_{iD}+\batteryvar_1+\batteryvar_2(\ell-k)$; depot mass that is not launched remains available at its level; and
existing commitments advance by one period. These rules define $\boldsymbol X(t+1)=\Phi_t\big(\boldsymbol X(t),\boldsymbol f(t),\boldsymbol w(t),\boldsymbol{\rtd}(t)\big)$ and preserve total fleet mass. %A distinct idling duration $\tauidle_{ii}$ can replace $\taurep_{ii}$ without changing the results.
Revenues and costs are counted when actions start; actions may finish after $T$, and all states have zero terminal value at $T+1$. 

Let $g_t(\boldsymbol f(t),\boldsymbol w(t),\boldsymbol{\rtd}(t))$ be the profit expression in \eqref{obj} with $f_{ik}p_{ij}$ replaced by $f_{ijk}$, evaluated at the period-$t$ decisions and primitives. The manager solves the following finite linear program:
\begin{equation}\label{eq:ns_problem}
\max\ \sum_{t=1}^{T}g_t \big(\boldsymbol f(t),\boldsymbol w(t),\boldsymbol{\rtd}(t)\big)
\quad\text{s.t.}\quad
\eqref{eq:ns_feasibility},\quad
\boldsymbol X(1)=\boldsymbol X^0,\quad
\boldsymbol X(t+1)=\Phi_t\big(\boldsymbol X(t),\boldsymbol f(t),\boldsymbol w(t),\boldsymbol{\rtd}(t)\big)
\end{equation}

\textbf{Structure of an optimal solution.} We next characterize the structure of an optimal solution to \eqref{eq:ns_problem}. For selected dual multipliers $\pi_{ik}^\star(t)$ and $\pi_{DK}^\star(t)$ (see Appendix~\ref{app:proof_appendix}), we extend \cref{def:battery_sensitivity} to define
$\Delta_{ik}(t):=\pi_{ik}^\star(t)-\pi_{i,k-1}^\star(t),\forall i\in\mathcal N, k\in \bracket{1,\ldots,K-1}.$ The next result provides the structure of optimal repositioning:

\begin{proposition}\label{cor:ns_repositioning}
In an optimal solution of \eqref{eq:ns_problem}, consider a period $t$, locations $i,i'\in\mathcal N$, and $j\in\mathcal N\setminus\bracket{i,i'}$ with $\Delta_{ik}(t+\taurep_{ji})<\Delta_{i'k}(t+\taurep_{ji'})$ for all $k\in\{1,\ldots,K-2\}$. Then, there exists a threshold $\bar\kappa_j(t)\in\mathcal K$ such that:
\begin{enumerate}[label=(\roman*)]
    \item vehicles do not reposition to $i'$ below the threshold in period $t$, i.e., $w^\star_{ji'k}(t)=0\;\forall k<\bar\kappa_j(t)$;
    \item vehicles do not reposition to $i$ above the threshold in period $t$, i.e., $w^\star_{jik}(t)=0\;\forall k>\bar\kappa_j(t)$.
\end{enumerate}
\end{proposition}

\cref{cor:ns_repositioning} is the natural extension of \cref{thm:threshold2}: it has the same structure, except that battery sensitivity is evaluated when the vehicle arrives at the destination, since that is when its remaining charge starts adding value. It shows that when the primitives are nonstationary, a destination can be attractive to high-SoC vehicles because its battery sensitivity is expected to rise, even if it is currently low. When battery sensitivities do not vary over time, in contrast, the condition of the result coincides with that of \cref{thm:threshold2}.

The next result extends \cref{thm:threshold_structure} to this nonstationary environment, under the additional condition that charging time does not depend on the initial SoC.

\begin{proposition}\label{thm:ns_rtd}
Suppose that $\taucharge_k=\taucharge_0$ for all $k\in\{0,\ldots,K-1\}$. There exists an optimal solution of \eqref{eq:ns_problem}, with some threshold $\kappa_i^\star(t)\in\{0,\ldots,K-1\}$ for each location $i\in\mathcal N$ and period $t$, such that vehicles RTD with a positive rate in period $t$ if their SoC $k\leq\kappa_i^\star(t)$ and do not return to the depot if $k>\kappa_i^\star(t)$. That is, $\rtd^\star_{ik}(t)>0$ for all $k\leq\kappa_i^\star(t)$ with $x^\star_{ik}(t)>0$ and $\rtd^\star_{ik}(t)=0$ for all $k>\kappa_i^\star(t)$.
\end{proposition}

\cref{thm:ns_rtd} generalizes \cref{thm:threshold_structure}, yielding an RTD threshold that is both location- and time-dependent. For instance, during a low-demand period that precedes a peak, the threshold may be higher, so that vehicles recharge and enter the peak fully charged; during the peak itself, it may be lower, so that vehicles stay on the road and capture more revenue. When the thresholds do not vary over time, this is the structure in \cref{thm:threshold_structure}.

\begin{remark}\label{rmk:ns_charging}
The assumption that the charging time does not depend on the initial SoC is essential for \cref{thm:ns_rtd}. When $\taucharge_k$ depends on $k$, a higher-SoC vehicle finishes charging sooner, so returning it to the depot may allow it to serve a demand peak that a lower-SoC vehicle would miss. RTD can therefore become more attractive as SoC increases, breaking the threshold structure. For the same reason, early launching, i.e., launching a vehicle before it is fully charged, may be optimal, so \cref{prop:launch_full_charge} no longer holds. In contrast, \cref{cor:ns_repositioning} does not rely on this assumption.
\end{remark}

\section{Conclusion}\label{sec:conclusion}

In this paper, we study how fleet managers should operate electric robotaxi fleets, where vehicle SoC is a central operational state variable. We develop a fluid model that captures service, repositioning, idling, and charging decisions in a network. We show that optimal charge-aware operations admit simple threshold structures. At each location, vehicles with SoC below a location-specific threshold should return to the depot, while only those above it may reposition. Moreover, battery sensitivity governs repositioning decisions: low-SoC vehicles should not be sent to more battery-sensitive destinations, nor high-SoC vehicles to less battery-sensitive ones. We then show that ignoring SoC can cause unbounded performance loss, and that both early RTD and charge-aware repositioning are essential to avoiding it. Using a case study calibrated with NYC trip records, we find that charge-aware control generates substantial gains over a charge-oblivious proxy, and that simple threshold-based heuristics deliver most of the value of a fully charge-aware solution. These structural features remain robust under destination-dependent fulfillment and a nonstationary environment.

Several directions remain open. Our depot operations model is intentionally stylized. In practice, depots may involve uncertain charging times and queuing. Incorporating these features would provide a richer understanding of how charging infrastructure and battery-aware fleet control should be jointly managed. On the theoretical side, it would be valuable to characterize when threshold-based heuristics admit provable approximation guarantees. %Third, solving the nonstationary and stochastic robotaxi control problem at scale, e.g., through mixed-integer programs that leverage the structural features identified here, remains an important direction.

% \input{sketch.tex}

% Bibliography
\bibliographystyle{plainnat}
\bibliography{refs}

\newpage
% Appendix

\begin{APPENDICES}{}
% Keep appendix hyperlink destinations distinct from those in the main text.
\renewcommand{\theHsection}{appendix.\Alph{section}}
\section{Time-Dependent Battery Consumption}\label{app:duration_energy}

Our base model assumes that every service or repositioning task consumes one battery level. We now allow battery consumption to depend on task duration and extend the threshold results in \cref{sec:results}.

\textbf{Setting.} We retain the primitives of \cref{sec:model}, except that a task of duration $t$ consumes $\energyfunction(t)$ battery levels, where $\energyfunction:[0,\infty)\to[0,\infty)$ is nondecreasing and concave; the base model corresponds to $\energyfunction\equiv1$. Thus, a service or repositioning trip from $i$ to $j$ consumes $e_{ij}:=\energyfunction(\taurep_{ij})\in\mathbb Z_{\ge1}$ battery levels and takes a vehicle from $(i,k)$ to $(j,k-e_{ij})$; similarly, a launch from the depot to $j$ consumes $e_{Dj}:=\energyfunction(\taurep_{Dj})$ levels. We normalize battery consumption so that each local trip consumes one level, i.e., $e_{ii}=1$ for all $i\in\mathcal N$.

A repositioning move to $j$ is feasible only if $k\ge e_{ij}$, and fulfillment at $i$ is feasible only if $k\ge\max_{j:p_{ij}>0}e_{ij}$. Accordingly, the flow-balance constraints become
\begin{equation}\label{eq:duration_node_balance}
\sum_{j\in\mathcal N}p_{ji}f_{j,k+e_{ji}}
+\sum_{j\in\mathcal N\cup\{D\}}w_{ji,k+e_{ji}}
=f_{ik}+\sum_{j\in\mathcal N}w_{ijk}+\rtd_{ik},
\qquad i\in\mathcal N,\quad 0\le k<K,
\end{equation}
where flows with out-of-range charge indices are zero. The objective, demand constraints, depot balance, and vehicle-time accounting in \eqref{lp:fluid} are unchanged.

\textbf{RTD and no-repositioning thresholds.} We first extend \cref{thm:threshold_structure,thm:no_reposition} to this setting.

\begin{corollary}
\label{cor:duration_thresholds}
Consider any optimal solution $(\boldsymbol f^\star,\boldsymbol w^\star,\boldsymbol{\rtd}^\star)$. For each location $i\in\mathcal N$, there exists a threshold $\kappa_i^\ast\in\mathcal K$ such that:
\begin{enumerate}[label=(\roman*)]
    \item vehicles do not return to the depot above the threshold, i.e., $\rtd^\star_{ik}=0$ for all $k>\kappa_i^\ast$;
    \item vehicles do not reposition below the threshold, i.e., $w^\star_{ijk}=0$ for all $j\in\mathcal N$ and $k<\kappa_i^\ast$.
\end{enumerate}
If the optimal solution satisfies strict complementary slackness, then vehicles RTD with a positive rate at every SoC $k\le\kappa_i^\ast$, i.e., $\rtd^\star_{ik}>0$ for all $k\le\kappa_i^\ast$.
\end{corollary}

\Cref{cor:duration_thresholds} extends \cref{thm:threshold_structure,thm:no_reposition}, with the thresholds $\kappa_i^\ast$ defined as in \eqref{eq:threshold_def_rtd}, and reduces to them when $\energyfunction\equiv1$. Intuitively, heterogeneous consumption changes which trips a vehicle can take at a given SoC and how much charge it has left afterwards, but not the tradeoff behind early RTD. As in \cref{thm:threshold_structure}, this tradeoff compares the continuation value of keeping a vehicle on the road with the value of RTD, and the former still increases sufficiently fast in SoC that RTD becomes less attractive at higher SoC levels.

\textbf{Charge-aware repositioning.} For destinations $i,i'$ and origin $j$, let
\[
\mathcal K_j:=\bigl\{k\in\{1,\ldots,K-1\}:k\ge\max\{e_{ji},e_{ji'}\}\bigr\}
\]
denote the charge levels at which both repositioning moves are feasible.

\begin{corollary}\label{cor:duration_repositioning}
In an optimal solution, consider $i,i'\in\mathcal N$ and $j\in\mathcal N\setminus\{i,i'\}$ with
\begin{equation}\label{eq:duration_sensitivity_order}
\Delta_{i,k+1-e_{ji}}<\Delta_{i',k+1-e_{ji'}},
\qquad\text{whenever }k,k+1\in\mathcal K_j.
\end{equation}
Then there exists a threshold $\bar\kappa_j\in\mathcal K$ such that:
\begin{enumerate}[label=(\roman*)]
    \item vehicles do not reposition to $i'$ below the threshold, i.e., $w^\star_{ji'k}=0$ for all $k<\bar\kappa_j$;
    \item vehicles do not reposition to $i$ above the threshold, i.e., $w^\star_{jik}=0$ for all $k>\bar\kappa_j$.
\end{enumerate}
\end{corollary}

\Cref{cor:duration_repositioning} is the counterpart of \cref{thm:threshold2} when repositioning trips consume different amounts of energy. Battery sensitivity is now compared at the charge levels reached after repositioning, since a vehicle departing from $j$ at level $k$ reaches $i$ at $k-e_{ji}$ and $i'$ at $k-e_{ji'}$. When $e_{ji}=e_{ji'}=1$, the condition reduces to the comparison used in the proof of \cref{thm:threshold2}. Proofs of both corollaries are in Appendix~\ref{app:proof_duration_energy}.

\section{Additional Simulations}

\subsection{Data Description and Calibration}\label{app:calibration}

\textit{Data extrapolation.}
We perform a numerical case study using the publicly available 2025 NYC Taxi and Limousine Commission (TLC) Trip Record Data. TLC classifies the service region in Manhattan into 64 taxi zones, upon which we add one external charging depot $D$ in Queens. This reflects the practical difficulty of locating a large robotaxi charging depot near Manhattan's demand center.

While the TLC trip records are censored, i.e., they contain only realized trips, we use them as a proxy for the latent demand patterns. Specifically, we aggregate raw trips into ordered origin--destination pairs $(i,j)$ and estimate the hourly demand rate $\lambda_i$ (the number of trips originating from zone $i$), destination probability $p_{ij}$ (the empirical fraction of trips from $i$ ending in $j$), and revenue $r_{ij}$ (the average observed fare for trips from $i$ to $j$). We set $\lambda_D=0$ and exclude passenger trips to or from the depot. For each pair of locations, we compute the centroid-to-centroid distance in miles and set the repositioning cost $c_{ij}$ to be $0.7$ times that distance. The value of \$0.7 per mile is chosen based on the quoted IRS 2024 business mileage rate of \$0.67 per mile \citep{irs2024mileage}.

\textit{Model Calibration.}
Motivated by the discussion in \cref{sec:introduction} that robotaxis may need to recharge after roughly six hours of operation, we set $K=24$, so that it takes 15 minutes to burn one battery level. We interpret $k=K$ as 80\% SoC and $k=0$ as 20\% SoC, so each level represents $(0.8-0.2)/24=0.025$ of battery capacity; this 20--80\% operating band reflects common battery-management practice, since fleet managers generally avoid charging vehicles to full and avoid recalling vehicles so late that they risk being stranded. Since each service or repositioning task consumes one level of battery, this calibration captures a 15-minute decision resolution. We thus restrict eligible service and repositioning arcs to pairs whose centroid distance is at most $30\times 0.25=7.5$ miles, using 30 mph as a stylized urban operating speed consistent with NACTO's urban speed guidance \citep{nacto_design_speed}; this cut-off threshold far exceeds the citywide average trip distance of 5 miles \citep[Figure~5]{tlc2024electrification}. When a vehicle idles, we assume it consumes energy three times more slowly than when actively driving, so one battery level is depleted after $45$ minutes. For depot visits, we assume a one-hour charging time regardless of the initial battery level, capturing charging as well as fixed overhead such as plug-in, cleaning, and safety checks; if higher-SoC vehicles could be charged faster, then timing RTD decisions would become even more important, strengthening the case for charge-aware control.

Finally, we set the baseline fleet size to $\size=7000$, which is consistent with a back-of-the-envelope calculation based on recent NYC Uber and Lyft volumes: reports based on TLC data indicate that Uber and Lyft together average on the order of 700{,}000 daily trips in NYC, and $700{,}000/(24\times 4)\approx 7{,}000$ vehicles if each vehicle completes about four trips per hour \citep{automarketplace2026}. We conduct a sensitivity analysis of fleet size in \cref{sec:sensitivity} to examine how this assumption affects the value of charge awareness.

\subsection{Approximating the Charge-Oblivious Benchmark}\label{app:approx}

For the baseline peak- and off-peak instances, we evaluate the charge-oblivious benchmark using Gurobi's nonlinear optimization routines. We obtain feasible solutions using its nonlinear barrier algorithm and compute upper bounds using spatial branch-and-bound, with a 30-minute time limit per solve. The baseline profit comparisons in \cref{sec:numerics} use the tightest upper bounds obtained and therefore provide conservative estimates of the gains from charge awareness. \Cref{fig:diff} shows the hourly revenue gain of the charge-aware solution at each location relative to the best feasible charge-oblivious solution obtained.

\begin{figure}[h]
    \centering
    {{\includegraphics[width=0.25\textwidth]{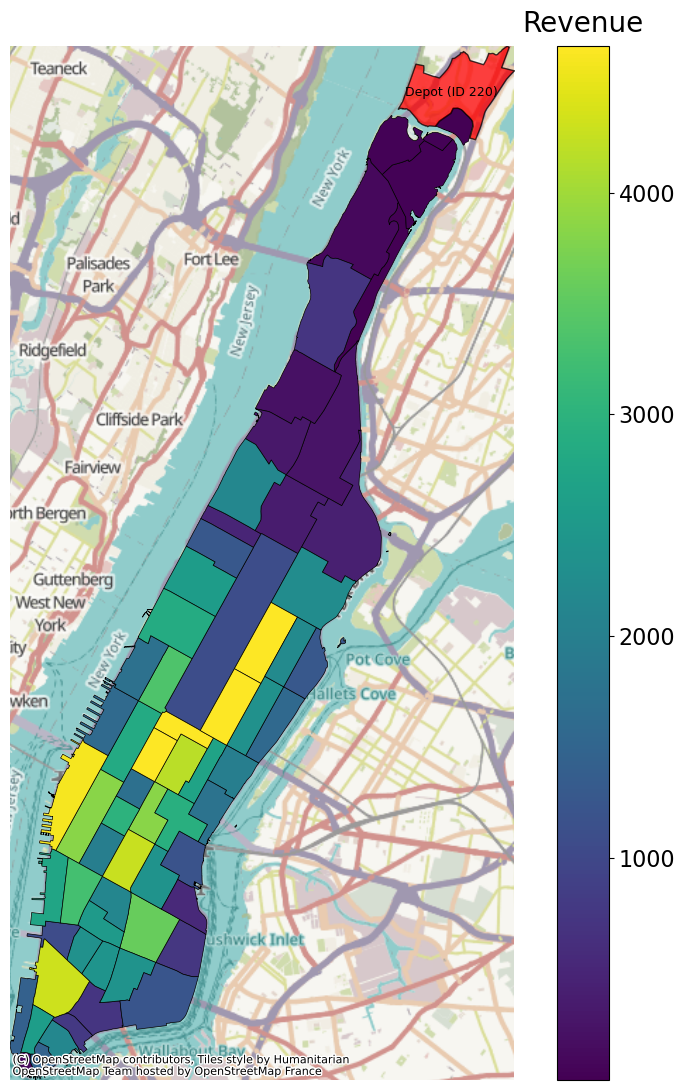} }}%
    \caption{Hourly revenue gain of the charge-aware solution relative to the best feasible charge-oblivious solution obtained.}
    \label{fig:diff}
\end{figure}

For computational scalability in the sensitivity analyses, we construct a tractable charge-oblivious proxy by fixing the repositioning proportions $\boldsymbol{\woblivious}$ based on the charge-aware solution $\boldsymbol w^\star$. Let
\begin{equation}\label{eq:w_approximation}
    \woblivious_{ij}
\;=\;
\frac{\sum_{k=1}^{K-1} w^\star_{ijk}}
{\sum_{k=1}^{K-1}\sum_{\ell\in\mathcal N} w^\star_{i\ell k}},
\qquad \forall i,j\in\mathcal N,
\end{equation}
with the convention that $\woblivious_{ij}=0$ whenever the corresponding denominator is zero. We then solve:
\[
\begin{aligned}
\max_{\boldsymbol f,\boldsymbol w,\boldsymbol \rtd}\quad
& \eqref{obj} \\
\text{s.t.}\quad
& \eqref{c:demand}-\eqref{eq:obl_no_RTD},
\quad w_{ijk}
=
\woblivious_{ij}\sum_{\ell\in\mathcal N} w_{i\ell k},
\quad \forall i,j\in\mathcal N,\; k=1,\ldots,K-1, \\
& \boldsymbol f,\boldsymbol w,\boldsymbol \rtd \ge 0 .
\end{aligned}
\]
Since $\{\woblivious_{ij}\}_{i,j\in\mathcal N}$ are now exogenous, the program above is an LP, while still ensuring that repositioning decisions cannot depend on the vehicle charge level. We denote its optimal objective value by $\widetilde V^o(\instance)$.

For both baseline instances, the proxy value is within $2\%$ of the corresponding upper bound. This small approximation gap suggests that using the proxy is unlikely to materially affect our qualitative findings in \cref{sec:numerics}.

\subsection{Constructing Charge-Aware Heuristics}\label{app:constructing_heuristics}

This appendix describes how we construct the charge-aware heuristics used in \cref{sec:heuristics}. We first state the binary admissibility restrictions that define the heuristics.

\textbf{RTD heuristic (used in $H_{+}$ and $H_2$).} Motivated by the no-RTD-above-threshold result in \cref{thm:threshold_structure}, we construct a simplified early-RTD rule using the thresholds $\kappa_i^\star$ from the fully charge-aware solution. Specifically, we partition locations into
\[
\mathcal N^{0}:=\{i\in\mathcal N:\kappa_i^\star=0\},
\qquad
\mathcal N^{+}:=\{i\in\mathcal N:\kappa_i^\star>0\}.
\]
Locations in $\mathcal N^{0}$ are those for which the fully charge-aware solution does not use early RTD. For these locations, the heuristic also rules out early RTD by imposing
\begin{equation}\label{eq:heur_rtd_no_early}
\rtd_{ik}=0,
\qquad
\forall i\in\mathcal N^{0},\; k\in\bracket{1,\ldots,K{-}1}.
\end{equation}
For locations in $\mathcal N^{+}$, the heuristic allows early RTD only at or below a common tunable threshold $\kappa_1$. That is, for a candidate $\kappa_1\in\bracket{0,\ldots,K{-}1}$, we impose
\begin{equation}\label{eq:heur_rtd_threshold}
\rtd_{ik}=0,
\qquad
\forall i\in\mathcal N^{+},\; k>\kappa_1.
\end{equation}
Thus, among locations where early RTD is relevant, all positive battery levels weakly below $\kappa_1$ are eligible for early RTD, while higher levels are not. Conditional on this binary rule, the RTD decisions are homogenized across battery levels, as described below.

\textbf{Charge-aware repositioning heuristic (used in $H_{+}$ and $H_1$).} Motivated by \cref{thm:threshold2}, we construct a simple charge-aware repositioning rule using the battery sensitivity proxy $b_i$ from \cref{sec:battery_sensitivity}. For a sensitivity cutoff $\delta$, define the set of high battery-sensitivity locations as
\[
\mathcal N^{h}(\delta):=\{i\in\mathcal N:b_i>\delta\}.
\]
The repositioning heuristic rules out these high-sensitivity destinations for low-SoC vehicles. Specifically, for a candidate pair $(\delta,\kappa_2)$, we impose
\begin{equation}\label{eq:heur_reposition_threshold}
w_{\ell i k}=0,
\qquad
\forall \ell\in\mathcal N,\; i\in\mathcal N^{h}(\delta),\; k<\kappa_2.
\end{equation}
Constraint \eqref{eq:heur_reposition_threshold} reflects the destination-exclusion principle in \cref{thm:threshold2}. Vehicles with $k\ge \kappa_2$ may still be repositioned to such locations, and vehicles at any charge level may be repositioned to locations outside $\mathcal N^{h}(\delta)$. Conditional on this binary restriction, the repositioning probabilities are homogenized separately for low- and high-SoC vehicles, as detailed below. Notice that one can apply the same treatment to design a charge-aware fulfillment heuristic, which we analyze in \cref{sec:ext_fulfillment}.

Given the binary restrictions for early RTD and charge-aware repositioning, we first solve \eqref{lp:fluid} subject to the additional constraints \eqref{eq:heur_rtd_no_early}--\eqref{eq:heur_reposition_threshold}. Let $(\widehat{\boldsymbol f},\widehat{\boldsymbol w},\widehat{\boldsymbol \rtd})$ denote an optimal solution to this program, which we leverage to construct charge-oblivious RTD and repositioning fractions, as we did in \eqref{eq:w_approximation}.

\noindent\textbf{Early RTD.} We impose the zero-RTD constraint at every state at which early RTD is ruled out by \eqref{eq:heur_rtd_no_early}--\eqref{eq:heur_rtd_threshold}. For the remaining admissible states, we compute the aggregate fraction of vehicles that RTD (across $k \leq \kappa_1$):
\[
\widehat\eta_i
:=
\frac{\sum_{k\in\bracket{1,\ldots,\kappa_1}}
\widehat\rtd_{ik}}
{\sum_{k\in\bracket{1,\ldots,\kappa_1}}
\left(
\widehat f_{ik}
+\widehat\rtd_{ik}
+\sum_{j\in\mathcal N}\widehat w_{ijk}
\right)},
\qquad i\in\mathcal N^{+}.
\]
When the denominator is zero, we set $\widehat\eta_i=0$. For $i\in\mathcal N^{+}$, this allows us to apply a charge-oblivious RTD control among $k \in \bracket{1,...,\kappa_1}$ and disallow early RTD when $k > \kappa_1$. 

\noindent\textbf{Charge-aware repositioning.} For repositioning, the binary rule separates battery levels into low- and high-SoC regimes. We therefore compute two repositioning rules for each origin $i\in\mathcal N$: the low-SoC rule aggregates repositioning flows over levels below $\kappa_2$, while the high-SoC rule aggregates repositioning flows over levels weakly above $\kappa_2$:
\[
\widehat w^{L}_{ij}
:=
\frac{\sum_{k\in\bracket{1,\ldots,\kappa_2-1}}\widehat w_{ijk}}
{\sum_{k\in\bracket{1,\ldots,\kappa_2-1}}\sum_{\ell\in\mathcal N}\widehat w_{i\ell k}},
\qquad 
\widehat w^{H}_{ij}
:=
\frac{\sum_{k\in\bracket{\kappa_2,\ldots,K-1}}\widehat w_{ijk}}
{\sum_{k\in\bracket{\kappa_2,\ldots,K-1}}\sum_{\ell\in\mathcal N}\widehat w_{i\ell k}},
\qquad j\in\mathcal N.
\]
When the corresponding denominator is zero, we set the distribution to zero. Notice that because $\widehat{\boldsymbol w}$ is obtained from the constrained LP, the low-SoC rule satisfies $\widehat w^L_{ij}=0\;\forall j\in\mathcal N^{h}(\delta)$.

Based on the computed $\widehat\eta_i$, $\widehat w^L_{ij}$ and $\widehat w^H_{ij}$, we solve the LP:
\begin{align}
\max_{\boldsymbol f,\boldsymbol w,\boldsymbol \rtd}\quad 
& \eqref{obj} \notag\\
\text{s.t.}\quad 
& \eqref{c:demand}-\eqref{c:vehicle_time},\ \eqref{eq:heur_rtd_no_early},\ \eqref{eq:heur_rtd_threshold},\ \eqref{eq:heur_reposition_threshold},\ \boldsymbol f,\boldsymbol w,\boldsymbol \rtd \ge 0 \notag\\
& \rtd_{ik}
=
\widehat\eta_i
\left(
 f_{ik}+\rtd_{ik}+\sum_{j\in\mathcal N}w_{ijk}
\right),
&& \forall i\in\mathcal N^{+},\; k\in\bracket{1,\ldots,\kappa_1}, \label{eq:rtd_heuristic}\\
& w_{ijk}
=
\widehat w^L_{ij}\sum_{\ell\in\mathcal N}w_{i\ell k},
&& \forall i,j\in\mathcal N,\; k\in\bracket{1,\ldots,\kappa_2-1}, \label{eq:repositioning_heuristic1}\\
& w_{ijk}
=
\widehat w^H_{ij}\sum_{\ell\in\mathcal N}w_{i\ell k},
&& \forall i,j\in\mathcal N,\; k\in\bracket{\kappa_2,\ldots,K{-}1}.\label{eq:repositioning_heuristic2}
\end{align}
In the program above, %the first proportionality constraint applies the same RTD rule only at states where early RTD is admissible; at all other positive levels, RTD is already forced to zero by \eqref{eq:heur_rtd_no_early}--\eqref{eq:heur_rtd_threshold}. The last two constraints assign two different charge-oblivious rules based on whether the SoC is above or below the threshold $\kappa_2$.  \eqref{eq:heuristic_final_lp} ensures that 
the early-RTD heuristic in \eqref{eq:rtd_heuristic} uses battery information only to determine whether early RTD is admissible at a state; conditional on admissibility, the same RTD fraction is applied across eligible levels. The charge-aware repositioning heuristic in \eqref{eq:repositioning_heuristic1}-\eqref{eq:repositioning_heuristic2} uses battery information only to distinguish low- and high-SoC regimes; conditional on the regime, the repositioning distribution is independent of the exact battery level. Thus, apart from the binary threshold rules, the policy follows charge-oblivious action fractions.

For the policies considered in \cref{sec:heuristics}, we conduct a grid search over $(\kappa_1,\delta,\kappa_2)$ to select the parameter vector that yields the highest objective value to the LP above. Then, the \textit{RTD + repositioning heuristic} corresponds exactly to the solution of this LP. \textit{Repositioning heuristic only} imposes the no-early-RTD constraint in \eqref{eq:obl_no_RTD} while allowing the repositioning heuristic in \eqref{eq:repositioning_heuristic1}-\eqref{eq:repositioning_heuristic2}. In contrast, \textit{RTD heuristic only} imposes the charge-oblivious repositioning constraint in \eqref{eq:obl_simplex} but allows the early RTD heuristic in \eqref{eq:rtd_heuristic}.

\subsection{Depot Capacity Constraint} \label{app:depot-cap} 

In practice, a depot may have limited charging, parking, cleaning, or maintenance capacity, which can restrict the ability of vehicles to return early for charging. We capture this by imposing a depot capacity constraint~$C$: 
\begin{equation}\label{eq:ct-depot-cap}
\sum_{k\in \mathcal{K}}\Big(\sum_{i\in\mathcal N} \rtd_{ik}\Big)\tau_k \le C.
\end{equation}
The left-hand side of \eqref{eq:ct-depot-cap} is the steady-state number of vehicles occupying the depot for charging, and it cannot exceed the depot capacity $C$.\footnote{Since \eqref{c:vehicle_time} is an inequality, the system may leave some vehicles parked, and hence unused, when depot capacity is scarce; parked vehicles do not count toward \eqref{eq:ct-depot-cap}.} Let $\eta\ge 0$ denote the dual multiplier associated with this constraint, which represents the shadow price of depot occupancy. It is straightforward to verify that the early RTD threshold structure in \cref{thm:threshold_structure} continues to hold under this additional constraint, because the additional term $-\eta\tau_k$ in the reduced cost of RTD below increases in $k$ at rate $\eta\batteryvar_2$: replacing $\batteryvar_2\nu^\star$ by $\batteryvar_2(\nu^\star+\eta)$ in \cref{lem:pi-monotone}(ii) and its proof, the proof of \cref{thm:threshold_structure} goes through unchanged. 

\begin{figure}[h]
    \centering
    \subfloat[\centering $C = 1000$]{{\includegraphics[width=0.3\textwidth]{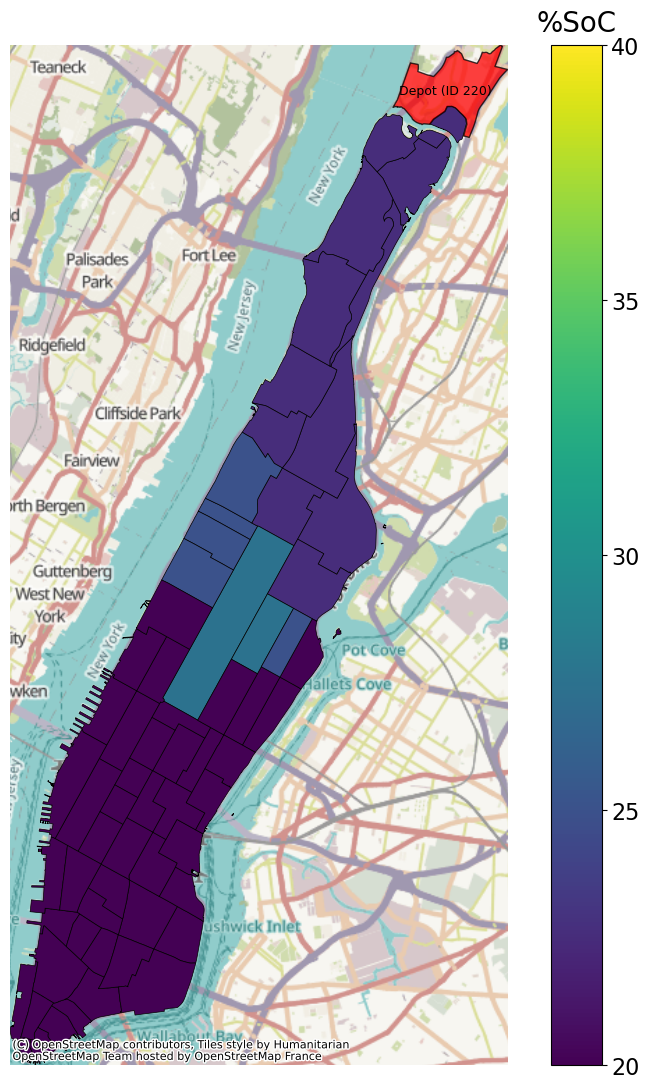} }}%
    \hspace{0cm}
    \subfloat[\centering $C = 900$]{{\includegraphics[width=0.3\textwidth]{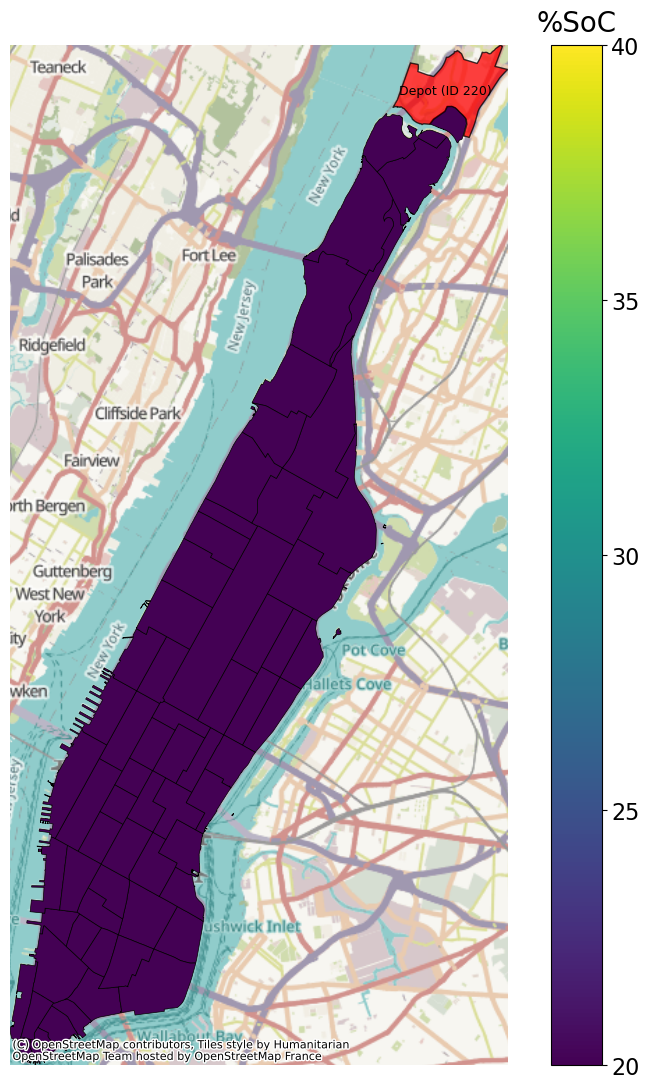} }}%
    \caption{Illustration of the early RTD thresholds under various depot capacities.}
    \label{fig:visualization_capacity}
\end{figure}

\Cref{fig:visualization_capacity} illustrates how depot capacity affects early RTD thresholds. When the depot capacity is relatively large, the optimal solution continues to send vehicles back early from selected locations. As depot capacity becomes tighter, however, the thresholds decrease substantially. In particular, when $C=900$, the thresholds collapse to the lowest safety charge across locations, indicating that the fleet manager no longer finds it optimal to use scarce depot capacity for proactive early RTD. This pattern can be explained through a reduced cost analysis. Among the reduced costs for service, repositioning, and RTD decisions, the depot-capacity constraint directly affects only RTD arcs. Specifically, the reduced cost of returning a vehicle from location $i$ at battery level $k$ becomes
\begin{equation*}%\label{eq:RC-w-toD-cap}
\reduced(\rtd_{ik})
=
- c_{iD}
-\nu\big(\taurep_{iD}+\tau_k\big)
-\pi_{ik}
-\zeta
-\eta\tau_k
\le 0.
\end{equation*}
Compared with the reduced cost in the unconstrained model, the extra term $-\eta\tau_k$ makes RTD less attractive.

\subsection{Impact of the Locations and Numbers of Depots}\label{app:multiple_depots}

We next investigate how the spatial configuration and number of charging facilities affect the operational decisions and value of charge awareness.

\textit{Impact of depot location.}
We first vary the depot location across different NYC neighborhoods. Consistent with \cref{fig:visualization_RTD}, early RTD thresholds tend to be higher for locations that are geographically closer to the depot. However, comparing across depot locations shows that the overall prevalence of early RTD depends critically on the depot's position relative to demand centers. As shown in \cref{fig:visualization_depot_location}, when the depot is located close to high-demand regions, early RTD is often unnecessary. In contrast, when the depot is farther from demand centers, early RTD becomes more important: higher thresholds allow the fleet manager to anticipate future charging needs and avoid costly low-SoC returns from inconvenient locations.

\begin{figure}[h]
    \centering
    \subfloat[\centering Depot close to demand center]{{\includegraphics[width=0.3\textwidth]{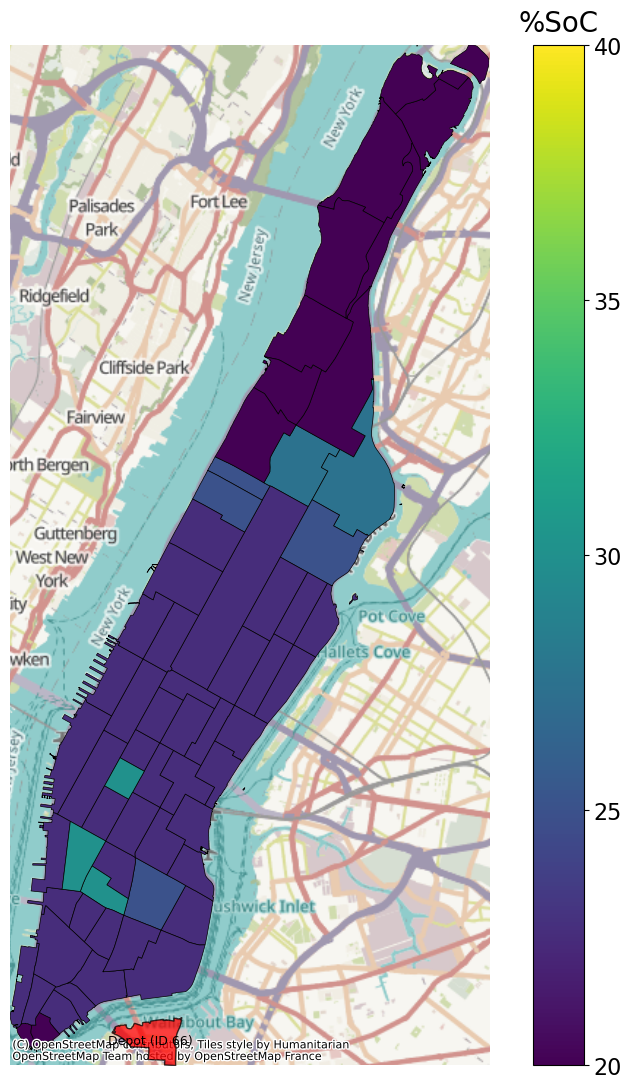} }}%
    \hspace{0cm}
    \subfloat[\centering Depot far from demand center]{{\includegraphics[width=0.39\textwidth]{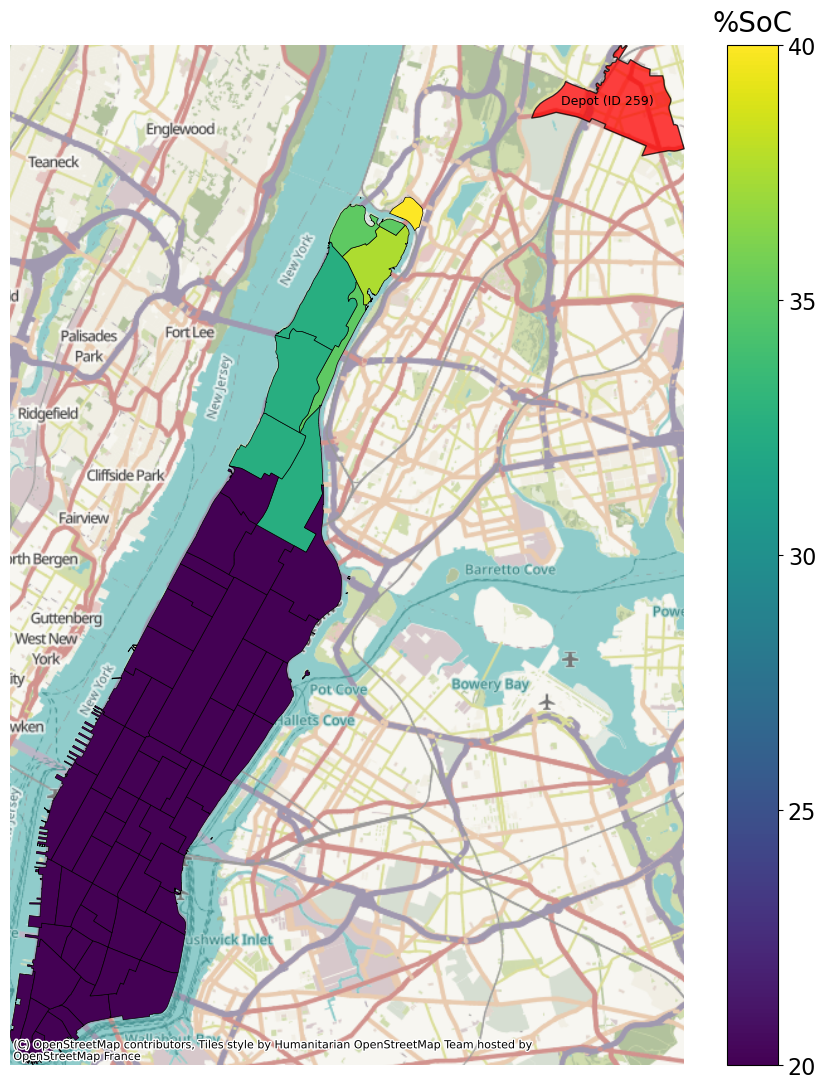} }}%
    \caption{Illustration of the early RTD thresholds under different depot locations.}
    \label{fig:visualization_depot_location}
\end{figure}

These differences in operational patterns translate directly into profit outcomes and the value of charge awareness. \cref{tab:depot_location_profit} reports profits under charge-aware and charge-oblivious policies for three representative depot locations. As the depot moves farther from the demand center, overall profit declines sharply, reflecting the increased difficulty of charging and repositioning vehicles. At the same time, the performance gap between charge-aware and charge-oblivious policies widens. This pattern indicates that distant depots raise the bar for effective operational control: anticipating charging needs and making battery-aware repositioning decisions become increasingly critical for maintaining system performance. In dense urban environments such as Manhattan—where constructing depots close to demand centers is often prohibitively expensive—these results underscore the practical importance of charge-aware fleet management.

\begin{table}[h]
\centering
\label{tab:depot_location_profit}
\renewcommand{\arraystretch}{1.2}
\begin{tabular}{c|ccc}
\hline
Depot zone &
Charge-aware profit &
Charge-oblivious profit &
Relative difference \\
\hline
66  & 145{,}402.00 & 141{,}161.77 & 3.00\% \\
220 & 137{,}182.58 & 119{,}248.43 & 15.04\% \\
116 & 130{,}267.17 & 109{,}726.19 & 18.72\% \\
\hline
\end{tabular}
\caption{Impact of depot location on profits and the value of charge awareness.}
\end{table}

\textit{Impact of the number of depots.}
Beyond depot location, the number of depots also plays a central role in fleet performance. To study this effect, we gradually add depots in zones $[220,116,246,66,112,119]$ to progressively expand spatial coverage, including into the Bronx and Brooklyn. The results are reported in \cref{fig:value_comparison}(d).

\subsection{Impact of Charging Time}\label{app:nonlinear}

The main body of the paper assumes an affine charging-time function $\tau_k=\batteryvar_1 + \batteryvar_2(K-k) \;k\in\mathcal K,$ where $\batteryvar_1$ captures a fixed setup time and $\batteryvar_2$ captures the marginal charging time for each battery level. In this appendix, we examine the impact of the charging-time specification on our model. We first vary $\batteryvar_1$ and $\batteryvar_2$ in the affine charging function, and then consider nonlinear charging functions.

\begin{figure}[h]%
    \centering
    \subfloat[Affine charging time\label{fig:charge_time_linear}]{{\includegraphics[width=0.45\textwidth]{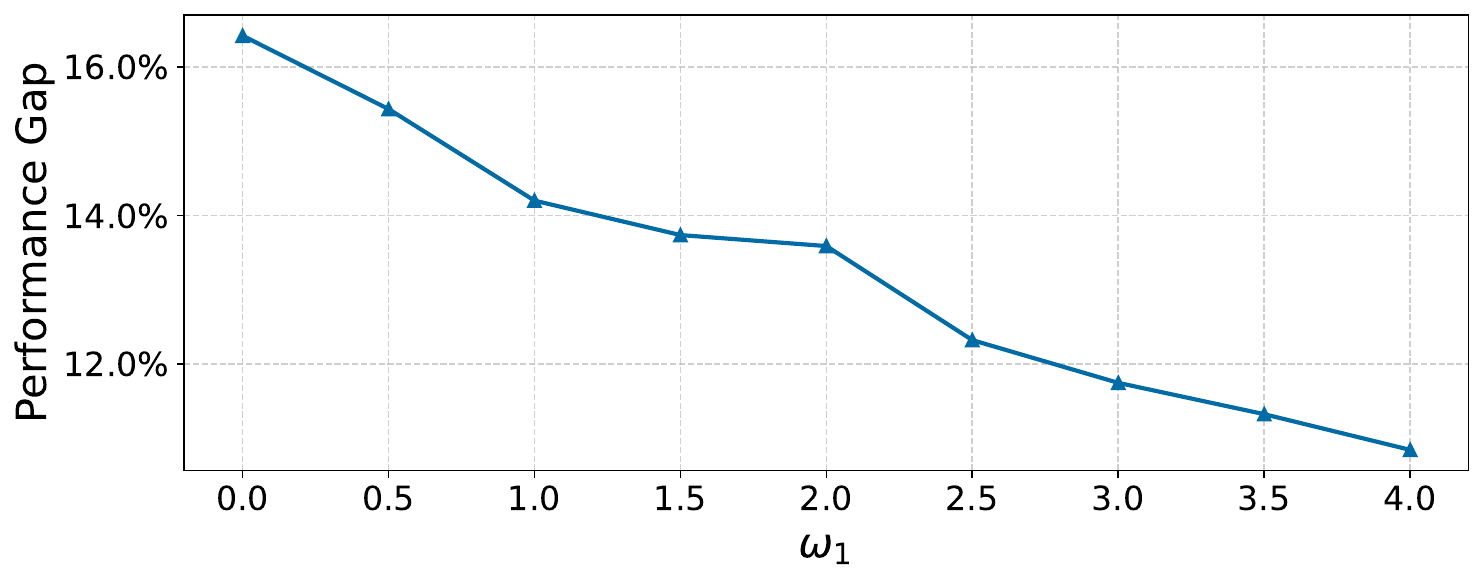}}}
    \hspace{0.5cm}
    \subfloat[Nonlinear charging time\label{fig:charge_time_nonlinear}]{{\includegraphics[width=0.45\textwidth]{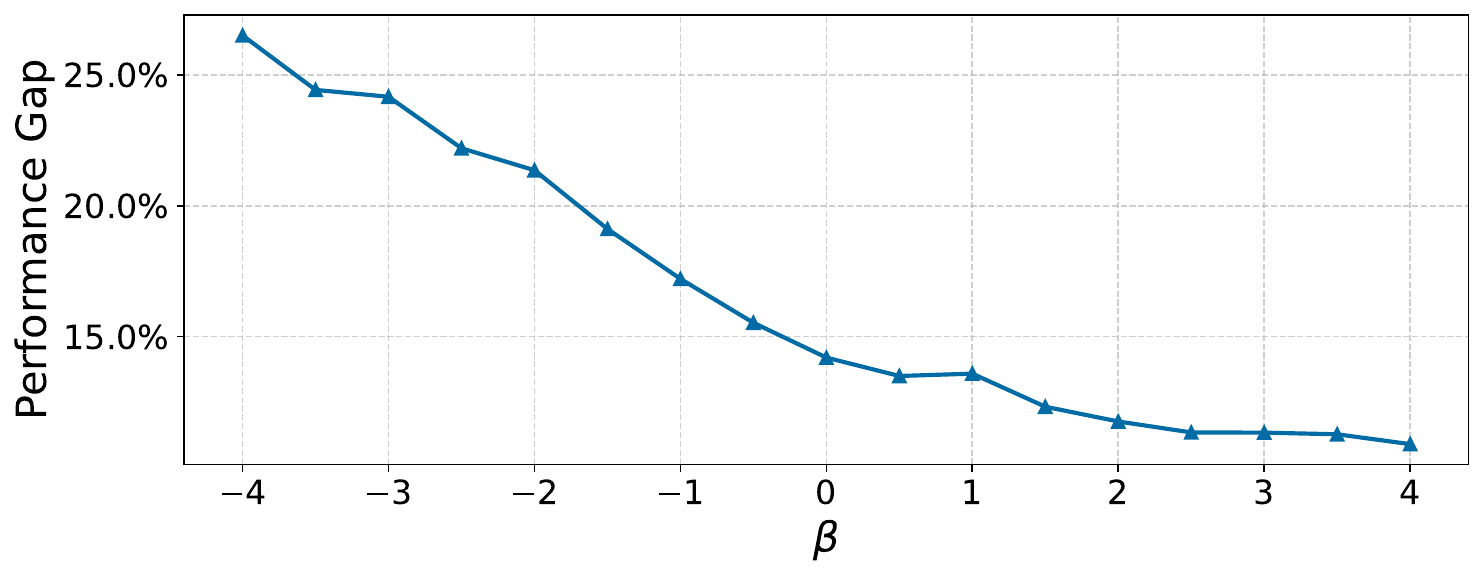}}}
    \caption{Performance gap between the charge-aware and charge-oblivious solutions under different charging-time functions.}
    \label{fig:charge_time}
\end{figure}

We begin with the affine case. %Since the structural results in \cref{sec:results} hold for any $\batteryvar_2\geq 0$ and $\batteryvar_1>0$, the RTD-threshold and battery-sensitivity results continue to apply. 
In the case study in \cref{sec:numerics}, we use a constant charging time $\tau_k=4\;\forall k$, i.e., $\batteryvar_1=4$ and $\batteryvar_2=0$. To vary the charging time function while holding $\tau_0 = 4$ fixed, we set
\[
\tau_k(\batteryvar_1)
=
\batteryvar_1+\frac{4-\batteryvar_1}{K}(K-k),
\qquad \batteryvar_1\in(0,4].
\]
Thus, decreasing $\batteryvar_1$ increases the slope $\batteryvar_2=(4-\batteryvar_1)/K$, making the charging time more sensitive to the vehicle's remaining SoC. As shown in \cref{fig:charge_time}(a), the performance gap between the charge-aware and charge-oblivious solutions increases as $\batteryvar_1$ decreases (i.e., as $\batteryvar_2$ increases). The intuition is that early RTD becomes more valuable when returning with a higher SoC saves not only future detours to the depot, but also charging time. For instance, when $\batteryvar_2$ is large, we find that RTD thresholds can become as high as $55\%$ SoC for zones close to the depot. A larger $\batteryvar_2$ similarly magnifies the difference in battery sensitivities across the locations and makes charge-aware repositioning more important. 

We next consider nonlinear charging times. Let $g_h(\beta)$ denote the marginal time required to charge from level $h-1$ to level $h$, and define
\[
g_h(\beta)
=
3\cdot
\frac{\exp(\beta h/K)}
{\sum_{\ell=1}^K \exp(\beta \ell/K)},
\qquad
\tau_k(\beta)
=
1+\sum_{h=k+1}^K g_h(\beta).
\]
This normalization keeps $\tau_0 = 4$. The parameter $\beta$ controls the curvature of the charging function: when $\beta=0$, we recover the affine charging function with $\batteryvar_1 = 1$ and $\batteryvar_2 = 3/K$; when $\beta>0$, marginal charging time is concentrated near the top of the battery ladder, and $\tau_k\to4$ for all $k<K$ as $\beta\to\infty$; when $\beta<0$, marginal charging time is concentrated near the bottom of the battery ladder, so returning with positive SoC substantially reduces the remaining charging time. 

While nonlinear charging times can preclude simple threshold structures for RTD and charge-aware repositioning---and sufficiently irregular charging curves can induce highly non-monotone decisions across SoC levels---both forms of charge-aware control remain important for operational efficiency. As shown in \cref{fig:charge_time}(b), smaller values of $\beta$ lead to a larger performance gap, sometimes exceeding $25\%$. This is accompanied by higher RTD thresholds and greater dispersion in battery sensitivity across locations. At the same time, even under strong curvature in the opposite direction, the performance gap remains above $10\%$. These findings show that the qualitative insights of the paper are robust to nonlinear charging functions.

\section{Proofs}

\subsection{Proofs of Results in \cref{sec:model}}\label{app:proof_model}

The following result formalizes the claim in \cref{sec:charge_aware} that it is without loss of optimality to launch vehicles only at full charge.

\begin{proposition}
\label{prop:launch_full_charge}
There exists an optimal solution $(\boldsymbol f^\star,\boldsymbol w^\star,\boldsymbol{\rtd}^\star)$ in which vehicles are launched from the depot only at full charge, i.e., $w^\ast_{Djk}=0$ for all $j\in\mathcal N$ and $k\in\{0,\dots,K-1\}$.
\end{proposition}

\begin{proof}{Proof of \cref{prop:launch_full_charge}.}
Consider any optimal solution of \eqref{lp:fluid} in which vehicles may be launched at any level. We shall construct from it a feasible solution with the same objective value that launches vehicles only at full charge, which proves the claim. Because every service or repositioning action lowers the SoC by one level, the flow decomposes into paths that start with a launch and end with an RTD. Shift every path that starts with a launch at level $k<K$ upward by $K-k$ levels. The shifted path uses only levels in $\bracket{1,\ldots,K-1}$, because its RTD level increases from some $r\le k-1$ to $r+K-k\le K-1$, and it has the same revenue, costs, travel times, and destinations; the demand constraints, which aggregate over charge levels, are unaffected. The charging term $\batteryvar_2\bigl(\sum_{j,k}k\,w_{Djk}-\sum_{i,k}k\,\rtd_{ik}\bigr)$ in \eqref{c:vehicle_time} is also unchanged, because the launch level and the RTD level both increase by $K-k$. Hence the shifted solution is feasible and optimal, and it launches vehicles only at full charge.
\end{proof}

\subsection{Proofs of Results in \cref{sec:results}} \label{app:results}

We organize this appendix as follows. In Appendix \ref{app:dual}, we first derive the dual of the primal LP and record the associated KKT conditions that will be used throughout the analysis. We then present several auxiliary lemmas in Appendix \ref{app:auxiliary_section3}. Building on these preliminaries, Appendix \ref{app:section_3_proof} contains the proofs of the main results, and the proofs of the corresponding corollaries are provided in Appendix \ref{app:section_3_corollary}.

\subsubsection{Dual Formulation and KKT Conditions}\label{app:dual}

\paragraph{Dual variables.}
For each constraint in the primal LP, we introduce the following multipliers:
\begin{align*}
&\sum_{k\in\mathcal K} f_{ik}\le \lambda_i,\quad \forall i\in\mathcal N
&&\longleftrightarrow\quad \mu_i\ge 0,\\[3pt]
&\sum_{j\in\mathcal N}f_{j,k+1}p_{ji}+\sum_{j\in\mathcal N\cup\depotset}w_{ji,k+1}
=
f_{ik}+\sum_{j\in\mathcal N} w_{ijk}+\rtd_{ik},
\quad \forall i\in\mathcal N,\; k\in\bracket{0,\ldots,K-1}
&&\longleftrightarrow\quad \pi_{ik}\ \textnormal{free},\\[3pt]
&f_{i0}=0,\quad f_{iK}=0,\qquad \forall i\in\mathcal N
&&\longleftrightarrow\quad \alpha_{i0},\alpha_{iK}\ \textnormal{free},\\[3pt]
&w_{ij0}=0,\qquad \forall i,j\in\mathcal N
&&\longleftrightarrow\quad \xi_{ij0}\ \textnormal{free},\\[3pt]
&w_{jiK}=0,\qquad \forall i,j\in\mathcal N
&&\longleftrightarrow\quad \xi_{jiK}\ \textnormal{free},\\[3pt]
&\rtd_{iK}=0,\qquad \forall i\in\mathcal N
&&\longleftrightarrow\quad \varrho_{iK}\ \textnormal{free},\\[3pt]
&\sum_{j\in\mathcal N} w_{DjK}
=
\sum_{k\in\mathcal K}\sum_{i\in\mathcal N} \rtd_{ik}
&&\longleftrightarrow\quad \zeta\ \textnormal{free},\\[3pt]
&w_{Djk}=0,\qquad \forall j\in\mathcal N,\; k\in\bracket{0,\ldots,K-1}
&&\longleftrightarrow\quad \gamma_{Djk}\ \textnormal{free},\\[3pt]
&\sum_{i,j\in\mathcal N}\sum_{k\in\mathcal K} f_{ik}p_{ij}\tauservice_{ij}
+\sum_{i\in\mathcal N\cup\depotset,\;j\in\mathcal N}\sum_{k\in\mathcal K} w_{ijk}\taurep_{ij}
+\sum_{i\in\mathcal N}\sum_{k\in\mathcal K}\rtd_{ik}\parenthesis{\taurep_{iD}+\tau_k}
\le\size
&&\longleftrightarrow\quad \nu\ge0.
\end{align*}

\paragraph{KKT conditions and reduced costs.}
Let $L$ denote the Lagrangian of the primal LP with the dual variables above. For every primal variable $x$, we write $\reduced(x)$ for the reduced cost of that variable.

\medskip
\noindent\emph{Service actions.}
For every $i\in\mathcal N$ and $k\in\bracket{1,\ldots,K-1}$,
\begin{equation}
\label{eq:RC-f}
\reduced(f_{ik})
:=
\sum_{j\in\mathcal N} r_{ij}p_{ij}
-\mu_i
-\nu\sum_{j\in\mathcal N} p_{ij}\tauservice_{ij}
-\pi_{ik}
+\sum_{j\in\mathcal N} p_{ij}\pi_{j,k-1}
\le 0.
\end{equation}
Moreover, at optimality
\[
f^\ast_{ik}>0\quad\Longrightarrow\quad \reduced(f^\ast_{ik})=0.
\]
% At the boundary levels,
% \begin{align}
% \label{eq:RC-f0}
% \sum_{j\in\mathcal N} r_{ij}p_{ij}-\mu_i-\alpha_{i0} &= 0,\qquad \forall i\in\mathcal N,\\
% \label{eq:RC-fK}
% \sum_{j\in\mathcal N} r_{ij}p_{ij}
% -\mu_i
% -\nu\sum_{j\in\mathcal N} p_{ij}\tauservice_{ij}
% +\sum_{j\in\mathcal N} p_{ij}\pi_{j,K-1}
% -\alpha_{iK} &= 0,\qquad \forall i\in\mathcal N.
% \end{align}

\medskip
\noindent\emph{Repositioning among demand locations.}
For every $i\neq j\in\mathcal N$ and $k\in\bracket{1,\ldots,K-1}$,
\begin{equation}
\label{eq:RC-w}
\reduced(w_{ijk})
:=
-c_{ij}
-\nu\taurep_{ij}
-\pi_{ik}
+\pi_{j,k-1}
\le 0.
\end{equation}
Again, at optimality
\[
w^\ast_{ijk}>0\quad\Longrightarrow\quad \reduced(w^\ast_{ijk})=0.
\]
For the stay-put variable $w_{iik}$, we have
\begin{equation}
\label{eq:RC-idle}
\reduced(w_{iik})
=
-c_{ii}-\nu\taurep_{ii}-\pi_{ik}+\pi_{i,k-1}\le 0.
\end{equation}

% For the boundary cases,
% \begin{align}
% \label{eq:RC-w0}
% -c_{ij}-\pi_{i0}-\xi_{ij0} &= 0,
% \qquad \forall i,j\in\mathcal N,\\
% \label{eq:RC-wK}
% -c_{ji}-\nu\taurep_{ji}+\pi_{i,K-1}-\xi_{jiK} &= 0,
% \qquad \forall i,j\in\mathcal N.
% \end{align}

\medskip
\noindent\emph{Launch from the depot.}
For every $j\in\mathcal N$,
\begin{equation}
\label{eq:RC-wD}
\reduced(w_{DjK})
:=
-c_{Dj}
-\nu\taurep_{Dj}
+\pi_{j,K-1}
+\zeta
\le 0,
\end{equation}
with
\[
w^\ast_{DjK}>0\quad\Longrightarrow\quad \reduced(w^\ast_{DjK})=0.
\]
% For the prohibited depot arcs at lower levels,
% \begin{equation}
% \label{eq:RC-wD_boundary}
% -c_{Dj}-\gamma_{Djk}=0,
% \qquad \forall j\in\mathcal N,\; k\in\bracket{0,\ldots,K-1}.
% \end{equation}

\medskip
\noindent\emph{RTD actions.}
For every $i\in\mathcal N$ and $k\in\bracket{0,\ldots,K-1}$,
\begin{equation}
\label{eq:RC-rtd}
\reduced(\rtd_{ik})
:=
-c_{iD}
-\nu\parenthesis{\taurep_{iD}+\tau_k}
-\pi_{ik}
-\zeta
\le 0.
\end{equation}
The variable $\rtd_{iK}$ is fixed at zero by \eqref{eq:node_balance_boundary}, so its reduced cost involves the free multiplier $\varrho_{iK}$ and imposes no restriction on $(\boldsymbol\mu,\nu,\zeta,\boldsymbol\pi)$; the same holds for $w_{Dj0}$, whose multiplier $\gamma_{Dj0}$ is free.
% Previous version (superseded by the boundary condition $\rtd_{iK}=0$ in \eqref{eq:node_balance_boundary}):
% For $k=K$, we have
% \begin{equation}\label{eq:RC-rtd-K}
% \reduced(\rtd_{iK})
% :=
% -c_{iD}
% -\nu\parenthesis{\taurep_{iD}+\tau_K}
% -\zeta
% \le 0.
% \end{equation}
Complementary slackness gives
\[
\rtd^\ast_{ik}>0\quad\Longrightarrow\quad \reduced(\rtd^\ast_{ik})=0.
\]

In addition to the KKT conditions above for positive primal flows, the demand constraints satisfy
\begin{equation}
\label{eq:CS_demand}
\mu^\ast_i\Big(\lambda_i-\sum_{k\in\mathcal K} f^\ast_{ik}\Big)=0,
\qquad \forall i\in\mathcal N.
\end{equation}

\paragraph{Selected optimal dual.} Since optimal dual solutions need not be unique, we fix a canonical selection in two steps. First, among all optimal dual solutions, we select one whose demand multipliers minimize $\sum_{i\in\mathcal N}\mu_i^2$. The set of optimal dual solutions is a nonempty polyhedron and $\sum_{i}\mu_i^2$ is strictly convex in $\boldsymbol\mu$, so the minimizing vector $\boldsymbol\mu^\star$ is unique; the fleet multiplier $\nu^\star$ is then determined by the dual objective, $\sum_{i}\lambda_i\mu^\star_i+\size\nu^\star=\OPT(\instance)$. Let $(\boldsymbol\mu^\star,\nu^\star,\zeta^\star,\tilde{\boldsymbol\pi})$ be any optimal dual solution with these demand and fleet multipliers, where $\zeta^\star$ denotes its depot multiplier and $\tilde{\boldsymbol\pi}$ its multipliers of the flow-balance constraints. Second, we replace $\tilde{\boldsymbol\pi}$ by the smallest values of these multipliers that are compatible with $(\boldsymbol\mu^\star,\nu^\star,\zeta^\star)$: let
\begin{equation}\label{eq:RTD_value}
R_{ik}:=-c_{iD}-\nu^\star\parenthesis{\taurep_{iD}+\tau_k}-\zeta^\star,\qquad i\in\mathcal N,\ k\in\bracket{0,\ldots,K-1},
\end{equation}
denote the value of RTD, and define recursively in $k$
\begin{equation}\label{eq:pi_recursion}
\pi^\star_{i0}:=R_{i0},\qquad
\pi^\star_{ik}:=\max\Bigl\{R_{ik},\ \sum_{j\in\mathcal N}p_{ij}\bigl(r_{ij}-\nu^\star\tauservice_{ij}+\pi^\star_{j,k-1}\bigr)-\mu^\star_i,\ \max_{j\in\mathcal N}\bigl\{-c_{ij}-\nu^\star\taurep_{ij}+\pi^\star_{j,k-1}\bigr\}\Bigr\}
\end{equation}
for $k\in\bracket{1,\ldots,K-1}$. The three terms are the values of RTD, service, and repositioning (including idling), so $\pi^\star_{ik}$ is the largest profit that a single vehicle at $(i,k)$ can earn up to and including its next RTD when vehicle time is priced at $\nu^\star$ and each request served at $i$ is charged $\mu^\star_i$.

\begin{lemma}\label{lem:canonical_dual}
$(\boldsymbol\mu^\star,\nu^\star,\zeta^\star,\boldsymbol\pi^\star)$ is an optimal dual solution, and $\pi^\star_{ik}\le\tilde\pi_{ik}$ for all $(i,k)$.
\end{lemma}

\begin{proof}{Proof.}
Dual feasibility of $\tilde{\boldsymbol\pi}$ means that the reduced costs in \eqref{eq:RC-f}, \eqref{eq:RC-w}, \eqref{eq:RC-idle}, and \eqref{eq:RC-rtd} are nonpositive, i.e., $\tilde\pi_{ik}$ is at least each term in \eqref{eq:pi_recursion} evaluated at $\tilde{\boldsymbol\pi}$; at $k=0$, only RTD is available. Induction on $k$ therefore gives $\pi^\star_{ik}\le\tilde\pi_{ik}$. By construction, $\boldsymbol\pi^\star$ satisfies the same reduced-cost inequalities. The only other dual constraints, \eqref{eq:RC-wD}, involve the variables $\boldsymbol\pi$ only through $+\pi_{j,K-1}$, so they continue to hold when $\tilde{\boldsymbol\pi}$ is replaced by the smaller $\boldsymbol\pi^\star$. Finally, the right-hand sides of \eqref{c:node_balance} and \eqref{c:depot_map} are zero, so the dual objective $\sum_{i}\lambda_i\mu^\star_i+\size\nu^\star$ does not involve the variables $\boldsymbol\pi$, and the new dual solution is optimal.
\end{proof}

Throughout the proofs of the results in \cref{sec:results}, $(\boldsymbol\mu^\star,\nu^\star,\zeta^\star,\boldsymbol\pi^\star)$ denotes such a selected optimal dual; the battery sensitivities $\Delta_{ik}$ and the thresholds $\kappa_i^\ast$ are defined from it. Because it is an optimal dual solution, complementary slackness holds between it and \emph{every} optimal primal solution. Statements that assume strict complementary slackness refer to this selected dual.

\subsubsection{Auxiliary Results}\label{app:auxiliary_section3}

\begin{lemma}[Properties of dual variables]\label{lem:pi-monotone} The selected optimal dual satisfies (i)--(ii), and every optimal solution satisfies (iii):
\begin{enumerate}[label=(\roman*)]
    \item $\nu^\star \geq 0$;
    \item For every $i\in\mathcal N$ and every $k\in\bracket{0,\ldots,K-2}$,
    $\pi^\star_{i,k+1}-\pi^\star_{ik}\ge \batteryvar_2 \nu^\star \ge 0.$
\item $w^\star_{iik}=0\;\forall i\in\mathcal N,\ \forall k\in\mathcal{K}.$
\end{enumerate}
\end{lemma}

\begin{proof}{Proof.}
\noindent\textbf{(i)} Constraint \eqref{c:vehicle_time} is an inequality constraint of a maximization problem, so its multiplier is nonnegative.

\medskip
\noindent\textbf{(ii)} By \eqref{eq:RTD_value}, $R_{i,k+1}-R_{ik}=\nu^\star(\tau_k-\tau_{k+1})=\batteryvar_2\nu^\star$. We show by induction on $k$ that $\pi^\star_{i,k+1}-\pi^\star_{ik}\ge\batteryvar_2\nu^\star$ for all $i\in\mathcal N$. For $k=0$, \eqref{eq:pi_recursion} gives $\pi^\star_{i1}\ge R_{i1}=R_{i0}+\batteryvar_2\nu^\star=\pi^\star_{i0}+\batteryvar_2\nu^\star$. Let $k\ge1$ and suppose that the claim holds at level $k-1$ for all locations. Each term in \eqref{eq:pi_recursion} at level $k+1$ exceeds the same term at level $k$ by at least $\batteryvar_2\nu^\star$: the RTD term by exactly $\batteryvar_2\nu^\star$, the service term by $\sum_{j}p_{ij}(\pi^\star_{jk}-\pi^\star_{j,k-1})$, and each repositioning term by $\pi^\star_{jk}-\pi^\star_{j,k-1}$, where the last two are at least $\batteryvar_2\nu^\star$ by the induction hypothesis. Hence so does their maximum.

\medskip
\noindent\textbf{(iii)} For $k\in\bracket{1,\ldots,K-1}$, \eqref{eq:RC-idle} and part (ii) give $\reduced(w_{iik})\le-c_{ii}-\nu^\star\taurep_{ii}-\batteryvar_2\nu^\star<0$, because $c_{ii}>0$ and $\nu^\star\ge0$. By complementary slackness, $w^\star_{iik}=0$ in every optimal solution; for $k\in\bracket{0,K}$, this holds by \eqref{eq:node_balance_boundary}.
\end{proof}

The next lemma records the property of the selected demand multipliers that drives the comparative statics in \cref{prop:comparison}. Recall from \cref{sec:early_rtd} that two locations $i,i'$ are identical in all primitives if $\lambda_i=\lambda_{i'}$ and rows $i$ and $i'$, as well as columns $i$ and $i'$, of every primitive coincide, i.e., $x_{i\ell}=x_{i'\ell}$ and $x_{\ell i}=x_{\ell i'}$ for each $x\in\{p,r,c,\taurep\}$ and every $\ell\in\mathcal N\cup\depotset$, including $\ell\in\bracket{i,i'}$. Under conditions (i)--(vi) of \cref{prop:comparison}, the equalities that a condition relaxes are replaced by the stated inequalities, and all others remain in force.

\begin{lemma}[Demand multipliers of the selected dual]\label{lem:mu_selection}
Let $i,i'\in\mathcal N$ be identical in all primitives except that one or more of the conditions (i)--(vi) of \cref{prop:comparison} hold, and define
\begin{equation}\label{eq:eta_def}
\eta:=(c_{i'D}-c_{iD})+\nu^\star(\taurep_{i'D}-\taurep_{iD})\ \ge0,
\qquad
\eta_s:=\sum_{j\in\mathcal N}p_{ij}\Bigl[(r_{i'j}-r_{ij})+\nu^\star\bigl(\tauservice_{ij}-\tauservice_{i'j}\bigr)\Bigr]\ \ge0.
\end{equation}
Then the selected optimal dual satisfies $\mu^\star_{i'}-\mu^\star_i\le\eta+\eta_s$.
\end{lemma}

\begin{proof}{Proof.}
Let $\delta:=\mu^\star_{i'}-\mu^\star_i$ and suppose, for contradiction, that $\delta>\eta+\eta_s$. We shall shift part of the demand multiplier of location $i'$ to that of location $i$ and show that this produces another optimal dual solution that the selection rule would have preferred, leading to a contradiction. Specifically, set $\theta:=\delta-\eta-\eta_s>0$, and define $\boldsymbol\mu'$ by $\mu'_i:=\mu^\star_i+\theta$, $\mu'_{i'}:=\mu^\star_{i'}-\theta$, and $\mu'_l:=\mu^\star_l$ for $l\notin\bracket{i,i'}$; since $\theta\le\delta$, we have $\mu'_{i'}\ge\mu^\star_{i'}-\delta=\mu^\star_i\ge0$, so $\boldsymbol\mu'\ge0$. Let $\boldsymbol\pi'$ denote the variables obtained from the recursion \eqref{eq:pi_recursion} with $(\boldsymbol\mu',\nu^\star,\zeta^\star)$ in place of $(\boldsymbol\mu^\star,\nu^\star,\zeta^\star)$. We shall show that $(\boldsymbol\mu',\nu^\star,\zeta^\star,\boldsymbol\pi')$ is a dual feasible solution whose objective is no larger than that of the selected dual and whose demand multipliers satisfy $\sum_{l}(\mu'_l)^2\le\sum_{l}(\mu^\star_l)^2$, which contradicts either the optimality or the selection of $(\boldsymbol\mu^\star,\nu^\star,\zeta^\star,\boldsymbol\pi^\star)$. To this end, we first prove by induction on $k$ that, for every $k\in\bracket{0,\ldots,K-1}$,
\begin{enumerate}[label=(\alph*)]
\item $\pi'_{lk}\le\pi^\star_{lk}$ for all $l\notin\bracket{i,i'}$;
\item $\pi'_{ik}+\pi'_{i'k}\le\pi^\star_{ik}+\pi^\star_{i'k}$;
\item $\max\{\pi'_{ik},\pi'_{i'k}\}\le\max\{\pi^\star_{ik},\pi^\star_{i'k}\}$.
\end{enumerate}
At $k=0$, all three statements hold with equality because $\pi'_{l0}=R_{l0}=\pi^\star_{l0}$ for every $l\in\mathcal N$. Fix $k\ge1$ and suppose that (a)--(c) hold at level $k-1$. Below, a term of \eqref{eq:pi_recursion} at level $k$ is said to be evaluated ``under $\boldsymbol\pi^\star$'' or ``under $\boldsymbol\pi'$'' according to whether the values $\pi^\star_{j,k-1}$ or $\pi'_{j,k-1}$, $j\in\mathcal N$, are used.

\emph{Step 1: the pair as seen from any location.} A location $l\in\mathcal N$ interacts with the pair $\bracket{i,i'}$ in \eqref{eq:pi_recursion} in two ways: through the destination sum $\sum_{j\in\mathcal N}p_{lj}\pi_{j,k-1}$ in its service term, and through its repositioning options toward $i$ and $i'$. We show that neither is larger under $\boldsymbol\pi'$ than under $\boldsymbol\pi^\star$. Since $i$ and $i'$ are identical in all primitives, $p_{li}=p_{li'}$, so the destination sum weights $\pi_{i,k-1}$ and $\pi_{i',k-1}$ equally, and
\[
\sum_{j\in\mathcal N}p_{lj}\pi'_{j,k-1}-\sum_{j\in\mathcal N}p_{lj}\pi^\star_{j,k-1}
=
\sum_{j\notin\bracket{i,i'}}p_{lj}\bigl(\pi'_{j,k-1}-\pi^\star_{j,k-1}\bigr)
+p_{li}\Bigl[\bigl(\pi'_{i,k-1}+\pi'_{i',k-1}\bigr)-\bigl(\pi^\star_{i,k-1}+\pi^\star_{i',k-1}\bigr)\Bigr]
\le0
\]
by (a) and (b). Likewise, $c_{li}=c_{li'}$ and $\taurep_{li}=\taurep_{li'}$ (for $l\in\bracket{i,i'}$, these are the within-pair equalities in the definition of identical locations, e.g., $c_{ii}=c_{ii'}$ and $\taurep_{ii}=\taurep_{ii'}$ for $l=i$), so the better of the two repositioning options toward the pair is worth $-c_{li}-\nu^\star\taurep_{li}+\max\{\pi'_{i,k-1},\pi'_{i',k-1}\}$ under $\boldsymbol\pi'$, which by (c) is at most its counterpart $-c_{li}-\nu^\star\taurep_{li}+\max\{\pi^\star_{i,k-1},\pi^\star_{i',k-1}\}$ under $\boldsymbol\pi^\star$; each repositioning option toward a destination $j\notin\bracket{i,i'}$ is worth weakly less under $\boldsymbol\pi'$ by (a). Hence, for every $l\in\mathcal N$, both the destination sum in the service term and the repositioning term at $(l,k)$ are weakly smaller under $\boldsymbol\pi'$ than under $\boldsymbol\pi^\star$.

\emph{Step 2: statement (a) at level $k$.} A location outside the pair is affected by the shift only through its destination sum and its repositioning options, both of which weakly decrease by Step~1; its own multiplier and its RTD term are unchanged. Formally, for $l\notin\bracket{i,i'}$ we have $\mu'_l=\mu^\star_l$ and the RTD term $R_{lk}$ does not involve the variables $\boldsymbol\pi$, so Step~1 gives $\pi'_{lk}\le\pi^\star_{lk}$.

\emph{Step 3: statements (b)--(c) at level $k$.} To prove (b) and (c), we express $\pi^\star_{ik}$, $\pi^\star_{i'k}$, $\pi'_{ik}$, and $\pi'_{i'k}$ through the service and repositioning terms at level $k$ and compare them directly. Let $V_{i'}$ and $V'_{i'}$ denote the service term at $(i',k)$ with the \emph{original} multiplier $\mu^\star_{i'}$, evaluated under $\boldsymbol\pi^\star$ and under $\boldsymbol\pi'$, respectively, and let $P_{i'}$ and $P'_{i'}$ (resp.\ $P_i$ and $P'_i$) denote the repositioning term at $(i',k)$ (resp.\ at $(i,k)$) under $\boldsymbol\pi^\star$ and under $\boldsymbol\pi'$; primes thus indicate evaluation under the shifted dual. Four facts relate these quantities. First, by Step~1, $V'_{i'}\le V_{i'}$, $P'_{i'}\le P_{i'}$, and $P'_i\le P_i$. Second, since $c_{ij}\ge c_{i'j}$ and $\taurep_{ij}\ge\taurep_{i'j}$ for all $j$, each repositioning option at $i$ is worth at most the corresponding option at $i'$, so $P_i\le P_{i'}$ and $P'_i\le P'_{i'}$. Third, $R_{ik}=R_{i'k}+\eta$ by \eqref{eq:RTD_value} and \eqref{eq:eta_def}. Fourth, since $p_{ij}=p_{i'j}$ for all $j$, the destination sums at $i$ and $i'$ coincide, so by the definitions of $\delta$, $\eta_s$, and $\theta$, the service term at $(i,k)$ equals $V_{i'}-\eta_s+\delta=V_{i'}+\eta+\theta$ under $\boldsymbol\pi^\star$ and $V'_{i'}-\eta_s+(\delta-\theta)=V'_{i'}+\eta$ under $\boldsymbol\pi'$, while the service term at $(i',k)$ equals $V'_{i'}+\theta$ under $\boldsymbol\pi'$. Therefore,
\[
\pi^\star_{ik}=\max\{R_{i'k}+\eta,\ V_{i'}+\eta+\theta,\ P_i\},
\qquad
\pi^\star_{i'k}=\max\{R_{i'k},\ V_{i'},\ P_{i'}\},
\]
\[
\pi'_{ik}=\max\{R_{i'k}+\eta,\ V'_{i'}+\eta,\ P'_i\},
\qquad
\pi'_{i'k}=\max\{R_{i'k},\ V'_{i'}+\theta,\ P'_{i'}\}.
\]
For (c), since $V'_{i'}\le V_{i'}$, $P'_i\le P'_{i'}\le P_{i'}$, and $\max\{\eta,\theta\}\le\eta+\theta$,
\[
\max\{\pi'_{ik},\pi'_{i'k}\}
\le\max\{R_{i'k}+\eta,\ V_{i'}+\max\{\eta,\theta\},\ P_{i'}\}
\le\max\{R_{i'k}+\eta,\ V_{i'}+\eta+\theta,\ P_{i'}\}
=\max\{\pi^\star_{ik},\pi^\star_{i'k}\},
\]
where the last equality uses $P_i\le P_{i'}$. For (b), we first show that, holding the values $\pi'_{j,k-1}$ fixed, moving the $\theta$ units of service value from $i'$ back to $i$ does not decrease the sum of the two maxima:
\begin{equation}\label{eq:shift_pair}
\max\{R_{i'k}+\eta,\ V'_{i'}+\eta,\ P'_i\}+\max\{R_{i'k},\ V'_{i'}+\theta,\ P'_{i'}\}
\ \le\
\max\{R_{i'k}+\eta,\ V'_{i'}+\eta+\theta,\ P'_i\}+\max\{R_{i'k},\ V'_{i'},\ P'_{i'}\}.
\end{equation}
We distinguish two cases.

\emph{Case 1: $V'_{i'}+\theta\le\max\{R_{i'k},P'_{i'}\}$.} Then $\max\{R_{i'k},V'_{i'}+\theta,P'_{i'}\}=\max\{R_{i'k},P'_{i'}\}=\max\{R_{i'k},V'_{i'},P'_{i'}\}$, and $\max\{R_{i'k}+\eta,V'_{i'}+\eta,P'_i\}\le\max\{R_{i'k}+\eta,V'_{i'}+\eta+\theta,P'_i\}$ because $\theta>0$. Adding the equality and the inequality gives \eqref{eq:shift_pair}.

\emph{Case 2: $V'_{i'}+\theta>\max\{R_{i'k},P'_{i'}\}$.} Then $\max\{R_{i'k},V'_{i'}+\theta,P'_{i'}\}=V'_{i'}+\theta$. Moreover, since $P'_i\le P'_{i'}$, $\max\{R_{i'k}+\eta,P'_i\}\le\max\{R_{i'k},P'_{i'}\}+\eta<V'_{i'}+\eta+\theta$, so $\max\{R_{i'k}+\eta,V'_{i'}+\eta+\theta,P'_i\}=V'_{i'}+\eta+\theta,$
and, by the same bound on $\max\{R_{i'k}+\eta,P'_i\}$, $\max\{R_{i'k}+\eta,V'_{i'}+\eta,P'_i\}\le\max\{R_{i'k},V'_{i'},P'_{i'}\}+\eta$. Therefore,
\[
\max\{R_{i'k}+\eta,V'_{i'}+\eta,P'_i\}+\max\{R_{i'k},V'_{i'}+\theta,P'_{i'}\}
\le\max\{R_{i'k},V'_{i'},P'_{i'}\}+\eta+V'_{i'}+\theta,
\]
and the right-hand side equals $\max\{R_{i'k}+\eta,V'_{i'}+\eta+\theta,P'_i\}+\max\{R_{i'k},V'_{i'},P'_{i'}\}$ by the two identities above, which is \eqref{eq:shift_pair}. Combining \eqref{eq:shift_pair} with $V'_{i'}\le V_{i'}$, $P'_i\le P_i$, and $P'_{i'}\le P_{i'}$,
\begin{align*}
\pi'_{ik}+\pi'_{i'k}
&\le\max\{R_{i'k}+\eta,\ V'_{i'}+\eta+\theta,\ P'_i\}+\max\{R_{i'k},\ V'_{i'},\ P'_{i'}\}\\
&\le\max\{R_{i'k}+\eta,\ V_{i'}+\eta+\theta,\ P_i\}+\max\{R_{i'k},\ V_{i'},\ P_{i'}\}
=\pi^\star_{ik}+\pi^\star_{i'k}.
\end{align*}
This completes the induction.

\emph{Step 4: dual feasibility and contradiction.} By construction, $\pi'_{lk}$ is at least each of the RTD, service, and repositioning terms in \eqref{eq:pi_recursion}, so $(\boldsymbol\mu',\nu^\star,\zeta^\star,\boldsymbol\pi')$ satisfies every dual constraint except possibly the launch constraints \eqref{eq:RC-wD}, which bound $\pi'_{l,K-1}$ from above. Since the launch legs of $i$ and $i'$ coincide, (a) and (c) at level $K-1$ are exactly what is needed. For $l\notin\bracket{i,i'}$, (a) gives $\pi'_{l,K-1}\le\pi^\star_{l,K-1}\le c_{Dl}+\nu^\star\taurep_{Dl}-\zeta^\star$; for the pair, $c_{Di}=c_{Di'}$ and $\taurep_{Di}=\taurep_{Di'}$, so (c) gives $\max\{\pi'_{i,K-1},\pi'_{i',K-1}\}\le\max\{\pi^\star_{i,K-1},\pi^\star_{i',K-1}\}\le c_{Di}+\nu^\star\taurep_{Di}-\zeta^\star$. Hence $(\boldsymbol\mu',\nu^\star,\zeta^\star,\boldsymbol\pi')$ is dual feasible, and its objective equals
\[
\sum_{l\in\mathcal N}\lambda_l\mu'_l+\size\nu^\star=\sum_{l\in\mathcal N}\lambda_l\mu^\star_l+\size\nu^\star-(\lambda_{i'}-\lambda_i)\,\theta,
\]
where $\lambda_i\le\lambda_{i'}$ under condition (iii) and $\lambda_i=\lambda_{i'}$ otherwise. If $\lambda_i<\lambda_{i'}$, the new dual solution has a strictly smaller objective than the optimal dual $(\boldsymbol\mu^\star,\nu^\star,\zeta^\star,\boldsymbol\pi^\star)$, a contradiction. If $\lambda_i=\lambda_{i'}$, the new dual solution is optimal, and
\[
\sum_{l\in\mathcal N}(\mu'_l)^2-\sum_{l\in\mathcal N}(\mu^\star_l)^2=2\theta(\mu^\star_i-\mu^\star_{i'})+2\theta^2=-2\theta(\eta+\eta_s)\le0.
\]
If $\eta+\eta_s>0$, this contradicts the minimality of $\sum_l\mu_l^2$ in the selection of the dual. If $\eta+\eta_s=0$, then $\boldsymbol\mu'\neq\boldsymbol\mu^\star$ and both dual solutions are optimal, so their average is optimal by convexity of the set of optimal dual solutions and has $\sum_l\bigl(\tfrac{\mu^\star_l+\mu'_l}{2}\bigr)^2<\sum_l(\mu^\star_l)^2$ by strict convexity, again a contradiction. Hence $\delta\le\eta+\eta_s$.
\end{proof}

% \begin{lemma}[Threshold structure for repositioning]\label{lem:no-repo-below-threshold}
% Let $(\boldsymbol f^\ast,\boldsymbol w^\ast,\boldsymbol \rtd^\ast)$ be an optimal primal solution with optimal duals $(\boldsymbol{\pi}^\ast,\nu^\ast,\zeta^\ast,\ldots)$, and let $\kappa_i^\ast$ be defined as in \eqref{eq:threshold_def_rtd}. Then for every $j\in\mathcal N$ and every $k\le \kappa_i^\ast$, we have
% \(w^\ast_{ijk}=0.\)
% \end{lemma}

\begin{lemma}\label{prop:battery_sensitivity_comparison}
Fix $k\ge 1$ and locations $i,i'\in\mathcal N$ that undertake the same actions at $k$ and $k-1$.
\begin{enumerate}[label=(\roman*)]
\item \emph{(Serve demand)} If $f^\ast_{ik},\ f^\ast_{i,k-1},\ f^\ast_{i'k},\ f^\ast_{i',k-1}>0,$ then
\begin{align}
\Delta_{ik}
=\sum_{j\in\mathcal N} p_{ij} \cdot \Delta_{j,k-1},
\qquad
\Delta_{i'k}
=\sum_{j\in\mathcal N} p_{i'j} \cdot \Delta_{j,k-1}.
\label{eq:delta_i_prime_service_expression}
\end{align}
\item \emph{(Reposition to $j$)} Assume that for some $j\in\mathcal N\cup\depotset$, $w^\ast_{ijk},\ w^\ast_{ij,k-1},\ w^\ast_{i'jk},\ w^\ast_{i'j,k-1}>0.$ Then
$$\Delta_{ik}=\Delta_{i'k}.$$
\end{enumerate}
% \item (Switching level: early RTD at $k{-}1$ but not at $k$) Assume that $w^\ast_{i\to D,k-1}, w^\ast_{i'\to D,k-1}>0$ and $w^\ast_{i\to D,k}= w^\ast_{i'\to D,k}=0$. Assume further that all primitives are identical across $i$ and $i'$ except $c_{i\to D}\geq c_{i'\to D}$ or $\taurep_{iD} \geq \taurep_{i'D}$. Then, $\Delta_{ik}\ \geq \ \Delta_{i'k}.$
\end{lemma}

\begin{proof}{Proof of \cref{prop:battery_sensitivity_comparison}.} 
\textbf{(i)} When $f^\ast_{ik}>0$ and $f^\ast_{i,k-1}>0$, the reduced-cost condition for service at level $k$ in \eqref{eq:RC-f} yields
\[
0
=
\sum_{j\in\mathcal N} r_{ij}p_{ij}
-\mu^\ast_i
-\nu^\ast\sum_{j\in\mathcal N}\tauservice_{ij}p_{ij}
-\pi^\ast_{ik}
+\sum_{j\in\mathcal N} p_{ij}\pi^\ast_{j,k-1}.
\]
Similarly, at level $k-1$,
\[
0
=
\sum_{j\in\mathcal N} r_{ij}p_{ij}
-\mu^\ast_i
-\nu^\ast\sum_{j\in\mathcal N}\tauservice_{ij}p_{ij}
-\pi^\ast_{i,k-1}
+\sum_{j\in\mathcal N} p_{ij}\pi^\ast_{j,k-2}.
\]
Subtracting the two equalities gives
\[
\Delta_{ik}
=
\sum_{j\in\mathcal N} p_{ij}\Big(\pi^\ast_{j,k-1}-\pi^\ast_{j,k-2}\Big) = \sum_{j\in\mathcal N} p_{ij} \cdot \Delta_{j,k-1}.
\]
The same argument applies to $i'$, proving \eqref{eq:delta_i_prime_service_expression}.

\textbf{(ii)} First consider the case $j\in\mathcal N$. Since $w^\ast_{ijk}>0$, the reduced-cost condition for repositioning in \eqref{eq:RC-w} implies $0
=
-c_{ij}-\nu^\ast\taurep_{ij}-\pi^\ast_{ik}+\pi^\ast_{j,k-1},$ or equivalently,
\[
\pi^\ast_{ik}=-c_{ij}-\nu^\ast\taurep_{ij}+\pi^\ast_{j,k-1}.
\]
Similarly, because $w^\ast_{ij,k-1}>0$,
\[
\pi^\ast_{i,k-1}=-c_{ij}-\nu^\ast\taurep_{ij}+\pi^\ast_{j,k-2}.
\]
Subtracting yields $
\Delta_{ik}=\pi^\ast_{j,k-1}-\pi^\ast_{j,k-2}.$ Repeating the same argument for $i'$ gives $\Delta_{i'k}=\pi^\ast_{j,k-1}-\pi^\ast_{j,k-2},$ and therefore $\Delta_{ik}=\Delta_{i'k}$.

Now consider the case $j=D$. The relevant reduced-cost condition is \eqref{eq:RC-rtd}: if $\rtd^\ast_{ik}>0$, then
\[
0
=
-c_{iD}-\nu^\ast\parenthesis{\taurep_{iD}+\tau_k}-\pi^\ast_{ik}-\zeta^\ast,
\]
while if $\rtd^\ast_{i,k-1}>0$, then $0
=
-c_{iD}-\nu^\ast\parenthesis{\taurep_{iD}+\tau_{k-1}}-\pi^\ast_{i,k-1}-\zeta^\ast.$
Subtracting these two equalities yields $\Delta_{ik}=\nu^\ast\parenthesis{\tau_{k-1}-\tau_k}.$
The same argument for $i'$ gives $\Delta_{i'k}=\nu^\ast\parenthesis{\tau_{k-1}-\tau_k}.$ This completes the proof.
\end{proof}

\begin{lemma}[Single crossing]\label{lem:single_crossing}
Let $a_k,b_k\le0$, $k\in\bracket{1,\ldots,K-1}$, be the reduced costs of two families of primal variables, and suppose that $G(k):=a_k-b_k$ satisfies $G(k+1)<G(k)$ for all $k\in\bracket{1,\ldots,K-2}$. Let $\bar\kappa:=\max\bigl(\{0\}\cup\{k\in\bracket{1,\ldots,K-1}:G(k)\ge0\}\bigr)$. Then $b_k<0$ for all $k<\bar\kappa$ and $a_k<0$ for all $k>\bar\kappa$.
\end{lemma}

\begin{proof}{Proof.}
For $k<\bar\kappa$, strict monotonicity gives $G(k)>G(\bar\kappa)\ge0$, so $b_k<a_k\le0$. For $k>\bar\kappa$, $G(k)<0$ by the definition of $\bar\kappa$, so $a_k<b_k\le0$.
\end{proof}

\subsubsection{Proofs of the Main Results}\label{app:section_3_proof}

\begin{proof}{Proof of \cref{thm:threshold_structure}.}
Fix $i\in\mathcal N$, let $(\boldsymbol f^\ast,\boldsymbol w^\ast,\boldsymbol \rtd^\ast)$ be any optimal solution, and let $(\boldsymbol\mu^\star,\nu^\star,\zeta^\star,\boldsymbol\pi^\star)$ be the selected optimal dual. We work with the reduced costs \eqref{eq:RC-rtd} of the RTD variables $\rtd_{ik}$, $k\in\bracket{0,\ldots,K-1}$.

\emph{Step 1: the RTD reduced costs are nonincreasing in $k$.} For $k\in\bracket{0,\ldots,K-2}$, \eqref{eq:RC-rtd} and \cref{lem:pi-monotone}(ii) give
\begin{equation}\label{eq:rc_rtd_monotone}
\reduced(\rtd_{i,k+1})-\reduced(\rtd_{ik})
=
-\nu^\star\tau_{k+1}-\pi^\star_{i,k+1}
-\bigl(-\nu^\star\tau_k-\pi^\star_{ik}\bigr)
=
-\Bigl(\pi^\star_{i,k+1}-\pi^\star_{ik}-\batteryvar_2\nu^\star\Bigr)
\le 0,
\end{equation}
where we used $\tau_k-\tau_{k+1}=\batteryvar_2$. Thus $\{\reduced(\rtd_{ik})\}_{k=0}^{K-1}$ is nonincreasing in $k$.

\emph{Step 2: the threshold.} Define
\begin{equation}\label{eq:threshold_def_rtd}
\kappa_i^\ast
:=
\max\Bigl\{\,k\in\bracket{0,\ldots,K-1}:\ \reduced(\rtd_{ik})=0\,\Bigr\},
\end{equation}
which is well defined because $\pi^\star_{i0}=R_{i0}$ by \eqref{eq:pi_recursion}, i.e., $\reduced(\rtd_{i0})=0$. For $k\le\kappa_i^\ast$, Step~1 gives $\reduced(\rtd_{ik})\ge\reduced(\rtd_{i,\kappa_i^\ast})=0$, and dual feasibility gives $\reduced(\rtd_{ik})\le0$; for $k>\kappa_i^\ast$, $\reduced(\rtd_{ik})\neq0$ by the definition of $\kappa_i^\ast$. Hence
\begin{equation}\label{eq:rtd_rc_zero}
\reduced(\rtd_{ik})=0\quad\forall k\in\bracket{0,\ldots,\kappa_i^\ast},
\qquad\text{and}\qquad
\reduced(\rtd_{ik})<0\quad\forall k\in\bracket{\kappa_i^\ast+1,\ldots,K-1}.
\end{equation}
In particular, RTD attains the maximum in \eqref{eq:pi_recursion} at $(i,k)$ if and only if $k\le\kappa_i^\ast$.

\emph{Step 3: conclusion.} Because the selected dual is an optimal dual solution, complementary slackness holds between it and every optimal solution. Hence \eqref{eq:rtd_rc_zero} implies $\rtd^\ast_{ik}=0$ for all $k\in\bracket{\kappa_i^\ast+1,\ldots,K-1}$, and $\rtd^\ast_{iK}=0$ by \eqref{eq:node_balance_boundary}. Thus $\rtd^\ast_{ik}=0$ for all $k>\kappa_i^\ast$, which proves the threshold structure. Moreover, \eqref{eq:rtd_rc_zero} gives $\reduced(\rtd_{ik})=0$ for all $k\le\kappa_i^\ast$, so under strict complementary slackness, $\rtd^\ast_{ik}>0$ for all $k\le\kappa_i^\ast$.
\end{proof}

\begin{proof}{Proof of \cref{prop:comparison}.}
Let $i,i'\in\mathcal N$ be identical in all primitives except that one or more of the conditions (i)--(vi) hold, and let $(\boldsymbol\mu^\star,\nu^\star,\zeta^\star,\boldsymbol\pi^\star)$ be the selected optimal dual. By \eqref{eq:threshold_def_rtd}, to prove $\kappa_i^\ast\ge \kappa_{i'}^\ast$ it suffices to show that for every $k\in\bracket{0,\ldots,K-1}$,
\begin{equation}\label{eq:rc-dominance}
\reduced(\rtd_{ik})\ \ge\ \reduced(\rtd_{i'k}).
\end{equation}
By \eqref{eq:RC-rtd}, for every $k\in\bracket{0,\ldots,K-1}$,
\[
\reduced(\rtd_{ik})
=
-c_{iD}-\nu^\star\bigl(\taurep_{iD}+\tau_k\bigr)-\pi^\star_{ik}-\zeta^\star,
\]
and similarly for $i'$. Hence \eqref{eq:rc-dominance} is equivalent to
\begin{equation}\label{eq:key-ineq}
-\,c_{iD}\;-\;\nu^\star\taurep_{iD}\;-\;\pi^\star_{ik}
\ \ge\
-\,c_{i'D}\;-\;\nu^\star\taurep_{i'D}\;-\;\pi^\star_{i'k},
\end{equation}
that is, with $\eta$ and $\eta_s$ as in \eqref{eq:eta_def}, to
\begin{equation}\label{eq:pi-bound}
\pi^\star_{ik}-\pi^\star_{i'k}\ \le\ \eta,
\qquad \forall k\in\bracket{0,\ldots,K-1}.
\end{equation}
For $k=0$, \eqref{eq:pi-bound} holds with equality because $\pi^\star_{l0}=R_{l0}$ for every $l\in\mathcal N$ and $R_{i0}-R_{i'0}=\eta$ by \eqref{eq:RTD_value}. Fix $k\in\bracket{1,\ldots,K-1}$ and compare the three terms of \eqref{eq:pi_recursion} at $(i,k)$ and at $(i',k)$; the values $\pi^\star_{j,k-1}$, $j\in\mathcal N$, entering these terms are common to both. First, the RTD terms satisfy $R_{ik}=R_{i'k}+\eta$. Second, since $p_{ij}=p_{i'j}$ for all $j\in\mathcal N$, the destination sums $\sum_j p_{ij}\pi^\star_{j,k-1}$ and $\sum_j p_{i'j}\pi^\star_{j,k-1}$ coincide, so the service terms satisfy
\[
\Bigl[\sum_{j}p_{ij}\bigl(r_{ij}-\nu^\star\tauservice_{ij}+\pi^\star_{j,k-1}\bigr)-\mu^\star_i\Bigr]
-\Bigl[\sum_{j}p_{i'j}\bigl(r_{i'j}-\nu^\star\tauservice_{i'j}+\pi^\star_{j,k-1}\bigr)-\mu^\star_{i'}\Bigr]
=-\eta_s+\bigl(\mu^\star_{i'}-\mu^\star_i\bigr)\le\eta,
\]
where the inequality is \cref{lem:mu_selection}. Third, since $c_{ij}\ge c_{i'j}$ and $\taurep_{ij}\ge\taurep_{i'j}$ for all $j\in\mathcal N$, each repositioning term $-c_{ij}-\nu^\star\taurep_{ij}+\pi^\star_{j,k-1}$ at $(i,k)$ is at most the corresponding term at $(i',k)$. Hence every term in the maximum that defines $\pi^\star_{ik}$ is at most the corresponding term in the maximum that defines $\pi^\star_{i'k}$ plus $\eta$, which yields \eqref{eq:pi-bound}. This establishes \eqref{eq:key-ineq} and \eqref{eq:rc-dominance}, and hence $\kappa_i^\ast\ge\kappa_{i'}^\ast$.
\end{proof}

\begin{proof}{Proof of \cref{thm:no_reposition}.}
Let $\kappa_i^\ast$ be defined in \eqref{eq:threshold_def_rtd}. For $j=i$, the claim follows from \cref{lem:pi-monotone}(iii), and for $k=0$ from \eqref{eq:node_balance_boundary}. Fix $j\neq i$ and $k\in\bracket{1,\ldots,\kappa_i^\ast-1}$, and let $W_{ijk}:=-c_{ij}-\nu^\star\taurep_{ij}+\pi^\star_{j,k-1}$ denote the value of repositioning from $(i,k)$ to $j$. Since $k\le\kappa_i^\ast$, \eqref{eq:rtd_rc_zero} gives $\reduced(\rtd_{ik})=0$, i.e., $\pi^\star_{ik}=R_{ik}$, so $\reduced(w_{ijk})=W_{ijk}-R_{ik}\le0$. By complementary slackness, it suffices to show that $W_{ijk}<R_{ik}$. Suppose to the contrary that $W_{ijk}=R_{ik}$.

\emph{Step 1: $\Delta_{jk}=\batteryvar_2\nu^\star$.} Since $k+1\le\kappa_i^\ast$, also $\pi^\star_{i,k+1}=R_{i,k+1}=R_{ik}+\batteryvar_2\nu^\star$. By \eqref{eq:pi_recursion}, $\pi^\star_{i,k+1}\ge W_{ij,k+1}=W_{ijk}+\Delta_{jk}=R_{ik}+\Delta_{jk}$. Hence $\Delta_{jk}\le\batteryvar_2\nu^\star$, and equality holds by \cref{lem:pi-monotone}(ii).

\emph{Step 2: RTD or repositioning to some $l\neq j$ attains the maximum in \eqref{eq:pi_recursion} at $(j,k-1)$.} As shown in the proof of \cref{lem:pi-monotone}(ii), the value of each action in \eqref{eq:pi_recursion} increases by at least $\batteryvar_2\nu^\star$ per level. Hence, if $\Delta_{jm'}=\batteryvar_2\nu^\star$ for all $m'\in\bracket{m+1,\ldots,k}$, every action that attains the maximum at $(j,m)$ also attains it at $(j,m')$ for all $m'\in\bracket{m,\ldots,k}$. By Step 1, this condition holds for $m=k-1$. Consider an action that attains the maximum at $(j,m)$; idling never does so by \cref{lem:pi-monotone}(iii), and at $m=0$ only RTD is available. If this action is RTD or repositioning to some $l\neq j$, it also attains the maximum at $(j,k-1)$, as claimed. If it is service, then service attains the maximum at both $(j,m)$ and $(j,m+1)$, so $\batteryvar_2\nu^\star=\Delta_{j,m+1}=\sum_{l\in\mathcal N}p_{jl}\Delta_{lm}$. Since $\Delta_{lm}\ge\batteryvar_2\nu^\star$ for all $l$ and $p_{jj}>0$, this forces $\Delta_{jm}=\batteryvar_2\nu^\star$, so the condition holds for $m-1$, and we repeat the argument there. Since $m$ decreases, the procedure terminates at some level $m\le k-1$ at which RTD or repositioning to some $l\neq j$ attains the maximum and $\Delta_{jm'}=\batteryvar_2\nu^\star$ for all $m'\in\bracket{m+1,\ldots,k}$. By the observation at the beginning of this step, that action also attains the maximum at $(j,k-1)$, which proves the claim of this step.

\emph{Step 3: Contradiction.} If RTD attains the maximum at $(j,k-1)$, then $\pi^\star_{j,k-1}=R_{j,k-1}$, and
\[
W_{ijk}=-(c_{ij}+c_{jD})-\nu^\star\bigl(\taurep_{ij}+\taurep_{jD}+\tau_{k-1}\bigr)-\zeta^\star<-c_{iD}-\nu^\star\bigl(\taurep_{iD}+\tau_k\bigr)-\zeta^\star=R_{ik}
\]
by the triangle inequality, $\tau_{k-1}\ge\tau_k$, and $\nu^\star\ge0$. If repositioning to some $l\neq j$ attains it, then $k\ge2$ and $W_{ijk}=-(c_{ij}+c_{jl})-\nu^\star(\taurep_{ij}+\taurep_{jl})+\pi^\star_{l,k-2}$. For $l\neq i$, the triangle inequality and \cref{lem:pi-monotone}(ii) give $W_{ijk}<-c_{il}-\nu^\star\taurep_{il}+\pi^\star_{l,k-1}\le\pi^\star_{ik}=R_{ik}$. For $l=i$, since $k-2\le\kappa_i^\ast$, we have $\pi^\star_{i,k-2}=R_{i,k-2}\le R_{ik}$, so $W_{ijk}\le R_{ik}-(c_{ij}+c_{ji})<R_{ik}$. Each case contradicts $W_{ijk}=R_{ik}$.
\end{proof}

\begin{proof}{Proof of \cref{prop:propagation_strong}.}
As stated in \cref{sec:reposition}, $i$ and $i'$ have residual demand, so $\mu^\star_i=\mu^\star_{i'}=0$ by \eqref{eq:CS_demand}. Since $i$ and $i'$ are identical in all primitives except the depot legs (in the sense defined in \cref{sec:early_rtd}), the service and repositioning terms of \eqref{eq:pi_recursion} coincide at $i$ and $i'$ for every $k\ge1$; let $C_k$ denote the largest of these common terms, i.e., the value of continuing operations at level $k$, so that $\pi^\star_{ik}=\max\{R_{ik},C_k\}$ and $\pi^\star_{i'k}=\max\{R_{i'k},C_k\}$. The RTD terms satisfy $R_{ik}=R_{i'k}+\eta$, with $\eta\ge0$ as defined in \eqref{eq:eta_def}. Hence, by the identity $\max\{a,b\}-\max\{a-\eta,b\}=\min\{\eta,[a-b]^+\}$,
\[
\pi^\star_{ik}-\pi^\star_{i'k}=\min\bigl\{\eta,[R_{ik}-C_k]^+\bigr\},\qquad k\in\bracket{1,\ldots,K-1},
\]
and $\pi^\star_{i0}-\pi^\star_{i'0}=\eta$. By \eqref{eq:RTD_value}, $R_{ik}-R_{i,k-1}=\batteryvar_2\nu^\star$, whereas $C_k-C_{k-1}\ge\batteryvar_2\nu^\star$ for $k\ge2$, as shown in the proof of \cref{lem:pi-monotone}(ii). Thus $R_{ik}-C_k$ is nonincreasing in $k\ge1$, so $\pi^\star_{ik}-\pi^\star_{i'k}$ is nonincreasing in $k\in\bracket{0,\ldots,K-1}$, which is equivalent to $\Delta_{ik}\le\Delta_{i'k}$ for all $k\in\bracket{1,\ldots,K-1}$.
\end{proof}

\begin{proof}{Proof of \cref{thm:threshold2}.}
Fix $j\in\mathcal N\setminus\bracket{i,i'}$ and, for $k\in\bracket{1,\ldots,K-1}$, let $G(k):=\reduced(w_{jik})-\reduced(w_{ji'k})$ denote the difference between the reduced costs of repositioning from $j$ to $i$ and from $j$ to $i'$ at level $k$. By \eqref{eq:RC-w}, the term $\pi^\star_{jk}$ of the origin cancels in this difference,
\[
G(k)=\bigl(\pi^\star_{i,k-1}-\pi^\star_{i',k-1}\bigr)-\bigl(c_{ji}-c_{ji'}\bigr)-\nu^\star\bigl(\taurep_{ji}-\taurep_{ji'}\bigr),
\]
so that $G(k+1)-G(k)=\Delta_{ik}-\Delta_{i'k}<0$ for all $k\in\bracket{1,\ldots,K-2}$. Applying \cref{lem:single_crossing} to the reduced costs $\reduced(w_{jik})$ and $\reduced(w_{ji'k})$, with $\bar\kappa_j$ the index defined there, gives $\reduced(w_{ji'k})<0$ for $k<\bar\kappa_j$ and $\reduced(w_{jik})<0$ for $k>\bar\kappa_j$, so complementary slackness gives the stated exclusions; \eqref{eq:node_balance_boundary} covers $k=0$. 
\end{proof}

\subsubsection{Corollaries of the Main Results}\label{app:section_3_corollary}

\begin{definition}\label{def:same_action}
Fix two locations $i,i'\in\mathcal N$ and a level $k\in\bracket{1,\ldots,K-1}$. We say that $i$ and $i'$ \textit{undertake the same action} at level $k$ if (1) both locations serve demand ($f^\ast_{ik},\,f^\ast_{i'k}>0$), (2) both locations return vehicles to the depot ($\rtd^\ast_{ik},\,\rtd^\ast_{i'k}>0$), or both locations reposition vehicles to some $j\in\mathcal N$ ($w^\ast_{ijk},\,w^\ast_{i'jk}>0$). Moreover, given a level $\hat k\in\bracket{1,\ldots,K-1}$, we say that $i$ and $i'$ have \textit{common service support} up to $\hat k$ if there exists some level $\bar{k}$ and a (possibly empty) set $\mathcal S_f = \bracket{\bar{k},\bar{k}+1,\ldots,\hat k}$ such that, for all $k\le\hat k$,
\[
f^\ast_{ik},\,f^\ast_{i'k}>0
\qquad\text{if and only if}\qquad
k\in\mathcal S_f.
\]
Similarly, we say that $i$ and $i'$ have \textit{common repositioning support} toward some $j\in\mathcal N$ up to $\hat k$ if there exists some level $\bar{k}$ and a (possibly empty) set $\mathcal S_w=\bracket{\bar k,\bar k+1,\ldots,\hat k}$ such that, for all $k\le\hat k$,
\[
w^\ast_{ijk},\,w^\ast_{i'jk}>0
\qquad\text{if and only if}\qquad
k\in\mathcal S_w.
\]
\end{definition}

\begin{proposition}[Comparative statics of battery-sensitivity order]\label{prop:propagation}
Consider two locations $i,i'\in\mathcal N$ with residual demand that undertake the same actions at every level $k\in\bracket{1,\ldots,\hat k}$ for some $\hat k\in\bracket{1,\ldots,K-1}$. We have $\Delta_{ik}\le \Delta_{i'k}$ for all $k\le\hat k$ if $i$ and $i'$ are identical in all primitives except one or more of the following:
\begin{enumerate}[label=(\roman*)]
    \item (Cheaper RTD) $c_{iD}\le c_{i'D}$;
    \item (Faster RTD) $\taurep_{iD}\le \taurep_{i'D}$;
    \item (Lower local demand) under common service support, $\lambda_i\le \lambda_{i'}$;
    \item (Lower trip revenue) under common service support, $r_{ij}\le r_{i'j}$ for some $j\in\mathcal N$;
    \item (Longer travel times) under common service/repositioning support toward $j$, $\tauservice_{ij}\ge \tauservice_{i'j}$ for some $j\in\mathcal N$;
    \item (Higher repositioning costs) under common repositioning support toward some $j\in\mathcal N$, $c_{ij}\ge c_{i'j}$.
\end{enumerate}
\end{proposition}

\begin{proof}{Proof of \cref{prop:propagation}.}
Items (i)--(ii) also follow from \cref{prop:propagation_strong}; the argument below covers all items jointly. Since $i$ and $i'$ have residual demand, $\mu^\star_i=\mu^\star_{i'}=0$. Because the values $\pi^\star_{l,k-1}$, $l\in\mathcal N$, entering the recursion \eqref{eq:pi_recursion} at $(i,k)$ and at $(i',k)$ are the same, the difference between the value of any given action at $(i,k)$ and its value at $(i',k)$ in \eqref{eq:pi_recursion} does not depend on $k$: for RTD, it equals $(c_{i'D}-c_{iD})+\nu^\star(\taurep_{i'D}-\taurep_{iD})\ge0$, which is positive only under (i)--(ii); for service, it equals $\sum_{l\in\mathcal N}p_{il}\bigl[(r_{il}-r_{i'l})-\nu^\star(\tauservice_{il}-\tauservice_{i'l})\bigr]\le0$, which is negative only under (iv)--(v); and for repositioning to $j$, it equals $-(c_{ij}-c_{i'j})-\nu^\star(\taurep_{ij}-\taurep_{i'j})\le0$, which is negative only under (v)--(vi). At every level $k\in\bracket{1,\ldots,\hat k}$, both locations take a common action with positive flow, so by complementary slackness this action attains the maximum in \eqref{eq:pi_recursion} at both locations, and $\pi^\star_{ik}-\pi^\star_{i'k}$ equals the value difference of that action; if several common actions exist, their value differences coincide. At $k=0$, $\pi^\star_{i0}-\pi^\star_{i'0}$ equals the RTD value difference.

Since $\Delta_{ik}-\Delta_{i'k}=(\pi^\star_{ik}-\pi^\star_{i'k})-(\pi^\star_{i,k-1}-\pi^\star_{i',k-1})$, it suffices to show that $\pi^\star_{ik}-\pi^\star_{i'k}$ is nonincreasing in $k$ on $\bracket{0,\ldots,\hat k}$. Suppose instead that $\pi^\star_{i,k+1}-\pi^\star_{i',k+1}>\pi^\star_{ik}-\pi^\star_{i'k}$ for some $k<\hat k$. By the previous paragraph, each side equals the value difference of a common action at the respective level. Since only the RTD value difference can be positive, either (a) RTD is a common action at $k+1$, or (b) the value difference at $k+1$ is nonpositive, in which case the value difference at $k$ is strictly negative, so that the common action at $k$ is service under one of (iv)--(v) or repositioning to $j$ under one of (v)--(vi). We show that both cases lead to a contradiction.

(a) If RTD is a common action at $k+1$, then RTD attains the maximum in \eqref{eq:pi_recursion} at $(i,k+1)$ and at $(i',k+1)$, so by \eqref{eq:rtd_rc_zero} it also attains the maximum at $(i,k)$ and at $(i',k)$. Hence both $\pi^\star_{i,k+1}-\pi^\star_{i',k+1}$ and $\pi^\star_{ik}-\pi^\star_{i'k}$ equal the RTD value difference, a contradiction.

(b) Suppose first that service is a common action at $k$ with a strictly negative value difference. Under (iv)--(v), $i$ and $i'$ have common service support, so service is a common action exactly at the levels of an upper interval $\bracket{\bar k,\ldots,\hat k}$. Since service is common at $k$, we have $k\ge\bar k$, hence $\bar k\le k+1\le\hat k$, and service is also a common action at $k+1$. By the previous paragraph, $\pi^\star_{i,k+1}-\pi^\star_{i',k+1}$ therefore equals the service value difference, as does $\pi^\star_{ik}-\pi^\star_{i'k}$, a contradiction. If instead repositioning to $j$ is a common action at $k$ with a strictly negative value difference, the same argument applies under (v)--(vi) with the common repositioning support toward $j$ in place of the common service support. Hence $\pi^\star_{ik}-\pi^\star_{i'k}$ is nonincreasing in $k$ on $\bracket{0,\ldots,\hat k}$.
\end{proof}

\begin{corollary}[Weak Battery-Sensitivity Order]
\label{cor:weak_order_implication}
Let $(\boldsymbol f^\ast,\boldsymbol w^\ast,\boldsymbol \rtd^\ast)$ be an optimal solution satisfying strict complementary slackness. Fix $i,i'\in\mathcal N$ and suppose the weak battery-sensitivity order holds:
\[
\Delta_{ik}\ \ge\ \Delta_{i'k}\qquad\forall\,k\in\bracket{1,\ldots,K-1}.
\]
Then for every origin $j\in\mathcal N$ there exists an index 
$\bar{\kappa}_j\in\mathcal K$ such that:
\begin{align*}
&\forall\,k\in\{1,\dots,\bar{\kappa}_j{-}1\}:
\qquad 
w^\ast_{jik}>0\ \Rightarrow\ w^\ast_{ji'k}>0,\\[4pt]
&\forall\,k\in\{\bar{\kappa}_j{+}1,\dots,K{-}1\}:
\qquad 
w^\ast_{ji'k}>0\ \Rightarrow\ w^\ast_{jik}>0.
\end{align*}
\end{corollary}

\begin{proof}{Proof.}
Fix $j$ and define $G(k):=\reduced(w_{jik})-\reduced(w_{ji'k})$ as in the proof of \cref{thm:threshold2}, so that $G(k+1)-G(k)=\Delta_{ik}-\Delta_{i'k}\ge0$ for all $k\in\bracket{1,\ldots,K-2}$. Let $\bar\kappa_j:=\max\bigl(\{0\}\cup\{k\in\bracket{1,\ldots,K-1}:G(k)\le0\}\bigr)$. For $k<\bar\kappa_j$, weak monotonicity gives $G(k)\le G(\bar\kappa_j)\le0$, so $\reduced(w_{jik})\le\reduced(w_{ji'k})\le0$. If $w^\ast_{jik}>0$, complementary slackness gives $\reduced(w_{jik})=0$, hence $\reduced(w_{ji'k})=0$, and strict complementary slackness gives $w^\ast_{ji'k}>0$. For $k>\bar\kappa_j$, $G(k)>0$ by the definition of $\bar\kappa_j$, so $\reduced(w_{ji'k})<\reduced(w_{jik})\le0$ and, by complementary slackness, $w^\ast_{ji'k}=0$; the stated implication holds trivially.
\end{proof}

\begin{corollary}[Single-Crossing on a Restricted Segment]
\label{cor:segment_single_crossing}
Let $(\boldsymbol f^\ast,\boldsymbol w^\ast,\boldsymbol \rtd^\ast)$ be an optimal solution satisfying strict complementary slackness, and fix an origin $j\in\mathcal N$. Let 
\[
\mathcal{S}=\{k_{\min},k_{\min}{+}1,\dots,k_{\max}\}
\subseteq \{\kappa_j^\ast{+}1,\dots,K{-}1\}
\]
be a contiguous set of levels on which $\Delta_{ik}>\Delta_{i'k}\;\forall\,k\in \mathcal S.$ Then there exists an index $\bar{\kappa}_j\in \mathcal S$ such that
\[
w^\ast_{jik}=0,\quad \forall k\in (k_{\min},\bar{\kappa}_j),
\qquad
w^\ast_{ji'k}=0,\quad \forall k\in (\bar{\kappa}_j,k_{\max}).
\]
\end{corollary}

\begin{proof}{Proof.}
Define $G(k)$ as in the proof of \cref{thm:threshold2}, so that $G(k+1)-G(k)=\Delta_{ik}-\Delta_{i'k}>0$ whenever $k,k+1\in\mathcal S$. Let $\bar\kappa_j:=\max\bigl(\{k_{\min}\}\cup\{k\in\mathcal S:G(k)\le0\}\bigr)\in\mathcal S$. For $k\in\mathcal S$ with $k_{\min}<k<\bar\kappa_j$, we have $G(\bar\kappa_j)\le0$ and hence $G(k)<0$, so $\reduced(w_{jik})<\reduced(w_{ji'k})\le0$, and complementary slackness gives $w^\ast_{jik}=0$. For $k\in\mathcal S$ with $\bar\kappa_j<k<k_{\max}$, $G(k)>0$ by the definition of $\bar\kappa_j$, so $w^\ast_{ji'k}=0$.
\end{proof}

\subsection{Proofs of Results in \cref{sec:importance}} \label{app:importance}

\subsubsection{Derivations of Charge-Aware and Charge-Oblivious Solutions} \label{app:derivation}

Throughout this section, we focus on the effects of cross-location repositioning costs and assume that idling carries no direct monetary cost, i.e., $c_{ii}=0\;\forall i\in\bracket{1,2}$. Recall that in the family $\instances$ vehicles are launched at full charge to location~$2$ only and do not idle at location~$2$. Moreover, it is easy to show that it is never optimal to reposition a vehicle empty from location~$2$ to location~$1$ instead of serving demand, so we assume throughout that no vehicle does so. Since $\lambda_1=0$ and $p_{21}=1$, every served trip departs from $2$ and ends at $1$. Let $m:=\lfloor K/2\rfloor$ and
\begin{equation}\label{eq:MT_star}
M^\star:=mr-(m-1)c_{12}-c_{1D}-c_{D2},
\qquad
T^\star:=m\tauservice_{21}+(m-1)\taurep_{12}+\taurep_{1D}+\taurep_{D2}+\tau_0,
\end{equation}
which are the per-cycle margin and duration of a cycle that is launched to $2$, alternates $m$ served trips with $m-1$ repositioning legs $1\to2$, and returns to the depot from $1$; we show in \cref{thm:aware_gap} that this cycle is optimal for the charge-aware problem. All results in this section rest on the following accounting for a single charging cycle, i.e., the sequence of actions of a unit of vehicle flow between its launch and its next RTD. We allow randomized cycles, in which case counts, margin, and duration are expectations.

\begin{lemma}[Cycle accounting]\label{lem:cycle_accounting}
Consider any (possibly randomized) cycle of a feasible solution for an instance $\instance\in\instances$. Let $\mathcal S$, $\mathcal R$, and $\mathcal U$ denote its expected numbers of served trips, repositioning legs $1\to2$, and idle levels at location~$1$, let $\mathcal Q$ denote the probability that it ends with an RTD from location~$2$, and let $M$ and $T$ denote its expected margin and duration. Then:
\begin{enumerate}[label=(\roman*)]
\item $\mathcal R=\mathcal S-1+\mathcal Q$, $\mathcal S+\mathcal R+\mathcal U\le K-1$ with equality if the cycle ends at SoC $0$, and $\mathcal S\le m$;
\item the margin and duration are
\begin{align}
M&=r\mathcal S-(\mathcal S-1)c_{12}-c_{1D}-c_{D2}-\bigl(c_{12}+c_{2D}-c_{1D}\bigr)\mathcal Q,\label{eq:cycle_margin}\\
T&=\mathcal S\tauservice_{21}+(\mathcal S-1)\taurep_{12}+\taurep_{1D}+\taurep_{D2}+\tau_0+\tauidle\,\mathcal U+\bigl(\taurep_{12}+\taurep_{2D}-\taurep_{1D}\bigr)\mathcal Q,\label{eq:cycle_time}
\end{align}
where $c_{12}+c_{2D}-c_{1D}>0$ and $\taurep_{12}+\taurep_{2D}-\taurep_{1D}>0$ are the additional cost and time of returning to the depot from location~$1$ via location~$2$ rather than directly;
\item $M\le\mathcal S\,M^\star/m$ and $T\ge\mathcal S\,T^\star/m$; if moreover $M^\star>0$, then $M\le T\,M^\star/T^\star$.
\end{enumerate}
\end{lemma}

\begin{proof}{Proof.}
Statement (i) expresses the fact that trips and repositioning legs must alternate and that the battery is the binding resource of a cycle. Consider first a deterministic cycle, which starts at $(2,K-1)$. A vehicle at location~$2$ either serves a trip, which ends at $1$, or returns to the depot, so every repositioning leg $1\to2$ is followed by a trip, except a terminal leg after which the vehicle returns from $2$; conversely, every trip after the first is preceded by a repositioning leg. Hence $\mathcal R=\mathcal S-1+\mathcal Q$. Every trip, repositioning leg, and idle level consumes exactly one battery level, and the vehicle has $K-1$ levels upon arrival, so $\mathcal S+\mathcal R+\mathcal U\le K-1$, with equality if the vehicle returns only at level $0$. Combining the two relations gives $2\mathcal S-1\le K-1$, i.e., $\mathcal S\le m$.

For (ii), since $c_{ii}=0$, the margin of the cycle is $r\mathcal S-c_{D2}-c_{12}\mathcal R-c_{1D}(1-\mathcal Q)-c_{2D}\mathcal Q$ and its duration is $\taurep_{D2}+\tauservice_{21}\mathcal S+\taurep_{12}\mathcal R+\tauidle\,\mathcal U+\taurep_{1D}(1-\mathcal Q)+\taurep_{2D}\mathcal Q+\tau_0$; substituting $\mathcal R=\mathcal S-1+\mathcal Q$ from (i) yields \eqref{eq:cycle_margin}--\eqref{eq:cycle_time}. For a randomized cycle, all relations hold for every realized path and are preserved by taking expectations, because the right-hand sides of \eqref{eq:cycle_margin}--\eqref{eq:cycle_time} are affine in $(\mathcal S,\mathcal U,\mathcal Q)$.

For (iii), the detour terms in (ii) only lower the margin and raise the duration, so
\[
M\le\mathcal S(r-c_{12})-\bigl(c_{1D}+c_{D2}-c_{12}\bigr),
\qquad
T\ge\mathcal S(\tauservice_{21}+\taurep_{12})+\bigl(\taurep_{1D}+\taurep_{D2}+\tau_0-\taurep_{12}\bigr).
\]
By the triangle inequality, $c_{1D}+c_{D2}-c_{12}>0$ and $\taurep_{1D}+\taurep_{D2}+\tau_0-\taurep_{12}>0$; since $\mathcal S\le m$, replacing each of these constants by $\mathcal S/m$ times itself weakly increases the right-hand side of the first inequality and weakly decreases that of the second, which gives $M\le\mathcal S M^\star/m$ and $T\ge\mathcal S T^\star/m$. Finally, suppose $M^\star>0$, so that $r>c_{12}$. Both bounds above are affine in $\mathcal S$ and coincide with $M^\star$ and $T^\star$ at $\mathcal S=m$, and $\mathcal S\le m$. Hence $M\le TM^\star/T^\star$ holds provided that the first bound decreases at least as fast as $M^\star/T^\star$ times the second when $\mathcal S$ decreases from $m$, i.e., provided that $r-c_{12}\ge(M^\star/T^\star)(\tauservice_{21}+\taurep_{12})$. This holds because $M^\star\le m(r-c_{12})$ and $T^\star\ge m(\tauservice_{21}+\taurep_{12})$: the margin per unit time of the whole cycle cannot exceed that of one trip together with the repositioning leg that enables it.
\end{proof}

We next consider a class of charge-oblivious policies parameterized by $\woblivious_{12}$ that we later show to be optimal: within a single charging cycle, a vehicle is launched from $D$ to location~$2$, serves demand from location~$2$ to $1$, and thereafter, whenever it is at location $1$ with positive battery, it either repositions to $2$ with probability $\woblivious_{12}$ or idles at $1$ until losing one battery level. Once the vehicle reaches battery level $k=0$, it returns to the depot from its current location. The next lemma characterizes the behavior of a vehicle under such a policy.

\begin{lemma}\label{lem:functions_of_p}
Consider a vehicle that evolves according to the charge-oblivious policy with parameter $\woblivious_{12}\in[0,1]$. Then, within a charging cycle,
\begin{itemize}
    \item the expected total number of served trips from location~$2$ to $1$ completed by this vehicle is
    \begin{align*}
        \mathcal S(\woblivious_{12})
        &:=
        1+\frac{\woblivious_{12}(K-2)}{1+\woblivious_{12}}-\frac{\woblivious_{12}(1-(-\woblivious_{12})^{K-2})}{(1+\woblivious_{12})^2},
    \end{align*} 
    \item the expected total number of repositioning moves from location~$1$ to $2$ executed by this vehicle is
    \begin{align*}
        \mathcal R(\woblivious_{12})
        &:=
        \frac{\woblivious_{12}(K-2)}{1+\woblivious_{12}}+\frac{\woblivious_{12}^2(1-(-\woblivious_{12})^{K-2})}{(1+\woblivious_{12})^2},
    \end{align*}
    \item the expected total number of idle levels at location $1$ undertaken by this vehicle is
    \begin{align*}
        \mathcal U(\woblivious_{12})
        &:=
        \frac{(1-\woblivious_{12})(K-2)}{1+\woblivious_{12}}
        +\frac{\woblivious_{12}(1-\woblivious_{12})(1-(-\woblivious_{12})^{K-2})}{(1+\woblivious_{12})^2},
    \end{align*}
    \item the probability that this vehicle returns to the depot from location $2$ (rather than $1$) is
    \begin{align*}
        \mathcal Q(\woblivious_{12})
        &:=
        \frac{\woblivious_{12}(1-(-\woblivious_{12})^{K-2})}{1+\woblivious_{12}}.
    \end{align*}
\end{itemize}
These functions satisfy $\mathcal S=1+\mathcal R-\mathcal Q$ and $\mathcal S+\mathcal R+\mathcal U=K-1$. Moreover, $\mathcal S(\woblivious_{12})$ is nondecreasing on $[0,1]$ with $\mathcal S(1)=m$; if $K\ge4$, then $\mathcal S(\woblivious_{12})<m$ for all $\woblivious_{12}<1$; and if $K$ is odd, then $\mathcal Q(\woblivious_{12})>0$ for all $\woblivious_{12}\in(0,1]$.
\end{lemma}

The next result characterizes both $\OPT(\instance)$ and $\voblivious(\instance)$ in closed form.

\begin{theorem}\label{thm:aware_gap}
For any instance $I \in \mathcal{I}$, with $M^\star$ and $T^\star$ as in \eqref{eq:MT_star},
\[
\OPT(\instance)=\max\bracket{0,\;\max\{M^\star,0\}\cdot\min\bracket{\frac{\lambda_2}{m},\frac{S}{T^\star}}},
\]
\[
\voblivious(\instance)
=\max\bracket{0,\max_{\woblivious_{12}\in[0,1]}\max\{M^o(\woblivious_{12}),0\} \cdot \min\bracket{\frac{\lambda_2}{\mathcal S(\woblivious_{12})},\frac{S}{T^o(\woblivious_{12})}}},
\]
where, with $\mathcal S,\mathcal R,\mathcal U,\mathcal Q$ as in \cref{lem:functions_of_p},
\begin{align}
M^o(\woblivious_{12})&:=r\mathcal S(\woblivious_{12})-c_{D2}-c_{12}\mathcal R(\woblivious_{12})-c_{1D}\bigl(1-\mathcal Q(\woblivious_{12})\bigr)-c_{2D}\mathcal Q(\woblivious_{12}),\label{eq:Mo_def}\\
T^o(\woblivious_{12})&:=\taurep_{D2}+\tauservice_{21}\mathcal S(\woblivious_{12})+\taurep_{12}\mathcal R(\woblivious_{12})+\tauidle\,\mathcal U(\woblivious_{12})+\taurep_{1D}\bigl(1-\mathcal Q(\woblivious_{12})\bigr)+\taurep_{2D}\mathcal Q(\woblivious_{12})+\tau_0\label{eq:To_def}
\end{align}
are the expected margin and duration of a cycle under the charge-oblivious policy with parameter $\woblivious_{12}$.
\end{theorem}

Both values are the product of a \emph{per-cycle margin} and a \emph{throughput term}: the first captures the net profit generated by a single vehicle over one charging cycle, and the second captures how many such cycles can be completed per unit time, subject to demand availability and fleet-time capacity. Applying \cref{lem:cycle_accounting}(ii) to the cycle of \cref{lem:functions_of_p} and using \eqref{eq:MT_star} gives
\begin{equation}\label{eq:obl_decomposition}
\begin{aligned}
M^o(\woblivious_{12})&=M^\star-(r-c_{12})\bigl(m-\mathcal S(\woblivious_{12})\bigr)-\bigl(c_{12}+c_{2D}-c_{1D}\bigr)\mathcal Q(\woblivious_{12}),\\
T^o(\woblivious_{12})&=T^\star-(\tauservice_{21}+\taurep_{12})\bigl(m-\mathcal S(\woblivious_{12})\bigr)+\tauidle\,\mathcal U(\woblivious_{12})+\bigl(\taurep_{12}+\taurep_{2D}-\taurep_{1D}\bigr)\mathcal Q(\woblivious_{12}).
\end{aligned}
\end{equation}
This decomposition makes the dilemma faced by a charge-oblivious policy explicit. Relative to the charge-aware cycle, it forgoes $m-\mathcal S(\woblivious_{12})$ trips, each worth $r-c_{12}$ net of the repositioning leg that enables it, and with probability $\mathcal Q(\woblivious_{12})$ it returns to the depot via location~$2$, which costs $c_{12}+c_{2D}-c_{1D}$ more and takes $\taurep_{12}+\taurep_{2D}-\taurep_{1D}$ longer than returning directly from $1$; in addition, it spends $\tauidle$ per idle level. By \cref{lem:functions_of_p}, $\mathcal S(\woblivious_{12})$ is nondecreasing and falls short of $m$ unless $\woblivious_{12}=1$, whereas $\mathcal Q(\woblivious_{12})$ is positive as soon as $\woblivious_{12}>0$. Hence a small $\woblivious_{12}$ forfeits trips and wastes vehicle time idling at $1$, whereas a large $\woblivious_{12}$ sends low-SoC vehicles to $2$, where they run out of battery and return with a detour: a charge-oblivious policy must trade off under-utilizing high-charge vehicles against over-sending low-charge vehicles to $2$.

\subsubsection{Proofs of the Main Results}\label{app:section_4_proof}

\begin{proof}{Proof of \cref{lem:functions_of_p}.}
After the launch and the first trip, the vehicle is at $(1,K-2)$, and from every state $(1,k)$ with $k\ge1$ it either idles, moving to $(1,k-1)$, or repositions, after which it serves a trip and moves to $(1,k-2)$ if $k\ge2$, or reaches $(2,0)$ and returns from $2$ if $k=1$. Starting from $(1,k)$, let $R_k$ denote the expected number of repositioning legs and $Q_k$ the probability of returning from $2$. Conditioning on the first action gives $R_0=Q_0=0$, $R_1=Q_1=\woblivious_{12}$, and, for $k\ge2$,
\[
R_k=(1-\woblivious_{12})R_{k-1}+\woblivious_{12}(1+R_{k-2}),
\qquad
Q_k=(1-\woblivious_{12})Q_{k-1}+\woblivious_{12}Q_{k-2},
\]
because idling contributes no repositioning leg and continues from $(1,k-1)$, whereas repositioning contributes one leg and continues from $(1,k-2)$. Each recursion determines its sequence uniquely from the two initial values, so it suffices to verify that
\[
R_k=\frac{\woblivious_{12}k}{1+\woblivious_{12}}+\frac{\woblivious_{12}^2\bigl(1-(-\woblivious_{12})^k\bigr)}{(1+\woblivious_{12})^2},
\qquad
Q_k=\frac{\woblivious_{12}\bigl(1-(-\woblivious_{12})^k\bigr)}{1+\woblivious_{12}}
\]
satisfy them: both initial values match; the constant sequence $1$ and the sequence $(-\woblivious_{12})^k$ both solve the homogeneous recursion, because $1=(1-\woblivious_{12})+\woblivious_{12}$ and $(-\woblivious_{12})^2=(1-\woblivious_{12})(-\woblivious_{12})+\woblivious_{12}$, so every combination of the two does as well; and the term $\woblivious_{12}k/(1+\woblivious_{12})$ accounts for the constant $\woblivious_{12}$ in the recursion for $R_k$. The remaining quantities follow by accounting. Every repositioning leg from a level $k\ge2$ is followed by a trip, whereas a leg from level $1$ is not, so the expected number of additional trips from $(1,k)$ is $R_k-Q_k$; and all $k$ levels are consumed, one per trip, repositioning leg, or idle level, so the expected number of idle levels is $k-R_k-(R_k-Q_k)$. Evaluating at $k=K-2$ and adding the first trip gives $\mathcal S=1+R_{K-2}-Q_{K-2}$, $\mathcal R=R_{K-2}$, $\mathcal U=(K-2)-\mathcal R-(\mathcal S-1)$, and $\mathcal Q=Q_{K-2}$, which are the stated expressions and identities.

For the monotonicity of $\mathcal S$, differentiating its closed form with respect to $\woblivious_{12}$ gives
\[
\frac{d\mathcal S(\woblivious_{12})}{d\woblivious_{12}}
=\frac{(K-3)+(K-1)\woblivious_{12}+(-\woblivious_{12})^{K-2}\bigl((K-1)+(K-3)\woblivious_{12}\bigr)}{(1+\woblivious_{12})^3}.
\]
If $K$ is even, then $(-\woblivious_{12})^{K-2}\ge0$ and the numerator is at least $K-3\ge0$. If $K$ is odd, then $(-\woblivious_{12})^{K-2}=-\woblivious_{12}^{K-2}$ and the numerator equals $(K-3)\bigl(1-\woblivious_{12}^{K-1}\bigr)+(K-1)\bigl(\woblivious_{12}-\woblivious_{12}^{K-2}\bigr)\ge0$ for $\woblivious_{12}\in[0,1]$. Hence $\mathcal S$ is nondecreasing. Substituting $\woblivious_{12}=1$ gives $\mathcal S(1)=1+\frac{K-2}{2}-\frac{1-(-1)^K}{4}=m$. If $K\ge4$ and $\woblivious_{12}<1$, the vehicle idles at its first two decisions at location~$1$ with probability $(1-\woblivious_{12})^2>0$; on such paths $\mathcal U\ge2$, and the two identities with $\mathcal R\ge\mathcal S-1$ give $2\mathcal S+1\le K-1$, i.e., $\mathcal S\le m-1$, so $\mathcal S(\woblivious_{12})<m$. Finally, if $K$ is odd and $\woblivious_{12}>0$, the path that repositions at every visit to location~$1$ has probability $\woblivious_{12}^{m}>0$ and makes its last leg from level $1$, so it ends at $2$ and $\mathcal Q(\woblivious_{12})>0$.
\end{proof}

\begin{proof}{Proof of \cref{thm:aware_gap}.}
\textbf{Step 1: Derivation of $\OPT(\instance)$.} Consider the cycle that is launched to $2$, serves a trip $2\to1$, repositions $1\to2$, serves again, and so on, until after the $m$-th trip it returns to the depot from $1$. Since $2m-1\le K-1$, this cycle is feasible; it contains $m$ trips and $m-1$ repositioning legs, never idles, and ends at $(1,K-2m)$ with $K-2m\in\bracket{0,1}$. By \cref{lem:cycle_accounting}(ii), its margin and duration are $M^\star$ and $T^\star$. Releasing it at rate $x$ serves demand at rate $mx$ and occupies fleet at rate $T^\star x$, so $x=\min\{\lambda_2/m,S/T^\star\}$ is feasible and $\OPT(\instance)\ge\max\{M^\star,0\}\min\{\lambda_2/m,S/T^\star\}$; parking the fleet gives $\OPT(\instance)\ge0$.

It remains to show that no feasible solution does better. Because every non-depot action lowers the SoC by one level, every feasible solution decomposes into cycles, and its profit, served demand, and vehicle time are the corresponding sums over cycles weighted by their rates. By \cref{lem:cycle_accounting}(iii), every cycle with $\mathcal S$ served trips, margin $M$, and duration $T$ satisfies $M\le\mathcal SM^\star/m$. If $M^\star\le0$, every cycle therefore has nonpositive margin and $\OPT(\instance)=0$. If $M^\star>0$, the lemma also gives $M\le TM^\star/T^\star$; summing the two bounds over cycles and using \eqref{c:demand} and \eqref{c:vehicle_time} gives $\OPT(\instance)\le M^\star\min\{\lambda_2/m,S/T^\star\}$, which proves the stated value of $\OPT(\instance)$.

\textbf{Step 2: Derivation of $\voblivious(\instance)$.} Consider any feasible charge-oblivious solution. In $\instances$ (see \cref{fig:3location}), vehicles are launched to $2$, and \eqref{eq:obl_no_RTD} rules out RTD at positive SoC. Hence vehicles at $(2,k)$ with $k\ge1$ serve, and vehicles at $(1,k)$ with $k\ge1$ reposition to $2$ or idle; by \eqref{eq:obl_simplex}, they do so in fixed proportions $\woblivious_{12}$ and $\woblivious_{11}=1-\woblivious_{12}$ at every level with positive flow. Thus, given $\woblivious_{12}$, every flow is proportional to the launch rate $x:=w_{D2K}$, and the flows per unit launch rate equal the expected numbers of the corresponding transitions of the policy in \cref{lem:functions_of_p}; conversely, every $\woblivious_{12}\in[0,1]$ and $x\ge0$ define a feasible charge-oblivious solution whenever the demand and fleet-size constraints hold. By \cref{lem:functions_of_p}, the expected margin and duration of a cycle are $M^o(\woblivious_{12})$ and $T^o(\woblivious_{12})$ as defined in \eqref{eq:Mo_def}--\eqref{eq:To_def}, so a release rate $x\ge0$ earns profit at rate $xM^o(\woblivious_{12})$, serves demand at rate $x\mathcal S(\woblivious_{12})$, and occupies fleet at rate $xT^o(\woblivious_{12})$. For fixed $\woblivious_{12}$, the optimal release rate is therefore $\min\{\lambda_2/\mathcal S(\woblivious_{12}),S/T^o(\woblivious_{12})\}$ if $M^o(\woblivious_{12})>0$ and $0$ otherwise. Maximizing over all charge-oblivious policies is equivalent to maximizing over $\woblivious_{12}\in[0,1]$, which yields the stated value of $\voblivious(\instance)$.
\end{proof}

\begin{proof}{Proof of \cref{prop:stylized_monotonicity}.}
We use the closed forms of \cref{thm:aware_gap} and the decomposition \eqref{eq:obl_decomposition}.

\textbf{Items (i)--(ii).} $\OPT(\instance)$ does not depend on $c_{2D}$, $\taurep_{2D}$, or $\tauidle$. For every fixed $\woblivious_{12}$, $M^o(\woblivious_{12})$ is weakly decreasing in $c_{2D}$ and does not depend on $\taurep_{2D}$ or $\tauidle$, whereas $T^o(\woblivious_{12})$ is weakly increasing in $\taurep_{2D}$ and $\tauidle$ and does not depend on $c_{2D}$, because $\mathcal Q(\woblivious_{12}),\mathcal U(\woblivious_{12})\ge0$. Hence $\max\{M^o(\woblivious_{12}),0\}\min\{\lambda_2/\mathcal S(\woblivious_{12}),S/T^o(\woblivious_{12})\}$ is weakly decreasing in each of the three parameters for every fixed $\woblivious_{12}$, and so is its maximum over $\woblivious_{12}$. Thus $\voblivious(\instance)$ is weakly decreasing, and the performance gain weakly increasing, in $c_{2D}$, $\taurep_{2D}$, and $\tauidle$.

\textbf{Item (iii).} We compare the fleet sizes at which the two policies cease to be fleet-constrained and become demand-constrained. By \cref{thm:aware_gap}, the charge-aware policy is fleet-constrained for $S\le\bar S^\star:=\lambda_2T^\star/m$ and demand-constrained for $S\ge\bar S^\star$, whereas the charge-oblivious policy with parameter $\woblivious_{12}$ is fleet-constrained for $S\le\bar S^o(\woblivious_{12}):=\lambda_2T^o(\woblivious_{12})/\mathcal S(\woblivious_{12})$. The key observation is that
\begin{equation}\label{eq:key_threshold_order}
\bar S^\star\le \bar S^o(\woblivious_{12}),\qquad \forall \woblivious_{12}\in[0,1],
\end{equation}
i.e., no charge-oblivious cycle uses less vehicle time per served trip than the charge-aware cycle: this is the inequality $T\ge\mathcal S\,T^\star/m$ of \cref{lem:cycle_accounting}(iii) applied to the cycle of \cref{lem:functions_of_p}, for which $\mathcal S=\mathcal S(\woblivious_{12})$ and $T=T^o(\woblivious_{12})$. For $S\le\bar S^\star$, \eqref{eq:key_threshold_order} implies that both the charge-aware policy and every fixed-$\woblivious_{12}$ charge-oblivious policy are fleet-constrained, so $\OPT(\instance)=\max\{M^\star,0\}S/T^\star$ and $\voblivious(\instance)=S\max\{0,\max_{\woblivious_{12}}M^o(\woblivious_{12})/T^o(\woblivious_{12})\}$, and their ratio is constant in $S$. For $S\ge\bar S^\star$, $\OPT(\instance)=\max\{M^\star,0\}\lambda_2/m$ is constant in $S$, whereas $\voblivious(\instance)$ is weakly increasing in $S$, because $S$ enters \cref{thm:aware_gap} only through the throughput term. Hence $\OPT(\instance)/\voblivious(\instance)$ is weakly decreasing in $S$.

\textbf{Unbounded gain.} Let $K=2m+1$ with $m\ge2$, fix $r>c_{12}>0$, $c_{1D}\in(0,r)$, and $c_{2D}$ with $c_{12}+c_{2D}>c_{1D}$, and choose $c_{21}$ and all travel times to satisfy the triangle inequality; the cost $c_{D2}$ is chosen below. By \eqref{eq:obl_decomposition}, $\voblivious(\instance)=0$ if and only if
\[
M^\star\ \le\ (r-c_{12})\bigl(m-\mathcal S(\woblivious_{12})\bigr)+\bigl(c_{12}+c_{2D}-c_{1D}\bigr)\mathcal Q(\woblivious_{12})
\qquad\text{for all }\woblivious_{12}\in[0,1],
\]
that is, if and only if the charge-aware margin never exceeds the value of the trips a charge-oblivious cycle forgoes plus its expected detour cost. The right-hand side is a continuous function of $\woblivious_{12}$ on $[0,1]$ that does not depend on $c_{D2}$ and is strictly positive: at $\woblivious_{12}=0$ it equals $(r-c_{12})(m-1)>0$, because $\mathcal S(0)=1$ and $\mathcal Q(0)=0$, and for $\woblivious_{12}>0$ its second term is positive by \cref{lem:functions_of_p}. Its minimum over $[0,1]$ is therefore positive. Choose $M^\star$ between $0$ and this minimum and set $c_{D2}:=r+(m-1)(r-c_{12})-c_{1D}-M^\star$, so that \eqref{eq:MT_star} holds with this $M^\star$. Since $M^\star\le(r-c_{12})(m-1)$, we have $c_{D2}\ge r-c_{1D}>0$ and $c_{1D}+c_{D2}\ge r>c_{12}$, so the instance belongs to $\instances$. For this instance, $\voblivious(\instance)=0<\OPT(\instance)$ by \cref{thm:aware_gap}, and hence $\sup_{\instance\in\instances}\OPT(\instance)/\voblivious(\instance)=+\infty$.
\end{proof}

\subsubsection{Corollaries of the Main Results}\label{app:section_4_corollary}

\begin{corollary}[Robustness under Symmetric Demand]\label{cor:robustness}
Consider a family of instances $\instances^s$ with symmetric demand primitives across all locations, i.e.,
\[
\lambda_1 = \lambda_2, 
\quad p_{12} = p_{21}, \text{and} \quad r_{12} = r_{21}.
\]
We have $\sup_{\instance \in \instances^s}\OPT(\instance)/ \voblivious(\instance) = +\infty.$
\end{corollary}

\begin{proof}{Proof of \cref{cor:robustness}.}
It suffices to construct symmetric-demand instances for which the ratio is arbitrarily large. Let $\mathcal N=\bracket{1,2}$, $K=2m+1$ with $m\ge1$, $\lambda_1=\lambda_2=\lambda>0$, $p_{12}=p_{21}=1$, $r_{12}=r_{21}=r>0$, and $\tau_k=\tau_0$ for all $k<K$. Choose costs such that
\begin{equation}\label{eq:sym_gap_cond1}
c_{D2}+c_{1D}<(2m-1)r,\qquad c_{2D}\ge2mr,\qquad c_{D1}\ge2mr,
\end{equation}
choose the remaining costs and all travel times to satisfy the triangle inequality, and let the idling times be $\taurep_{11}=\taurep_{22}=\tauidle$, which the triangle inequality does not constrain. Since vehicles are launched at full charge, every cycle uses at most $2m$ levels after its launch, so it serves at most $2m$ trips and earns revenue at most $2mr$.

\emph{Charge-aware value.} The cycle that is launched to $2$, serves $2m-1$ alternating trips, and returns to the depot from $(1,1)$ earns $(2m-1)r-c_{D2}-c_{1D}>0$ per cycle, serves at most $m$ trips from each location, and never idles. Hence $\OPT(\instance)\ge\bigl((2m-1)r-c_{D2}-c_{1D}\bigr)\min\{\lambda/m,S/T\}>0$, where the duration $T$ of this cycle does not depend on $\tauidle$.

\emph{Charge-oblivious value.} Consider any, possibly randomized, cycle of a feasible solution that satisfies \eqref{eq:obl_no_RTD}, and let $M^s$ and $T^s$ denote its expected margin and duration. Paths launched to $1$ earn at most $2mr-c_{D1}\le0$. Paths launched to $2$ that never idle use only trips and empty moves between $1$ and $2$, each of which switches the location, until they reach charge level $0$ after $2m$ such moves; they therefore end at $2$ and earn at most $2mr-c_{2D}\le0$. Hence only paths that idle at least once can earn a positive profit, so $M^s\le2mr\,\PP{\text{idle}}$, whereas $T^s\ge\tauidle\,\PP{\text{idle}}$. Since every feasible solution is a mixture of such cycles that satisfies \eqref{c:vehicle_time}, $\voblivious(\instance)\le2mrS/\tauidle$. As $\OPT(\instance)$ does not depend on $\tauidle$, letting $\tauidle\to\infty$ yields $\sup_{\instance\in\instances^s}\OPT(\instance)/\voblivious(\instance)=+\infty$. The same argument applies to the benchmark in \cref{cor:robustness_reposition}(1), which imposes only \eqref{eq:obl_no_RTD}.
\end{proof}

\begin{proof}{Proof of \cref{cor:robustness_reposition}, part~(1).}
If $K$ is even, the charge-aware cycle in the proof of \cref{thm:aware_gap} ends at $(1,0)$ without early RTD, so $\voblivious_1(\instance)=\OPT(\instance)$ and the claims hold trivially; we therefore assume $K=2m+1$. Since $\OPT(\instance)$ is given by \cref{thm:aware_gap}, it suffices to characterize $\voblivious_1(\instance)$. As in Step~2 of the proof of \cref{thm:aware_gap}, we may restrict attention to solutions in which vehicles at $2$ with positive SoC serve, so after its first trip a vehicle is at $(1,2m-1)$ and, since early RTD is not allowed, continues until its SoC reaches $0$. Because repositioning decisions may now depend on the SoC, a cycle is described by two quantities: the number $a\in\bracket{0,\ldots,m-1}$ of repositioning legs $1\to2$ that are followed by a trip, and the indicator $b\in\bracket{0,1}$ of a terminal repositioning leg at SoC $1$, after which the vehicle returns to the depot from $2$ rather than from $1$. A cycle of type $(a,b)$ therefore serves $\mathcal S_1(a,b):=a+1$ trips, makes $\mathcal R_1(a,b):=a+b$ repositioning legs, idles $\mathcal U_1(a,b):=2m-1-2a-b$ levels, and returns from $2$ with probability $\mathcal Q_1(a,b):=b$. Its margin $M^1(a,b)$ and duration $T^1(a,b)$ are given by \eqref{eq:Mo_def}--\eqref{eq:To_def} with these quantities in place of $\mathcal S,\mathcal R,\mathcal U,\mathcal Q$, and \cref{lem:cycle_accounting}(ii) gives, as in \eqref{eq:obl_decomposition},
\begin{equation}\label{eq:cycle_type_decomposition}
\begin{aligned}
M^1(a,b)&=M^\star-(r-c_{12})(m-1-a)-\bigl(c_{12}+c_{2D}-c_{1D}\bigr)b,\\
T^1(a,b)&=T^\star-(\tauservice_{21}+\taurep_{12})(m-1-a)+\tauidle\,(2m-1-2a-b)+\bigl(\taurep_{12}+\taurep_{2D}-\taurep_{1D}\bigr)b.
\end{aligned}
\end{equation}
Decomposing the steady-state flow into such cycles with release rates $x_{ab}\ge0$,
\begin{equation}
\label{eq:voblivious1_closed_form}
\voblivious_1(\instance)
=
\max_{\{x_{ab}\ge 0\}}
\Bigl\{\sum_{a,b} M^1(a,b)x_{ab}:\ 
\sum_{a,b} \mathcal S_1(a,b)x_{ab}\le \lambda_2,\ 
\sum_{a,b} T^1(a,b)x_{ab}\le S\Bigr\},
\end{equation}
which is the analogue of the closed form for $\voblivious(\instance)$ in \cref{thm:aware_gap}, except that the benchmark now optimizes over charge-dependent deterministic cycle types rather than over a charge-independent repositioning probability $\woblivious_{12}$.

\emph{Monotonicity properties.} $\OPT(\instance)$ does not depend on $c_{2D}$, $\taurep_{2D}$, or $\tauidle$. By \eqref{eq:cycle_type_decomposition}, for every cycle type, $M^1(a,b)$ is weakly decreasing in $c_{2D}$ and does not depend on $\taurep_{2D}$ or $\tauidle$, whereas $T^1(a,b)$ is weakly increasing in $\taurep_{2D}$ and $\tauidle$ and does not depend on $c_{2D}$; maximizing over the release rates preserves these properties, so $\voblivious_1(\instance)$ is weakly decreasing in each of the three parameters. This proves (i)--(ii). For (iii), \cref{lem:cycle_accounting}(iii) applied to a cycle of type $(a,b)$ gives $T^1(a,b)\ge\mathcal S_1(a,b)T^\star/m$, i.e., no cycle type uses less vehicle time per served trip than the charge-aware cycle. Hence, with $\bar S^\star:=\lambda_2T^\star/m$ as in the proof of \cref{prop:stylized_monotonicity}, every feasible $\{x_{ab}\}$ with $S\le\bar S^\star$ satisfies $\sum_{a,b}\mathcal S_1(a,b)x_{ab}\le(m/T^\star)\sum_{a,b}T^1(a,b)x_{ab}\le mS/T^\star\le\lambda_2$: the demand constraint in \eqref{eq:voblivious1_closed_form} is slack, both $\OPT(\instance)$ and $\voblivious_1(\instance)$ scale linearly with $S$, and their ratio is constant. For $S\ge\bar S^\star$, $\OPT(\instance)$ is constant in $S$, whereas $\voblivious_1(\instance)$ is weakly increasing in $S$ because a larger fleet only enlarges the feasible region of \eqref{eq:voblivious1_closed_form}. Hence $\OPT(\instance)/\voblivious_1(\instance)$ is weakly decreasing in $S$, which proves (iii).

\emph{Unbounded gain.} Fix $r>c_{12}>c_{1D}>0$, choose $M^\star\in(0,r-c_{12}]$, set $c_{D2}:=r+(m-1)(r-c_{12})-c_{1D}-M^\star$ so that \eqref{eq:MT_star} holds with this $M^\star$, and choose $c_{2D}$ large enough that $c_{12}+c_{2D}-c_{1D}\ge M^\star$, together with $c_{21}$ and all travel times satisfying the triangle inequality. Since $r-M^\star\ge c_{12}>c_{1D}$, we have $c_{D2}>0$ and $c_{1D}+c_{D2}>c_{12}$, so the instance belongs to $\instances$. By \eqref{eq:cycle_type_decomposition}, $M^1(a,0)\le M^\star-(r-c_{12})\le0$ for $a\le m-2$, $M^1(m-1,0)=M^\star>0$, and $M^1(a,1)\le M^\star-(c_{12}+c_{2D}-c_{1D})\le0$ for all $a$: the only profitable cycle type is $(m-1,0)$, which differs from the charge-aware cycle only by idling once at location~$1$ instead of returning early, so that $T^1(m-1,0)=T^\star+\tauidle$. Hence \eqref{eq:voblivious1_closed_form} gives $\voblivious_1(\instance)=M^\star\min\{\lambda_2/m,S/(T^\star+\tauidle)\}$, whereas $\OPT(\instance)=M^\star\min\{\lambda_2/m,S/T^\star\}$ by \cref{thm:aware_gap}. For $\lambda_2\ge mS/T^\star$, both policies are fleet-constrained and $\OPT(\instance)/\voblivious_1(\instance)=(T^\star+\tauidle)/T^\star$, which grows without bound as $\tauidle\to\infty$. This proves that $\sup_{\instance\in\instances}\OPT(\instance)/\voblivious_1(\instance)=+\infty$.
\end{proof}

\begin{proof}{Proof of \cref{cor:robustness_rtd}, part~(2).}
\emph{Monotonicity.} For fixed fractions in \eqref{eq:obl_simplex}, the program that defines $\voblivious_2(\instance)$ is a linear program, and $\voblivious_2(\instance)$ is the supremum of its optimal value over all fractions. Increasing $c_{2D}$ weakly decreases the objective of every feasible solution, and increasing $\taurep_{2D}$ weakly shrinks the feasible set, because RTD flows from $2$ enter \eqref{obj} and \eqref{c:vehicle_time} with nonnegative coefficients. Hence $\voblivious_2(\instance)$ is weakly decreasing in $c_{2D}$ and $\taurep_{2D}$, while $\OPT(\instance)$ does not depend on them by \cref{thm:aware_gap}. This proves property (i) of \cref{prop:stylized_monotonicity} for ${\OPT(\instance)}/{\voblivious_2(\instance)}$.

\emph{Unbounded gap.} We first describe the cycles of this benchmark. By the charge-oblivious condition, a vehicle at location~$1$ with positive SoC repositions to $2$, idles, or returns to the depot with probabilities $\woblivious_{12}$, $\woblivious_{11}$, and $1-\woblivious_{12}-\woblivious_{11}$ that do not depend on its SoC. A vehicle at location~$2$ with positive SoC serves: idling at $2$ is not available in $\instances$, empty moves $2\to1$ are ruled out as explained at the beginning of this section, and similarly one can show that it is never optimal to return a vehicle from location~$2$ at positive SoC instead of serving demand.

Let $K=2m+1$ with $m\ge2$, $d:=r-c_{12}>0$, and $\gamma\in(0,d)$, and choose
\begin{equation}\label{eq:rtd_unbounded_cond1}
r+(m-2)d+\gamma\le c_{D2}+c_{1D}<r+(m-1)d,
\end{equation}
so that $M^\star=r+(m-1)d-c_{D2}-c_{1D}\in(0,d-\gamma]$ and $\OPT(\instance)>0$ by \cref{thm:aware_gap}. Let $B:=\binom{2m-1}{m-1}$ and $\omega_0:=\min\bigl\{\frac{1}{2(2m-1)},\bigl(\frac{\gamma}{4BM^\star}\bigr)^{1/(m-1)}\bigr\}$, and choose $c_{2D}\ge c_{1D}+M^\star\bigl(1+B\omega_0^{-m}\bigr)$. The remaining primitives can be chosen to satisfy the triangle inequality.

Fix the fractions at location~$1$ and consider the resulting randomized cycle. As in the proof of \cref{thm:aware_gap}, every feasible solution is a mixture of such cycles, so it suffices to show that the expected margin $M^o_2$ of every cycle is nonpositive. We classify the paths of the cycle by where they end and by the number $\mathcal S\le m$ of trips they serve, and bound the margin of each class by \cref{lem:cycle_accounting}(ii), which gives a margin of at most $M^\star-(r-c_{12})(m-\mathcal S)$ for a path that ends at $1$ and at most $M^\star-(c_{12}+c_{2D}-c_{1D})$ for a path that ends at~$2$:
\begin{itemize}
\item Paths that end at $1$ with $\mathcal S\le m-1$ have margin at most $M^\star-d\le-\gamma$ by \eqref{eq:rtd_unbounded_cond1}. They include the paths that never reposition, i.e., idle or return at each visit to location~$1$; such paths serve exactly one trip, since every further trip requires a repositioning leg, and they end at $1$. Since the vehicle visits location~$1$ at most $2m-1$ times, they occur with probability at least $(1-\woblivious_{12})^{2m-1}$.
\item Paths that end at $1$ with $\mathcal S=m$ have margin at most $M^\star$. Each trip after the first requires a repositioning leg, so such a path repositions at least $m-1$ times among at most $2m-1$ decisions at location~$1$, each of which repositions independently with probability $\woblivious_{12}$; a union bound over the $B$ choices of $m-1$ of these decisions gives probability at most $B\woblivious_{12}^{m-1}$.
\item Paths that end at $2$ have margin at most $M^\star-(c_{12}+c_{2D}-c_{1D})<0$. They occur with probability at least $\woblivious_{12}^m$, because with this probability the vehicle repositions at each of its at most $m$ visits to location~$1$ and then ends at $2$.
\end{itemize}
Since $c_{12}+c_{2D}-c_{1D}\ge c_{2D}-c_{1D}$, it follows that
\[
M^o_2\le B\woblivious_{12}^{m-1}M^\star-\gamma(1-\woblivious_{12})^{2m-1}-\bigl(c_{2D}-c_{1D}-M^\star\bigr)\woblivious_{12}^m,
\]
where $c_{2D}-c_{1D}-M^\star\ge BM^\star\omega_0^{-m}$ by the choice of $c_{2D}$. If $\woblivious_{12}\le\omega_0$, then $(1-\woblivious_{12})^{2m-1}\ge1-(2m-1)\woblivious_{12}\ge\frac12$, so $M^o_2\le B\omega_0^{m-1}M^\star-\frac{\gamma}{2}\le0$ by the choice of $\omega_0$. If $\woblivious_{12}>\omega_0$, then $M^o_2\le BM^\star-\bigl(c_{2D}-c_{1D}-M^\star\bigr)\omega_0^m\le0$. Hence $\voblivious_2(\instance)=0<\OPT(\instance)$, which proves that $\sup_{\instance\in\instances}\OPT(\instance)/\voblivious_2(\instance)=+\infty$.
\end{proof}

\subsection{Proofs of Results in \cref{sec:numerics}}\label{app:proof_numerics}

To establish the relationship between demand and vehicle value, we replace the demand vector $\boldsymbol{\lambda}$ by $z\boldsymbol{\lambda}$ for some scalar $z>0$ and leave all other primitives unchanged. Let $\nu^\star(z)$ denote an optimal dual multiplier associated with the vehicle-time constraint \eqref{c:vehicle_time} in the corresponding fluid LP. The next result shows that the shadow value of vehicle time is weakly increasing in demand. 

\begin{proposition}[Vehicle shadow value increases in demand]\label{prop:demand_shadow_price}
For any $0<z_1<z_2$, there exist optimal dual multipliers $\nu^\star(z_1)$ and $\nu^\star(z_2)$ such that $\nu^\star(z_1)
\;\le\;
\nu^\star(z_2).$
\end{proposition}

\begin{proof}{Proof of \cref{prop:demand_shadow_price}.}
Let $V(z,S)$ denote the optimal value of the fluid LP when the demand vector is $z\boldsymbol\lambda$ and the right-hand side of \eqref{c:vehicle_time} is $S$, and write $V_0(S):=V(1,S)$ for the baseline-demand value function. We use two properties of $V$. First, $V(z,S)=zV_0(S/z)$ for all $z,S>0$: if $(\boldsymbol f,\boldsymbol w,\boldsymbol\rtd)$ is feasible under demand $z\boldsymbol\lambda$ and fleet size $S$, then $(\boldsymbol f,\boldsymbol w,\boldsymbol\rtd)/z$ is feasible under demand $\boldsymbol\lambda$ and fleet size $S/z$, because the demand and fleet-size constraints scale by $1/z$ while the flow-balance constraints are homogeneous, and the objective scales by $1/z$ as well; the reverse mapping gives the opposite inequality. Second, $V_0$ is concave: for $S_1,S_2>0$ and $\theta\in[0,1]$, a convex combination of optimal solutions at $S_1$ and $S_2$ is feasible at $\theta S_1+(1-\theta)S_2$, because all constraints are linear, and it attains $\theta V_0(S_1)+(1-\theta)V_0(S_2)$.

Now fix $0<z_1<z_2$, so that $S/z_1>S/z_2$, and for $i\in\{1,2\}$ let $g_i$ be an optimal multiplier of \eqref{c:vehicle_time} in the baseline-demand problem with fleet size $S/z_i$. By LP duality, $g_i$ is a supergradient of the concave function $V_0$ at $S/z_i$, so $V_0(S/z_2)\le V_0(S/z_1)+g_1(S/z_2-S/z_1)$ and $V_0(S/z_1)\le V_0(S/z_2)+g_2(S/z_1-S/z_2)$. Adding the two inequalities gives $0\le(g_2-g_1)(S/z_1-S/z_2)$, hence $g_1\le g_2$. Finally, the problem with demand $z_i\boldsymbol\lambda$ and fleet size $S$ is obtained from the baseline-demand problem with fleet size $S/z_i$ by multiplying all nonzero right-hand sides by $z_i$, which leaves the dual feasible region unchanged and multiplies the dual objective by $z_i$; hence $g_i$ remains an optimal multiplier of \eqref{c:vehicle_time} in that problem, and setting $\nu^\star(z_i):=g_i$ gives $\nu^\star(z_1)=g_1\le g_2=\nu^\star(z_2)$.
\end{proof}

\subsection{Proofs of Results in \cref{sec:extensions}}\label{app:proof_appendix}

\subsubsection{Proofs of Results in \cref{sec:ext_fulfillment}}\label{app:proof_fulfillment}

Throughout this part, the LP is the destination-dependent LP of \cref{sec:ext_fulfillment}, and the dual formulation of Appendix~\ref{app:dual} applies with $f_{ik}p_{ij}$ replaced by $f_{ijk}$ and \eqref{c:demand} by \eqref{c:demand_match}. We first state and prove the two corollaries referenced in \cref{sec:ext_fulfillment}, which extend \cref{thm:threshold_structure,thm:no_reposition,thm:threshold2} to this setting.

\begin{corollary}
\label{thm:threshold_structure_match}
There exists an optimal solution $(\boldsymbol f^\star,\boldsymbol w^\star,\boldsymbol{\rtd}^\star)$ in which, for each location $i\in\mathcal N$, there exists a threshold $\kappa_i^\star\in\mathcal K$ such that:
\begin{enumerate}[label=(\roman*)]
    \item vehicles do not return to the depot if their SoC $k>\kappa_i^\star$, i.e., $\rtd^\star_{ik}=0$ for all $k>\kappa_i^\star$;
    \item vehicles at location $i$ neither serve nor reposition if their SoC $k<\kappa_i^\star$, i.e., $f^\star_{ijk}=w^\star_{ijk}=0$ for all $j\in\mathcal N$ and $k<\kappa_i^\star$; in particular, $\rtd^\star_{ik}>0$ at every level $k\le\kappa_i^\star$ with positive vehicle inflow.
\end{enumerate}
\end{corollary}

\begin{proof}{Proof of \cref{thm:threshold_structure_match}.} We construct an optimal solution with the stated structure. Among all optimal solutions, choose one, denoted $(\boldsymbol f^\star,\boldsymbol w^\star,\boldsymbol{\rtd}^\star)$, that maximizes \[\Phi(\boldsymbol f,\boldsymbol w,\boldsymbol{\rtd}):=\sum_{i,j\in\mathcal N}\sum_{k\in\mathcal K} k f_{ijk}+\sum_{i,j\in\mathcal N}\sum_{k\in\mathcal K} k w_{ijk};\] a maximizer exists because the optimal face is a nonempty polytope, as all flows are bounded by \eqref{c:vehicle_time}.

We claim that, in this solution, there is no location $i$, no pair of levels $h>k$, and no destination $j\in\mathcal N$ such that $\rtd^\star_{ih}>0$ and either $f^\star_{ijk}>0$ or $w^\star_{ijk}>0$. Suppose, to the contrary, that such an inversion exists, and let $a$ denote the non-RTD arc at $(i,k)$ that carries positive flow. Let $\delta:=h-k>0$ and choose $0<\varepsilon\le\min\{\rtd^\star_{ih},\text{flow on }a\}$. By flow decomposition, the $\varepsilon$-flow that starts with $a$ can be represented as a path through the state network,
\begin{equation}\label{eq:path} (i,k)\to(j,k-1)\to\cdots\to(l_T,r)\to D, \end{equation}
whose final step is an RTD from some state $(l_T,r)$; the path may include destination-dependent service arcs, empty repositioning arcs, and stay-put arcs.

We now perform a swap. First, remove $\varepsilon$ units from the path and instead send them directly to the depot from $(i,k)$ by increasing $\rtd_{ik}$ by $\varepsilon$. Second, remove $\varepsilon$ units from $\rtd^\star_{ih}$ and let them follow the path in \eqref{eq:path}, but starting from $(i,h)$ rather than $(i,k)$. Since every on-road action consumes one battery level, the shifted path is feasible and terminates with an RTD from $(l_T,r+\delta)$. The exchange preserves the flow-balance constraints and the demand constraints \eqref{c:demand_match}, which aggregate over battery levels, and it preserves the vehicle-time constraint, because the additional charging time of the RTD from $(i,k)$ instead of $(i,h)$ is offset by the reduced charging time of the RTD from $(l_T,r+\delta)$ instead of $(l_T,r)$. Revenues and costs are unchanged, so the new solution is optimal. However, $\Phi$ strictly increases, because every non-RTD arc of \eqref{eq:path}, of which there is at least one, is now used $\delta$ levels higher. This contradicts the choice of the solution, so no inversion exists.

For each $i\in\mathcal N$, define $\kappa_i^\star:=\max\{k\in\mathcal K:\rtd^\star_{ik}>0\}$, with $\kappa_i^\star:=0$ if the set is empty. Then $\rtd^\star_{ik}=0$ for all $k>\kappa_i^\star$. Because no inversion exists, $f^\star_{ijk}=w^\star_{ijk}=0$ for all $j$ and $k<\kappa_i^\star$, so by flow balance all vehicles that reach $(i,k)$ with $k<\kappa_i^\star$ return to the depot, i.e., $\rtd^\star_{ik}>0$ whenever the inflow into $(i,k)$ is positive; at $k=\kappa_i^\star$, positivity holds by definition.
\end{proof}

\begin{corollary}
\label{thm:threshold2_match}
In an optimal solution $(\boldsymbol f^\ast,\boldsymbol w^\ast,\boldsymbol{\rtd}^\ast)$, consider locations $i,i'\in\mathcal N$ with $\Delta_{ik}<\Delta_{i'k}$ for all $k\in\{1,\ldots,K-1\}$. Then, for every $j\in\mathcal N\setminus\bracket{i,i'}$, there exists a threshold $\bar{\kappa}_j\in\mathcal{K}$ such that:
\begin{enumerate}[label=(\roman*)]
    \item vehicles do not reposition to $i'$ below the threshold, i.e., $w^\ast_{ji'k}=0\;\forall k<\bar\kappa_j$;
    \item vehicles do not reposition to $i$ above the threshold, i.e., $w^\ast_{jik}=0\;\forall k>\bar\kappa_j$.
\end{enumerate}
\end{corollary}

\begin{proof}{Proof of \cref{thm:threshold2_match}.} Replacing $f_{ik}p_{ij}$ by $f_{ijk}$ changes neither the empty-vehicle arcs nor their reduced costs, so \eqref{eq:RC-w} holds verbatim for the dual of the destination-dependent LP. The proof of \cref{thm:threshold2}, which uses only \eqref{eq:RC-w}, \cref{lem:single_crossing}, and complementary slackness, therefore applies without change.
\end{proof}

\begin{proof}{Proof of \cref{thm:threshold_f}.} 
Fix an origin $j\in\mathcal N\setminus\bracket{i,i'}$ and compare the reduced costs of fulfilling demand from $j$ to $i$ and from $j$ to $i'$ at battery level $k$. We have:
\[
\reduced(f^\ast_{jik})
=
r_{ji}
-u_{ji}
-\nu\tauservice_{ji}
-\pi^\ast_{jk}
+\pi^\ast_{i,k-1} \text{ and }
\reduced(f^\ast_{ji'k})
=
r_{ji'}
-u_{ji'}
-\nu\tauservice_{ji'}
-\pi^\ast_{jk}
+\pi^\ast_{i',k-1}.
\]
Define the reduced-cost difference $G_j(k)
:=
\reduced(f^\ast_{jik})-\reduced(f^\ast_{ji'k}).$ Then
\begin{align*}
G_j(k)
&=
(r_{ji}-u_{ji})-(r_{ji'}-u_{ji'})
-\nu(\tauservice_{ji}-\tauservice_{ji'})
+\pi^\ast_{i,k-1}-\pi^\ast_{i',k-1}.
\end{align*}
Therefore, for adjacent battery levels,
\begin{align*}
G_j(k+1)-G_j(k)
&=
\bigl(\pi^\ast_{ik}-\pi^\ast_{i,k-1}\bigr)
-
\bigl(\pi^\ast_{i'k}-\pi^\ast_{i',k-1}\bigr) =
\Delta_{ik}-\Delta_{i'k}.
\end{align*}
By assumption, $G_j(k+1)<G_j(k)$ for all $k\in\bracket{1,\ldots,K-2}$. Applying \cref{lem:single_crossing} to the reduced costs $\reduced(f_{jik})$ and $\reduced(f_{ji'k})$, with $\widetilde\kappa_j$ the index defined there, gives that $\reduced(f_{ji'k})<0$ for $k<\widetilde\kappa_j$ and $\reduced(f_{jik})<0$ for $k>\widetilde\kappa_j$, so complementary slackness gives the stated exclusions; \eqref{eq:node_balance_boundary} covers $k=0$.
\end{proof}

\begin{proof}{Proof of \cref{prop:service_dominates_reposition}.} Let $u^\star_{ij}\ge 0$ denote the optimal dual variable associated with the demand constraint \eqref{c:demand_match}, and keep the dual variable definitions of all other constraints the same. At battery level $k\in\mathcal{K}$, the reduced cost of fulfilling demand from $i$ to $j$ is \begin{equation}\label{eq:rc_f} \reduced(f^\star_{ijk})=r_{ij}-u^\star_{ij}-\nu^\star\tauservice_{ij}-\pi^\star_{ik}+\pi^\star_{j,k-1}. \end{equation} By comparison, the reduced cost of repositioning an empty vehicle from $i$ to $j$ at battery level $k$ is \begin{equation}\label{eq:rc_w} \reduced(w^\star_{ijk})=-c_{ij}-\nu^\star\taurep_{ij}-\pi^\star_{ik}+\pi^\star_{j,k-1}. \end{equation} Thus, \begin{equation}\label{eq:rc_compare_fw} \reduced(f^\star_{ijk})-\reduced(w^\star_{ijk})=r_{ij}+c_{ij}-u^\star_{ij}. \end{equation} Since the demand constraint for $(i,j)$ is slack, complementary slackness implies that $u^\star_{ij}=0$. Thus, \[\reduced(f^\star_{ijk})-\reduced(w^\star_{ijk})=r_{ij}+c_{ij}>0.\] Since reduced-cost feasibility implies that $\reduced(f^\star_{ijk})\le 0$, we obtain $\reduced(w^\star_{ijk})<0.$ By complementary slackness, this proves $w^\ast_{ijk}=0.$ \end{proof}

\subsubsection{Proofs of Results in \cref{sec:ext_nonstationary}}\label{app:proof_nonstationary}

\paragraph{Dual variable selection.} Throughout, early launching is allowed, as described in \cref{sec:ext_nonstationary}. The multipliers in \cref{sec:ext_nonstationary} are selected as follows. Let $\mu_{ij}^\star(t)$ be optimal multipliers of the demand constraints in \eqref{eq:ns_feasibility}, i.e., the value of one additional request from $i$ to $j$ in period $t$. We next define $\pi^\star_{ik}(t)$ and $\pi^\star_{D\ell}(t)$ by backward induction from period $T$, with zero value after $T$. Let
\[
R_{ik}(t):=-c_{iD}(t)+\max_{\ell\in\{\max\{k,1\},\ldots,K\}}\pi^\star_{D\ell}\bigl(t+\taurep_{iD}+\batteryvar_1+\batteryvar_2(\ell-k)\bigr)
\]
denote the value of an RTD. For $t\le T$, $\pi^\star_{i0}(t)=R_{i0}(t)$, while for $1\le k<K$,
\begin{equation}\label{eq:ns_bellman}
\pi^\star_{ik}(t)=\max\left\{
\begin{aligned}
&R_{ik}(t),\\
&\max_{j\in\mathcal N}\bigl\{r_{ij}(t)-\mu_{ij}^\star(t)+\pi^\star_{j,k-1}(t+\tauservice_{ij})\bigr\},\\
&\max_{j\in\mathcal N}\bigl\{-c_{ij}(t)+\pi^\star_{j,k-1}(t+\taurep_{ij})\bigr\},
\end{aligned}\right.
\end{equation}
and, at the depot,
\[
\pi^\star_{D\ell}(t)=\max\left\{\pi^\star_{D\ell}(t+1),\ \max_{j\in\mathcal N}\bigl\{-c_{Dj}(t)+\pi^\star_{j,\ell-1}(t+\taurep_{Dj})\bigr\}\right\}.
\]
Values of committed mass are those at its next available state and completion time. By construction, no action can earn more than these values, so every reduced cost is nonpositive; in particular,
\begin{equation}\label{eq:ns_reduced_costs}
\reduced(w_{jik}(t))=-c_{ji}(t)-\pi^\star_{jk}(t)+\pi^\star_{i,k-1}(t+\taurep_{ji})\le0.
\end{equation}
Moreover, let $(\boldsymbol\mu^\star,\tilde{\boldsymbol\pi})$ be any optimal dual solution with the selected $\boldsymbol\mu^\star$. By the same argument as in the proof of \cref{lem:canonical_dual}, the variables $\tilde{\boldsymbol\pi}$ satisfy the same inequalities as $\boldsymbol\pi^\star$ and are therefore, by backward induction from period $T$, at least as large as $\boldsymbol\pi^\star$; moreover, replacing $\tilde{\boldsymbol\pi}$ by $\boldsymbol\pi^\star$ preserves optimality. 

\begin{proof}{Proof of \cref{cor:ns_repositioning}.}
Define
\[
G^j_{i,i'}(t,k):=\reduced(w_{jik}(t))-\reduced(w_{ji'k}(t)).
\]
Subtracting the two repositioning reduced costs in \eqref{eq:ns_reduced_costs} cancels the term $\pi^\star_{jk}(t)$ of the origin. Taking adjacent differences in $k$ cancels the movement costs, which do not depend on $k$, so for $k=1,\ldots,K-2$,
\begin{equation}\label{eq:ns_single_crossing}
G^j_{i,i'}(t,k+1)-G^j_{i,i'}(t,k)
=\Delta_{ik}(t+\taurep_{ji})-\Delta_{i'k}(t+\taurep_{ji'}).
\end{equation}
Use the single-crossing argument of \cref{thm:threshold2}. By \eqref{eq:ns_single_crossing}, $G^j_{i,i'}(t,k)$ is strictly decreasing in $k$. Set
\[
\bar\kappa_j(t):=\max\bigl(\{0\}\cup\{k\in\{1,\ldots,K-1\}:G^j_{i,i'}(t,k)\ge0\}\bigr).
\]
Below this threshold, $G^j_{i,i'}(t,k)>0$, so dual feasibility implies $\reduced(w_{ji'k}(t))<0$. Above it, $G^j_{i,i'}(t,k)<0$, so $\reduced(w_{jik}(t))<0$. Complementary slackness gives the stated exclusions; at $k=0$ they follow from the boundary restrictions. The argument uses only the repositioning reduced costs, so it does not depend on how vehicles are launched from the depot.
\end{proof}

\begin{proof}{Proof of \cref{thm:ns_rtd}.}
Since $\taucharge_k=\batteryvar_1+\batteryvar_2(K-k)$, the condition $\taucharge_k=\taucharge_0$ means $\batteryvar_2=0$. Then a vehicle that returns from $(i,k)$ in period $t$ can be launched at any level $\ell\ge\max\{k,1\}$ from period $t+\taurep_{iD}+\batteryvar_1$ onward, so a lower-SoC vehicle has every launch option of a higher-SoC one at the same times.

We construct an optimal solution with the stated structure. Among all optimal solutions of \eqref{eq:ns_problem}, choose one, denoted $(\boldsymbol f^\star,\boldsymbol w^\star,\boldsymbol{\rtd}^\star,\boldsymbol x^\star)$, that maximizes
\[
\Phi(\boldsymbol f,\boldsymbol w,\boldsymbol{\rtd},\boldsymbol x):=\sum_{t=1}^{T}\sum_{i\in\mathcal N}\sum_{k=1}^{K-1}k\bigl(x_{ik}(t)-\rtd_{ik}(t)\bigr),
\]
the SoC-weighted mass that takes a non-RTD action at a demand location; a maximizer exists because the optimal face is a nonempty polytope, as all masses are bounded by the fleet size.

We claim that, in this solution, there are no location $i$, period $t$, and levels $k<h$ such that $\rtd^\star_{ih}(t)>0$ and $x^\star_{ik}(t)-\rtd^\star_{ik}(t)>0$, i.e., such that some mass returns from SoC $h$ while some mass at the lower SoC $k$ takes a non-RTD action. Suppose, to the contrary, that such an inversion exists. Then $k\ge1$, since vehicles with zero SoC can only return. Choose $0<\varepsilon\le\min\{\rtd^\star_{ih}(t),\,x^\star_{ik}(t)-\rtd^\star_{ik}(t)\}$. By flow decomposition in the time-expanded network, we can follow $\varepsilon$ units of the non-RTD mass at $(i,k)$ in period $t$ along their subsequent actions, including their returns to the depot and relaunches, until the end of the horizon, and likewise $\varepsilon$ units of the mass returning from $(i,h)$ in period $t$.

We now perform a swap. First, let the $\varepsilon$ units at $(i,h)$ take the non-RTD action of the $\varepsilon$ units at $(i,k)$ and thereafter replicate, action by action and period by period, the path of those units, launching at a level at least as high whenever that path is relaunched. This is feasible: until their first return to the depot, the units hold exactly $h-k$ more charge than the units they replicate; whenever the replicated path is relaunched at level $\ell$ and they arrive at the depot with charge $q'$, they are relaunched at level $\max\{\ell,q'\}\le K$. They therefore hold at least as much charge as the units they replicate at every point of the path, so every on-road action remains feasible, and because charging time does not depend on SoC, they are relaunched at the same times and to the same destinations. Second, let the $\varepsilon$ units at $(i,k)$ return to the depot in period $t$ instead and, by the observation above, be relaunched at the same time and level as the $\varepsilon$ units from $(i,h)$ were, after which they replicate the path of those units exactly. All actions occur at the same locations and periods as before, and the RTD cost $c_{iD}(t)$ does not depend on SoC. Hence revenues, costs, and the demand served for every origin--destination pair and period are unchanged, so the new solution is feasible and optimal. However, $\Phi$ strictly increases: the non-RTD action at $(i,k)$ in period $t$ is now taken at SoC $h>k$, every subsequent non-RTD action of the replicated path is taken at a level at least as high as before, and the replication of the other path leaves its contribution to $\Phi$ unchanged. This contradicts the choice of the solution, so no inversion exists.

For each $i\in\mathcal N$ and $t$, define $\kappa_i^\star(t):=\max\{k:\rtd^\star_{ik}(t)>0\}$, with $\kappa_i^\star(t):=0$ if no vehicle returns from $i$ in period $t$ (in which case no vehicle is at $(i,0)$ in period $t$). Then $\rtd^\star_{ik}(t)=0$ for all $k>\kappa_i^\star(t)$ by definition. Because no inversion exists, all mass at $(i,k)$ with $k<\kappa_i^\star(t)$ returns to the depot, i.e., $\rtd^\star_{ik}(t)=x^\star_{ik}(t)>0$ whenever $x^\star_{ik}(t)>0$; at $k=\kappa_i^\star(t)$, positivity holds by definition. This gives the stated structure.
\end{proof}

\subsection{Proofs of Results in Appendix~\ref{app:duration_energy}}\label{app:proof_duration_energy}

\begin{proof}{Proof of \cref{cor:duration_thresholds}.}
Select optimal multipliers $(\boldsymbol\mu^\star,\nu^\star,\zeta^\star)$ as in Appendix~\ref{app:dual}, and retain the RTD value $R_{ik}$ in \eqref{eq:RTD_value}. The canonical continuation values satisfy \eqref{eq:pi_recursion} with each successor level $k-1$ replaced by $k-e_{ij}$, i.e.,
\begin{equation}\label{eq:duration_pi_recursion}
\begin{aligned}
\pi^\star_{ik}&=\max\Bigl\{R_{ik},F_{ik},\max_{j\in\mathcal N}W_{ijk}\Bigr\},\qquad
W_{ijk}=-c_{ij}-\nu^\star\taurep_{ij}+\pi^\star_{j,k-e_{ij}},\\
F_{ik}&=\sum_{j\in\mathcal N}p_{ij}\bigl(r_{ij}-\nu^\star\tauservice_{ij}+\pi^\star_{j,k-e_{ij}}\bigr)-\mu_i^\star,
\end{aligned}
\end{equation}
where $W_{iik}$ is the value of idling and infeasible actions have value $-\infty$. In particular, $\pi^\star_{i0}=R_{i0}$. Every successor level is smaller than $k$, so the recursion is well defined. The proof of \cref{lem:canonical_dual} therefore applies verbatim, with the launch constraints now involving $+\pi^\star_{j,K-e_{Dj}}$, so the canonical dual remains optimal and is complementary to every optimal primal solution.

Write $q:=\batteryvar_2\nu^\star\ge0$, $h_i(k):=\pi^\star_{ik}-qk$, and $A_i:=R_{ik}-qk$, which is independent of $k$ by \eqref{eq:RTD_value}. Subtracting $qk$ in \eqref{eq:duration_pi_recursion} expresses $h_i(k)$ as the maximum of the constant $A_i$ and action values with charge-independent rewards and continuation terms $h_j(k-e_{ij})$. Induction on $k$ shows that every $h_i$ is nondecreasing: each action feasible at $k$ remains feasible at $k+1$, and its adjusted value weakly increases. Hence, as in \cref{lem:pi-monotone}(ii),
\begin{equation}\label{eq:duration_value_bound}
\pi^\star_{i,k+1}-\pi^\star_{ik}\ge q,
\qquad
\reduced(\rtd_{ik})=A_i-h_i(k).
\end{equation}
Since $h_i(0)=A_i$, the set of levels at which $h_i(k)=A_i$, i.e., at which RTD attains the maximum in \eqref{eq:duration_pi_recursion}, is a nonempty initial interval; its last index is $\kappa_i^\ast$ in \eqref{eq:threshold_def_rtd}. Complementary slackness gives $\rtd^\star_{ik}=0$ for $k>\kappa_i^\ast$, and strict complementary slackness, when imposed with this selected dual, gives $\rtd^\star_{ik}>0$ for $k\le\kappa_i^\ast$. Moreover, idling has strictly negative reduced cost, since $\reduced(w_{iik})=W_{iik}-\pi^\star_{ik}\le-c_{ii}-\nu^\star\taurep_{ii}-q<0$.

It remains to exclude repositioning below $\kappa_i^\ast$. Let $a_{ij}:=-c_{ij}-\nu^\star\taurep_{ij}-qe_{ij}$, so that $a_{ij}+h_j(k-e_{ij})$ is the adjusted value of repositioning from $(i,k)$ to $j$.
Suppose that a feasible repositioning action $i\to j$, $j\ne i$, has zero reduced cost at $k<\kappa_i^\ast$, and put $m:=k-e_{ij}$. Since $h_i(k)=h_i(k+1)=A_i$,
\[
A_i=a_{ij}+h_j(m)\le a_{ij}+h_j(m+1)\le h_i(k+1)=A_i.
\]
Thus $h_j(m)=h_j(m+1)$. We claim that RTD or repositioning to some $l\ne j$ attains the maximum at $(j,m)$. As in Step~2 of the proof of \cref{thm:no_reposition}, we descend from $(j,m)$. Suppose that $h_j$ is constant on $\{r,\ldots,m+1\}$, initially with $r=m$. Any action attaining the maximum at $(j,r)$ also attains it at all these higher levels, since its adjusted value is nondecreasing. If service attains it, equality of its values at $r$ and $r+1$ gives
\[
0=\sum_{l:p_{jl}>0}p_{jl}\bigl[h_l(r+1-e_{jl})-h_l(r-e_{jl})\bigr].
\]
Every summand is nonnegative. Since $p_{jj}>0$ and $e_{jj}=1$, this forces $h_j(r)=h_j(r-1)$, extending the constant interval down one level. Repeat at $r-1$. This procedure eventually reaches an optimal RTD or repositioning action, because idling is strictly suboptimal and at level zero only RTD is feasible. That action also attains the maximum at $(j,m)$, proving the claim.

If RTD attains the maximum there, then $h_j(m)=A_j$ and, by the triangle inequalities,
\[
a_{ij}+A_j-A_i
=-(c_{ij}+c_{jD}-c_{iD})
-\nu^\star(\taurep_{ij}+\taurep_{jD}-\taurep_{iD})-qe_{ij}<0,
\]
contradicting $a_{ij}+h_j(m)=A_i$. Otherwise, let $j\to l$ be an optimal repositioning action at $(j,m)$. If $l\ne i$, concavity and nonnegativity of $\energyfunction$ imply $\energyfunction(a)+\energyfunction(b)\ge\energyfunction(a+b)+\energyfunction(0)\ge\energyfunction(a+b)$ for $a,b\ge0$, which, together with monotonicity of $\energyfunction$ and the travel-time triangle inequality, yields $e_{il}\le e_{ij}+e_{jl}$. The direct move $i\to l$ is therefore feasible, and the cost and time triangle inequalities give $a_{ij}+a_{jl}<a_{il}$. Since $h_l$ is nondecreasing,
\[
A_i=a_{ij}+a_{jl}+h_l(k-e_{ij}-e_{jl})
<a_{il}+h_l(k-e_{il})\le h_i(k)=A_i,
\]
a contradiction. If $l=i$, the contradiction instead follows from $a_{ij}+a_{ji}<0$ and $h_i(k-e_{ij}-e_{ji})\le h_i(k)=A_i$. Thus every feasible repositioning action at $k<\kappa_i^\ast$ has strictly negative reduced cost. Complementary slackness completes the proof.
\end{proof}

\begin{proof}{Proof of \cref{cor:duration_repositioning}.}
Retain the notation $q=\batteryvar_2\nu^\star$ and the action values in \eqref{eq:duration_pi_recursion}. On $\mathcal K_j$, the difference between the two repositioning reduced costs is
\[
G(k):=W_{jik}-W_{ji'k}
=\pi^\star_{i,k-e_{ji}}-\pi^\star_{i',k-e_{ji'}}
-(c_{ji}-c_{ji'})-\nu^\star(\taurep_{ji}-\taurep_{ji'}).
\]
For $k,k+1\in\mathcal K_j$, \eqref{eq:duration_sensitivity_order} gives
\[
G(k+1)-G(k)
=\Delta_{i,k+1-e_{ji}}-\Delta_{i',k+1-e_{ji'}}<0.
\]
Thus the single-crossing argument of \cref{lem:single_crossing} applies on the common feasible range.

Suppose first that $\mathcal K_j\ne\varnothing$, and put $m:=\max\{e_{ji},e_{ji'}\}$. If $e_{ji}>e_{ji'}$ and $G(m)<0$, then the move to $i$ is infeasible below $m$ and has strictly negative reduced cost at every level in $\mathcal K_j$. Taking $\bar\kappa_j=0$ gives both exclusions. Otherwise, set
\[
\bar\kappa_j:=\max\bigl(\{m-1\}\cup\{k\in\mathcal K_j:G(k)\ge0\}\bigr).
\]
Strict monotonicity of $G$ and complementary slackness give the stated exclusions on $\mathcal K_j$. Below $m$, the move to $i'$ is infeasible if $e_{ji}\le e_{ji'}$. In the remaining case, $e_{ji}>e_{ji'}$ and $G(m)\ge0$. If $w^\star_{ji'\ell}>0$ for some $\ell<m$, complementary slackness and \eqref{eq:duration_value_bound}, together with $\pi^\star_{j\ell}\ge R_{j\ell}=R_{jm}-q(m-\ell)$, imply
\[
W_{ji'm}\ge W_{ji'\ell}+q(m-\ell)
=\pi^\star_{j\ell}+q(m-\ell)\ge R_{jm}.
\]
But the move to $i$ from level $m=e_{ji}$ arrives at level zero, where $\pi^\star_{i0}=R_{i0}$. Consequently,
\[
W_{jim}-R_{jm}
=-(c_{ji}+c_{iD}-c_{jD})
-\nu^\star(\taurep_{ji}+\taurep_{iD}-\taurep_{jD})-qm<0
\]
by the triangle inequalities. This gives $G(m)<0$, a contradiction. Hence $w^\star_{ji'k}=0$ for all $k<m$ as well, proving the global exclusions.

Finally, if $\mathcal K_j=\varnothing$, at least one move is infeasible at every level. Take $\bar\kappa_j=0$ when $e_{ji}\ge K$, and $\bar\kappa_j=K$ otherwise, in which case $e_{ji'}\ge K$. The boundary restrictions cover $k=0,K$.
\end{proof}

\end{APPENDICES}

\end{document}